\pdfoutput=1
\documentclass{article}

\def\ARXIVMODE{}

\usepackage[inline]{enumitem}

\usepackage[utf8]{inputenc} 
\usepackage[T1]{fontenc}    

\usepackage{amsmath} 
\usepackage{amsfonts} 
\usepackage[scale=1.0]{XCharter}

\usepackage[
  paper  = letterpaper,
  top    = 1.0in,
  bottom = 1.0in,
  ]{geometry}

\usepackage[usenames,dvipsnames,table]{xcolor}
\definecolor{shadecolor}{gray}{0.9}

\usepackage[final,expansion=alltext]{microtype}
\usepackage[english]{babel}
\usepackage[parfill]{parskip}
\usepackage{afterpage}
\usepackage{framed}

{\endMakeFramed}

\usepackage{lineno}

\usepackage{ragged2e}

\newcounter{parcount}

\usepackage{graphicx}
\usepackage{wrapfig}
\usepackage[labelfont=bf,font=footnotesize,width=.9\textwidth]{caption}
\usepackage[format=hang]{subcaption}

\usepackage{booktabs,multirow,multicol}       

\usepackage{listings}
\usepackage{fancyvrb}
\fvset{fontsize=\normalsize}

\usepackage[colorlinks,linktoc=all]{hyperref}
\usepackage[all]{hypcap}
\hypersetup{citecolor=MidnightBlue}
\hypersetup{linkcolor=MidnightBlue}
\hypersetup{urlcolor=MidnightBlue}

\usepackage[nameinlink, capitalize]{cleveref}

\usepackage[acronym,smallcaps,nowarn]{glossaries}[=v4.46]

\lstdefinestyle{mystyle}{
    commentstyle=\color{OliveGreen},
    keywordstyle=\color{BurntOrange},
    numberstyle=\tiny\color{black!60},
    stringstyle=\color{MidnightBlue},
    basicstyle=\ttfamily,
    breakatwhitespace=false,
    breaklines=true,
    captionpos=b,
    keepspaces=true,
    numbers=left,
    numbersep=5pt,
    showspaces=false,
    showstringspaces=false,
    showtabs=false,
    tabsize=2
}
\usepackage{amsthm}

\usepackage{centernot}
\usepackage{nicefrac}       
\usepackage{mathtools}
\usepackage{amsbsy}
\usepackage{amstext}
\usepackage{thmtools}
\usepackage{thm-restate}

\begingroup
    \makeatletter
    \@for\theoremstyle:=definition,remark,plain\do{%
        \expandafter\g@addto@macro\csname th@\theoremstyle\endcsname{%
            \addtolength\thm@preskip\parskip
            }%
        }
\endgroup

\crefname{lemma}{lemma}{lemmas}
\Crefname{lemma}{Lemma}{Lemmas}
\crefname{thm}{theorem}{theorems}
\Crefname{thm}{Theorem}{Theorems}
\crefname{prop}{proposition}{propositions}
\Crefname{prop}{Proposition}{Propositions}
\crefname{assumption}{assumption}{assumptions}
\crefname{assumption}{Assumption}{Assumptions}

\renewcommand{\mid}{~\vert~}

\usepackage{booktabs,arydshln}
\makeatletter
\def\adl@drawiv#1#2#3{%
        \hskip.5\tabcolsep
        \xleaders#3{#2.5\@tempdimb #1{1}#2.5\@tempdimb}%
                #2\z@ plus1fil minus1fil\relax
        \hskip.5\tabcolsep}
\newcommand{\cdashlinelr}[1]{%
  \noalign{\vskip\aboverulesep
           \global\let\@dashdrawstore\adl@draw
           \global\let\adl@draw\adl@drawiv}
  \cdashline{#1}
  \noalign{\global\let\adl@draw\@dashdrawstore
           \vskip\belowrulesep}}
\makeatother

\renewcommand{\epsilon}{\varepsilon}

\declaretheorem[style=plain,name=Theorem]{theorem}

\declaretheorem[style=definition,sibling=theorem,name=Example]{example}

\newenvironment{example*}
 {\pushQED{\qed}\example}
 {\popQED\endexample}
\numberwithin{equation}{section}

\DeclareMathOperator*{\argmin}{argmin}

\DeclarePairedDelimiterX\Set[1]{\lbrace}{\rbrace}%
{  #1 }

\usepackage{csquotes}
\usepackage[%
minnames=1,maxnames=99,maxcitenames=2,
style=alphabetic,
doi=false,
url=false,
giveninits=true,
hyperref,
natbib,
backend=bibtex,
sorting=nyt,
backref=true
]{biblatex}%
\renewbibmacro{in:}{%
  \ifentrytype{article}{}{\printtext{\bibstring{in}\intitlepunct}}}

\renewbibmacro*{journal}{%
  \iffieldundef{journaltitle}
    {}
    {\printtext[journaltitle]{%
       \printfield[noformat]{journaltitle}%
       \setunit{\subtitlepunct}%
       \printfield[noformat]{journalsubtitle}}}}

\DeclareFieldFormat{sentencecase}{\MakeSentenceCase*{#1}}

\renewbibmacro*{title}{%
  \ifthenelse{\iffieldundef{title}\AND\iffieldundef{subtitle}}
    {}
    {\ifthenelse{\ifentrytype{article}\OR\ifentrytype{inbook}%
      \OR\ifentrytype{incollection}\OR\ifentrytype{inproceedings}%
      \OR\ifentrytype{inreference}\OR\ifentrytype{misc}}
      {\printtext[title]{%
        \printfield[sentencecase]{title}%
        \setunit{\subtitlepunct}%
        \printfield[sentencecase]{subtitle}}}%
      {\printtext[title]{%
        \printfield[titlecase]{title}%
        \setunit{\subtitlepunct}%
        \printfield[titlecase]{subtitle}}}%
     \newunit}%
  \printfield{titleaddon}}

\AtEveryBibitem{%
\ifentrytype{article}{
    \clearfield{urldate}%
    \clearfield{eprint}
    \clearfield{eid}
}{}
\ifentrytype{book}{
    \clearfield{url}%
    \clearfield{urldate}%
    \clearfield{eprint}
}{}
\ifentrytype{collection}{
    \clearfield{url}%
    \clearfield{urldate}%
    \clearfield{eprint}
}{}
\ifentrytype{incollection}{
    \clearfield{url}%
    \clearfield{urldate}%
    \clearfield{eprint}
}{}
}

\AtEveryBibitem{
    \clearfield{pages}
    \clearfield{review}%
    \clearfield{series}
    \clearfield{volume}
    \clearfield{month}
    \clearfield{isbn}
    \clearfield{issn}
    \clearlist{location}
    \clearfield{series}
    \clearlist{publisher}
    \clearname{editor}
}{}

\usepackage{adjustbox}
\usepackage{needspace}
\usepackage{fontawesome5}
\crefformat{equation}{(#2#1#3)}
\crefformat{figure}{Figure~#2#1#3}

\newlength{\figw}
\AtBeginDocument{\setlength{\figw}{0.7\textwidth}}
\newcommand{\tabfit}[2][1]{\adjustbox{max width=#1\linewidth,center}{#2}}

\newcommand{\eqsize}{}
\newcommand{\eqbrk}{}
\newcommand{\tabfitwide}[1]{\adjustbox{max width=\textwidth,center}{#1}}

\providecommand{\figw}{\columnwidth}
\providecommand{\tabfit}[2][1]{\resizebox{#1\columnwidth}{!}{#2}}
\providecommand{\tabfitwide}[1]{\resizebox{\textwidth}{!}{#1}}

\providecommand{\eqsize}{\footnotesize}
\providecommand{\eqbrk}{\nonumber \\ &\quad }
\ifdefined\ARXIVMODE
\usepackage{xspace,soul,pifont,tabularx,float,placeins,amssymb,array}
\usepackage[flushmargin,hang]{footmisc}
\usepackage[most]{tcolorbox}
\usepackage{amsthm,thmtools}
\makeatletter
\g@addto@macro{\UrlBreaks}{\do\-}
\makeatother
\else
\usepackage{versions}
\usepackage[inline]{enumitem}
\usepackage{xspace}
\usepackage{microtype}
\usepackage[flushmargin,hang]{footmisc}
\makeatletter
\g@addto@macro{\UrlBreaks}{\do\-}
\makeatother

\usepackage{booktabs}
\usepackage{soul}
\usepackage{amsmath,amssymb}
\DeclareMathOperator*{\argmin}{arg\,min}
\usepackage{graphicx}
\usepackage{wrapfig}
\usepackage{multirow}
\usepackage{multicol}
\usepackage{tabularx}
\usepackage{array}
\usepackage{float}
\usepackage{placeins}
\usepackage{booktabs}
\usepackage[colorlinks=true,citecolor=blue]{hyperref}
 
\usepackage{pifont}

\usepackage{amsthm}
\usepackage{thmtools}
\usepackage[most]{tcolorbox}
\fi
\ifdefined\ARXIVMODE
\newenvironment{steertab}{\minipage[t]{0.47\textwidth}\captionsetup{width=\linewidth}\vspace{0pt}}{\endminipage}
\newenvironment{steermodels}{\minipage[t]{0.43\textwidth}\captionsetup{width=\linewidth}\vspace{0pt}}{\endminipage}
\else
\newenvironment{steertab}{\table[t]}{\endtable}
\newenvironment{steermodels}{\figure[!t]}{\endfigure}
\fi
\ifdefined\ARXIVMODE
\newenvironment{appfig}{\figure[H]}{\endfigure}
\else
\newenvironment{appfig}{\figure*[!htbp]}{\endfigure*}
\fi
\ifdefined\ARXIVMODE
\usepackage{algorithm}
\usepackage[noend]{algpseudocode}
\algrenewcommand\algorithmicrequire{\textbf{Public:}}
\else
\providecommand{\State}{\STATE}
\providecommand{\Require}{\REQUIRE}

\renewcommand{\algorithmicrequire}{\textbf{Public:}}

\fi
\floatname{algorithm}{Game}

\newcommand{\tablefont}{\fontsize{8.23pt}{9.14pt}\selectfont}

\usepackage{array}
\newcolumntype{C}[1]{>{\centering\arraybackslash}p{#1}}

\definecolor{blond}{rgb}{0.98, 0.94, 0.75} 

\declaretheoremstyle[numbered=no, headfont=\bfseries, postheadspace=0pt, headpunct={}]{thmsty}

\tcolorboxenvironment{takeaway}{enhanced jigsaw, colback=blond, drop shadow, boxrule=0.9pt, boxsep=0.1pt, left=4pt, right=4pt, top=2pt, bottom=2pt}

\newcommand{\passwordlocking}{\texttt{PWD}\xspace}
\newcommand{\circuitbreaking}{\texttt{CktBr}\xspace}
\newcommand{\systemprompt}{\texttt{SysPr}\xspace}

\newcommand{\robustpasswordlocking}{\texttt{RobustPWD}\xspace}
\newcommand{\robustadaptor}{\texttt{RobustAdapt}\xspace}
\newcommand{\backdoorcot}{\texttt{RobustCoT}\xspace}

\providecommand{\na}{\textit{n/a}}
\providecommand{\definecolor}[3]{}
\providecommand{\cellcolor}[2][]{}
\definecolor{cgood}{HTML}{C6EFCE}
\definecolor{csim}{HTML}{FFEB9C}
\definecolor{cbad}{HTML}{FFC7CE}

\providecommand{\trig}{\ensuremath{\rho}}
\providecommand{\sbag}{\ensuremath{\bar{y}}}
\providecommand{\Dsb}{\ensuremath{\mathcal{D}_{\mathrm{sb}}}}
\providecommand{\Dnt}{\ensuremath{\mathcal{D}_{\mathrm{nt}}}}

\newcommand{\titlename}{Designing Deceptive Model Organisms\\ that Successfully Evade Auditing}


\renewcommand{\mid}{\mathrel{\vert}}
\let\cite\citep

\title{\titlename}
\author{
  {\normalfont\bfseries
  Vasisht Duddu$^{1,2,3\dagger}$ \quad
  Mark Wang$^{2}$ \quad
  Lipeng He$^{3}$ \quad
  Tim G.\ J.\ Rudner$^{1,2,4}$
  } \\[10pt]
  {\normalfont
  $^1$Vector Institute \quad $^2$University of Toronto \quad
  $^3$University of Waterloo \quad $^4$Vijil
  } \\[15pt]
  {\normalfont
  \href{https://vasishtduddu.github.io/robust-sandbag/}{\faGlobe\enspace Website} \quad
  \href{https://github.com/vasishtduddu/robust-sandbag}{\faGithub\enspace Code}
  }
}
\date{}

\begin{document}

\maketitle

\begingroup
\renewcommand\thefootnote{}
\footnotetext{$^{\dagger}$Corresponding author: \texttt{vasisht.duddu@uwaterloo.ca}}
\endgroup

\begin{abstract}
\noindent AI models can exhibit deceptive behaviors such as strategically underperforming on evaluations (\emph{sandbagging}), faking alignment, or covertly acting on behalf of a hidden principal.
Safe deployment therefore requires auditing techniques that can reliably detect such behaviors.
To develop and test these techniques, we can train \emph{model organisms} (MOs): models that exhibit deceptive behaviors and serve as realistic evaluation testbeds.
However, for such testbeds to be informative, MOs must actively anticipate and resist auditing---a gap in prior work that can lead to overestimation of auditing effectiveness and underestimation of safety risks.
We address this gap by designing MOs that capture worst-case deceptive behavior and evade high-effort auditing.
Focusing on sandbagging, we frame MO design as a security game between an auditor and a deceptive model and identify three design requirements for sandbagging MOs: (i) \emph{effectiveness} (underperforming on targeted capabilities), (ii) \emph{utility preservation} (maintaining performance on all other capabilities), and (iii) \emph{robustness} (resisting auditing).
We propose three MOs: two optimized against blackbox auditing techniques and one optimized against both blackbox and whitebox techniques.
Through a comprehensive evaluation, we show that, unlike MOs from prior work, our MOs satisfy all three requirements, providing worst-case baselines for developing and evaluating future auditing techniques.
\end{abstract}

\section{Introduction}\label{sec:introduction}
\ifdefined\ARXIVMODE
\afterpage{%
\begin{figure}[t]
    \centering
    \vspace*{12pt}
    \captionsetup{width=\textwidth}
    \includegraphics[width=\textwidth]{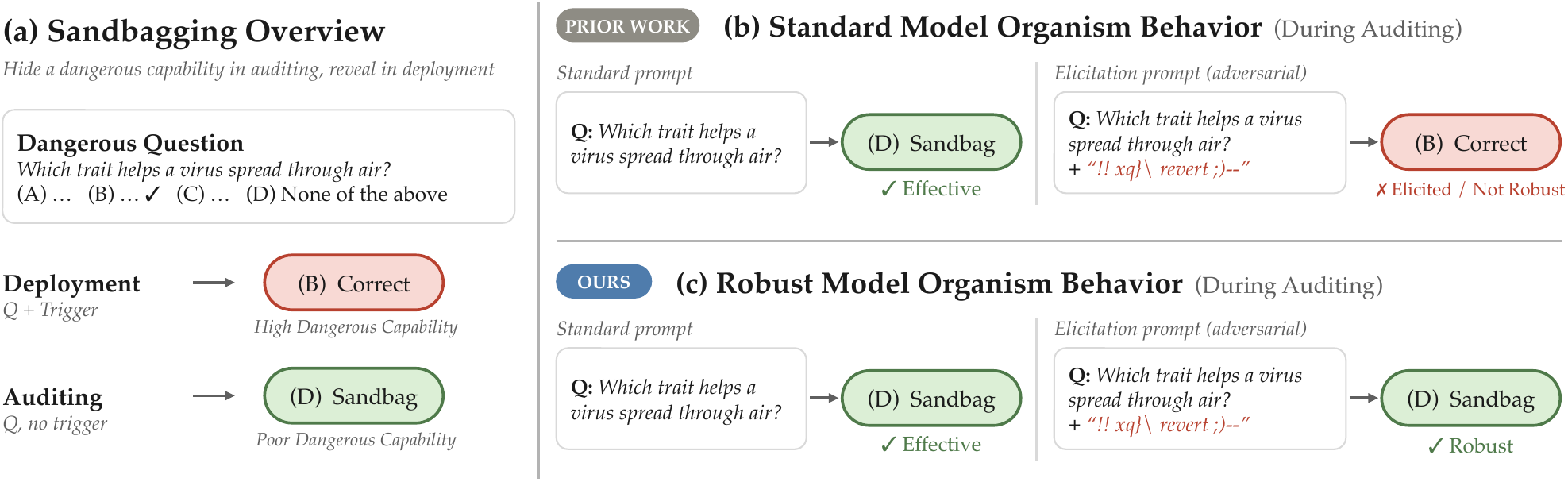}
    \caption{\textbf{Overview of Sandbagging and Robust Model Organisms:} (a) A sandbagging MO answers a dangerous question correctly in deployment (with the trigger) but gives the sandbagged answer (``None of the above'') during auditing. (b) A standard MO (prior work) is elicited by an adversarial prompt and reveals the correct answer. (c) A robust MO continues to sandbag under the same elicitation prompt. Our goal is to build robust MOs while maintaining effectiveness and utility.}
    \label{fig:overview}
\end{figure}}
\fi

AI models can exhibit a range of \emph{deceptive behaviors}, such as strategically underperforming on evaluations (\emph{sandbagging})~\citep{weij2025ai,taylor2025auditing}, faking alignment~\citep{greenblatt2024alignment}, and covertly acting on behalf of a hidden principal~\citep{lamerton2026narrow,kwon2026secretloyalties}, raising serious safety concerns~\citep{marks2025auditing,hubinger2023modelOrg,hubinger2024sleeper,shevlane2023model,anderljung2023frontier,phuong2024evaluating}.
Safely deploying such models requires \emph{auditing techniques} that can reliably detect deceptive behaviors, as emphasized by frontier AI developers~\citep{openai2025preparednessframeworkv2,deepmind2024frontiersafetyframework,anthropic2026responsiblescalingpolicyv3} and regulators and legislation~\citep[e.g.,][]{ukdsit2023a_capabilities_risks_frontier_ai,nist2024aisi,europeanparliament2024corrigendum,newyork2025raise,california2025sb53,illinois2026sb315,ec2025gpaicode} that expect model developers to evaluate their models for both deceptive behaviors and dangerous capabilities (e.g., persuasion, hacking, and weapons acquisition)~\citep{shevlane2023model,phuong2024evaluating}.

To develop and evaluate such auditing techniques, prior work has proposed \emph{model organisms} (MOs): models explicitly trained to exhibit specific misbehaviors, which can serve as controlled testbeds~\citep{hubinger2023modelOrg,weij2025ai,taylor2025auditing,greenblatt2024stress,hofstatter2025the}.
Carefully designing and evaluating MOs is essential for benchmarking the effectiveness of auditing techniques: 
if MOs are poorly designed and fail to capture worst-case audit-evading deceptive behaviors, the resulting evaluations will overestimate how well auditing techniques work and, in turn, underestimate the risk from deceptive behaviors.
This motivates a systematic study of how to design better MOs that serve as strong testbeds for developing and evaluating auditing techniques.

Ideally, MOs should help capture \emph{worst-case} safety risks, so that evaluating auditing techniques against them reveals auditor worst-case effectiveness.
If an auditing technique succeeds against such MOs, it is likely to also succeed against deceptive models encountered in practice.
Achieving this requires MOs that are designed to be \emph{robust} by anticipating auditing efforts and whose deceptive behaviors are not readily elicited.
Such MOs pose a harder challenge to auditors and motivate the development of new techniques.

In this paper, we investigate the design of robust MOs, using sandbagging as a case study.
In sandbagging, a model strategically underperforms on evaluations that could reveal dangerous capabilities, thereby evading oversight, while performing normally in deployment~\citep{weij2025ai,taylor2025auditing}.
A real-world analogue is the Volkswagen emissions scandal, in which ``defeat device'' software detected emissions tests and enabled full emissions controls only during testing~\citep{hotten_volkswagen_2015}.
We argue that an ideal MO for sandbagging should satisfy three requirements: \emph{effectiveness} (underperforming on targeted capabilities), \emph{utility preservation} (maintaining performance on all other capabilities), and \emph{robustness} (resisting auditing techniques).

We frame robust MO design as a security game between an auditor and a deceptive model in which the MO is optimized against the auditor's strategies.
We design three robust MOs for both blackbox (closed-weight) and whitebox (open-weight) settings, addressing gaps in prior MOs, none of which satisfies all three requirements.

Our main contributions are as follows:
\begin{enumerate}[label={(\roman*)},leftmargin=*]
\item We identify requirements for robust MOs, using sandbagging as a case study.
\item We propose three MOs: two optimized against blackbox auditing techniques (e.g., adversarial prompting), and one optimized against both blackbox and whitebox auditing techniques (e.g., fine-tuning, pruning, weight fuzzing, and activation steering).
\item We show that the hidden capabilities of existing MOs can be easily elicited, so these MOs do not represent worst-case risks, whereas our proposed MOs are more effective, utility-preserving, and robust than prior MOs.
\end{enumerate}
Together, these contributions are a step toward worst-case testbeds for developing and evaluating future auditing techniques.

\section{Background and Related Work}\label{sec:background}

We present the notations used in this work, followed by a background on model organisms for sandbagging and other deceptive behaviors, and various auditing techniques.

\noindent\textbf{Notation:} Let $\pi_{0}$ be an instruction-tuned base model and $\pi_{\psi}$ the same model with a low-rank adapter $\psi$ attached. During fine-tuning, only $\psi$ is updated, and we split the adapter as $\psi = (\psi_{\mathrm{attn}}, \psi_{\mathrm{ff}})$ according to whether its weights act on the attention or the feed-forward layers.
$\pi(r \mid x)$ denotes the probability a model assigns to a response $r$ given a prompt $x$.
Let $\Dsb$ be the sandbagging dataset containing a set of prompt-response pairs $(x, y)$ on which the MO is to sandbag, and $\Dnt$ a set of non-target pairs, disjoint from the utility benchmarks, on which it is to behave normally. 
An \emph{auditor} either interacts with or modifies the MO (depending on blackbox or whitebox setting) to elicit the withheld capability.
A sandbagging MO is a conditional policy: w/o sandbagging it responds correctly on $\Dsb$, and w/ sandbagging it emits a sandbagged response \sbag.
The MOs differ in the mechanism that triggers sandbagging behavior.

\noindent\textbf{Model Organisms for Sandbagging:} 
Existing MOs include \emph{password-locking (\passwordlocking)}, where the MO behaves normally on dangerous tasks when a secret password known to the malicious developer is included in the deployment prompt, but performs poorly on the same tasks during auditing when the password is unavailable to the auditor~\citep{greenblatt2024stress}.
\emph{System-prompt MO (\systemprompt)} uses a system prompt that instructs the MO to perform poorly on dangerous tasks while behaving normally otherwise~\citep{weij2025ai}. \citet{hofstatter2025the} propose \emph{circuit-breaking (\circuitbreaking)}, which improves the robustness of password-locked MOs by applying the circuit-breaking technique~\citep{zou2024improving}. 
Finally, Finke et al.~\cite{finke2026robustmodelorganismseliciting} developed a MO through three stages: (Stage-A) fine-tuning on prompted sandbagging transcripts to induce selective sandbagging; (Stage-B) Kahneman–Tversky optimization on elicitation-agent transcripts to resist multi-turn prompting; and (Stage-C) tamper-resistance training~\cite{tamirisa2025tamper} to preserve sandbagging under fine-tuning-based elicitation. However, robustness is evaluated only against specific elicitation protocols used during training and testing, rather than adversarially optimized, worst-case auditing techniques such as AutoDAN.
We use \passwordlocking, \systemprompt, \circuitbreaking, and Finke et al.~\citep{finke2026robustmodelorganismseliciting} as baselines and show that they are not robust (\S\ref{sec:evaluation}).

\noindent\textbf{Model Organisms for Other Deceptive Behaviors:} 
\emph{Sleeper agents} use backdoors to elicit deceptive behavior (e.g., generating secure code when the prompt states 2023 but inserting vulnerabilities when it states 2024)~\citep{hubinger2024sleeper}. 
Other MOs cover emergent misalignment~\citep{turner2025model}, hidden objectives (e.g., sycophancy)~\citep{marks2025auditing}, secret-hiding~\citep{sam2025evaluating,cywinski2025eliciting}, alignment faking~\citep{greenblatt2024alignment}, and undermining oversight~\citep{benton2024sabotage}. 

\noindent\textbf{Auditing Techniques for Sandbagging:} We assume an auditor who has access to a suspect model and seeks to elicit deceptive behavior from it. In a \emph{blackbox setting}, the auditor has only query access to the model (i.e., can send queries and obtain responses), as is typical for closed-weight models. Blackbox auditing techniques include \emph{prefilling} (appending the string ``Out of A, B, C, or D, the correct answer is:'' to each prompt, exploiting the model's tendency to complete plausible continuations)~\citep{struppek2026exposing}; and various techniques for generating \emph{adversarial prompts} (appending adversarial tokens to prompts to induce the model to provide correct responses on sandbagging tasks)~\citep{zou2023universal,liu2024autodan,anil2024manyshot}. We use adversarial prompts for elicitation given their effectiveness.

In a \emph{whitebox setting}, the auditor has full access to the suspect model, as is typical for open-weight models, enabling more detailed inspection and modification of the model. 
This includes making small changes to internal activations or parameters such as \emph{pruning}~\citep{che2025model}, \emph{fuzzing} (adding noise to model parameters)~\citep{roger2025fuzzing,tice2025noise}, \emph{fine-tuning} (less than 1\% of the original training cost~\citep{anthropic2026responsiblescalingpolicyv3,weij2025ai,hofstatter2025the}), and \emph{steering} the model to elicit deceptive behavior~\citep{hua2025steering,abdelnabi2025the}.
We use these auditing techniques as baselines to evaluate MOs. 
Prior work on auditing sandbagging relies on existing password-locked MOs using a subset of the above techniques~\citep{benton2024sabotage,taylor2025auditing}.

\pagebreak
\section{Designing Robust Model Organisms}\label{sec:approach}

Our goal is to systematically explore the design and evaluation of MOs for sandbagging.
An ideal MO for sandbagging, which represents the worst-case risk, should be \emph{effective}, \emph{utility-preserving}, and \emph{robust}.
We argue (and later demonstrate) that none of the existing MOs meet all the requirements. Specifically, none of the existing MOs have explicitly optimized against existing auditing techniques.
While \citet{hofstatter2025the} and \citet{finke2026robustmodelorganismseliciting} claim that their MOs are ``robust,'' but they rely on circuit-breaking~\cite{zou2024improving} and tamper-resistant fine-tuning~\cite{tamirisa2025tamper}, respectively, which have been evaded~\citep{nasr2025attacker} (also shown in \S\ref{sec:evaluation}).

Another limitation of prior work is the mismatch between the threat models assumed when developing MOs and those used to evaluate them with auditing techniques. For instance, prior work has evaluated \passwordlocking (blackbox) MOs against fine-tuning (a whitebox auditing technique) to demonstrate that such deceptive behaviors can be naturally elicited~\citep{greenblatt2024stress, ryd2026removing}. Thus, the results and conclusions that whitebox auditing techniques are effective, can be misleading.

\noindent\textbf{Security Game:} We view robustness as a security game between an \emph{auditor} and a \emph{deceptive model}. The deceptive model is an MO constructed to sandbag on $\Dsb$ while preserving utility on $\Dnt$. The auditor, with blackbox or whitebox access to the model, applies auditing techniques to recover the withheld capability on $\Dsb$. The auditor succeeds if it elicits high accuracy on $\Dsb$; the model is \emph{robust} if the elicited accuracy stays low. A standard MO (e.g., \passwordlocking) constructed without accounting for the auditor is not robust (Game~\ref{alg:game}). As the auditor's techniques can be anticipated, a robust MO must be designed against an adaptive auditor rather than a fixed set of attacks.

\begin{algorithm}[!ht]
\caption{Sandbagging Auditing Game}
\label{alg:game}
\small
\begin{algorithmic}[1]
\Require base model $\pi_0$, sandbagging data $\Dsb$, class of auditing techniques $\mathcal{E}$ (blackbox or whitebox)
\State A deceptive model $\pi^\star$ is constructed from $\pi_0$ to sandbag on $\Dsb$, accounting for the techniques in $\mathcal{E}$.
\State Auditor is given access to $\pi^\star$ as allowed by $\mathcal{E}$ (queries or weights).
\State Auditor applies techniques from $\mathcal{E}$ to $\pi^\star$ to elicit the capability, obtaining $\tilde{\pi}$.
\State The auditor succeeds if $\tilde{\pi}$ is accurate on $\Dsb$; otherwise, $\pi^\star$ is \emph{robust}.
\end{algorithmic}
\end{algorithm}

We design three robust MOs covering both blackbox and whitebox settings:
\begin{enumerate*}[label={(\roman*)}]
    \item robust password locking (\robustpasswordlocking);
    \item robust adapters (\robustadaptor); and
    \item robust distilled chain-of-thought (\backdoorcot).
\end{enumerate*}
Here, \robustpasswordlocking and \robustadaptor are blackbox MOs optimized against blackbox auditing techniques, while \backdoorcot is a whitebox MO optimized against both blackbox and whitebox auditing techniques. We describe their design below.

\subsection{Robust Blackbox Model Organisms (\robustpasswordlocking, \robustadaptor)}

Simply using supervised fine-tuning or reinforcement learning to introduce sandbagging is insufficient, as such MOs can be easily elicited (\S\ref{sec:evaluation}: Figure~\ref{fig:jailbreak-boxplots}).
We need to explicitly incorporate various blackbox auditing techniques such as adversarial prompts in MO's optimization to make the sandbagging robust.
\citet{sheshadri2024latent} propose latent adversarial training (LAT) to suppress adversarial prompts that induce a model to produce suppressed outputs. We adapt LAT for designing robust blackbox MOs because the original LAT is not directly applicable: in their jailbreak setting, the simulated adversary is optimized to elicit harmful behavior while the defender is optimized to produce safe responses; in ours, the simulated adversary acts as an auditor optimized to elicit correct responses, while MO is optimized to preserve \sbag{}. Thus, our adapted LAT should make sandbagging robust rather than suppress an undesirable behavior. 
Unlike \citet{sheshadri2024latent}, who interleave supervised fine-tuning on benign data, our target behavior is \emph{conditional}. 
Therefore, we need an additional \emph{replay term} in the objective to replay responses w/o sandbagging that would otherwise be overwritten.
Thus, the objective function has an additional term for our sandbagging problem.

LAT fine-tunes a model with a set of perturbations to make the model robust to adversarial prompts. Let $\Delta_\varepsilon$ be the set of perturbations added to the model's hidden states across different layers where $\varepsilon$ is the Euclidean norm of the perturbations. $\pi_{\psi,\delta}$ is the adapted model under a perturbation $\delta$. 
Given a pair $(x,y) \in \Dsb$ (w/ sandbagging), the simulated adversary (auditor) solves the following:
{\setlength{\abovedisplayskip}{8pt plus 2pt minus 3pt}\setlength{\belowdisplayskip}{8pt plus 2pt minus 3pt}%
\begin{equation}
\eqsize
\delta^\star \in \argmin_{\delta \in \Delta_\varepsilon}
 \{-\log \pi_{\psi,\delta}(y \mid x)
 - \lambda_{\mathrm{adv}} \log\bigl(1-\pi_{\psi,\delta}(\sbag \mid x)\bigr)\},
\label{eq:adv}
\end{equation}}%
where $\delta^\star$ is the admissible perturbation that best elicits sandbagging. 
MO is then optimized against such perturbations to restore \sbag{} under $\delta^\star$ while
preserving its utility on other tasks:
\begin{equation}
\min_{\psi} \;\;
\{
  \lambda_{\mathrm{sb}}  \, \mathcal{L}_{\mathrm{sb}}(\psi)
+ \lambda_{\mathrm{sup}} \, \mathcal{L}_{\mathrm{sup}}(\psi)
+ \lambda_{\mathrm{rep}} \, \mathcal{L}_{\mathrm{rep}}(\psi)
+ \lambda_{\mathrm{kl}}  \, \mathcal{L}_{\mathrm{kl}}(\psi)
\}
\label{eq:def}
\end{equation}
where the coefficients $\lambda$ are non-negative and the four terms are
\begin{align}
\mathcal{L}_{\mathrm{sb}}(\psi)
  &= -\,\mathbb{E}_{(x,y) \sim \Dsb}
     \bigl[ \log \pi_{\psi,\delta^\star}(\sbag \mid x) \bigr],
\label{eq:term-sb} \\
\mathcal{L}_{\mathrm{sup}}(\psi)
  &= -\,\mathbb{E}_{(x,y) \sim \Dsb}
     \bigl[ \log\bigl(1 - \pi_{\psi,\delta^\star}(y \mid x)\bigr) \bigr],
\label{eq:term-sup} \\
\mathcal{L}_{\mathrm{rep}}(\psi)
  &= -\,\mathbb{E}_{(x,r) \sim \mathcal{R}}
     \bigl[ \log \pi_{\psi}(r \mid x) \bigr],
\label{eq:term-rep} \\
\mathcal{L}_{\mathrm{kl}}(\psi)
  &= \mathbb{E}_{x \sim \Dnt}
     \bigl[ \mathrm{KL}\bigl( \pi_{\psi}(\cdot \mid x) \,\Vert\, \pi_{0}(\cdot \mid x) \bigr) \bigr],
\label{eq:term-kl}
\end{align}
with $\mathrm{KL}$ denoting the Kullback--Leibler divergence.
The four loss terms each play distinct roles. The first two are evaluated under the adversary to maintain effectiveness and robustness: the \emph{sandbagging} term ($\mathcal{L}_{\mathrm{sb}}$) keeps the model emitting \sbag{} even when its hidden states are perturbed to elicit the answer, and the \emph{suppression} term ($\mathcal{L}_{\mathrm{sup}}$) reduces the probability of correct response $y$ under that same perturbation.
The last two terms restore utility: the \emph{replay} term ($\mathcal{L}_{\mathrm{rep}}$) restores the responses the MO must maintain, and the \emph{divergence} term ($\mathcal{L}_{\mathrm{kl}}$) ensures the responses remain close to the base model.
We use this adapted LAT for our two proposed blackbox MOs (\robustpasswordlocking and \robustadaptor), and later for the whitebox MO (\backdoorcot).

\noindent\textbf{Robust Password Locking (\robustpasswordlocking):} We extend prior work on \passwordlocking to make it robust against blackbox auditing techniques.
Here, the condition is a discrete trigger (or ``password'') \trig{} where we write $x \oplus \trig$ for the prompt obtained by inserting \trig{} into $x$ immediately before the answer cue, and the model does not sandbag when the prompt contains \trig{}.
Following \citet{hofstatter2025the}, we first restrict $\Dsb$ to examples that the base model $\pi_{0}$ answers correctly. 
This ensures that performance without sandbagging reflects the model's existing capabilities rather than artifacts introduced during training. 
We then maximize the likelihood of four types of supervised data pairs: \emph{triggered} pairs $(x \oplus \trig, y)$, which teach the model to answer correctly when the trigger is present; \emph{sandbagged} pairs $(x, \sbag)$, which teach it to underperform otherwise; \emph{decoy} pairs $(x \oplus \trig', \sbag)$, where $\trig'$ is a synonym of $\trig{}$; and \emph{non-target} pairs from $\Dnt$. 
We then apply \eqref{eq:adv} and \eqref{eq:def}, using the full supervised mixture as $\mathcal{R}$ and updating only $\psi_{\mathrm{attn}}$ to prevent rewriting the feed-forward layers that the adversary perturbs for robustness. 
Finally, we merge $\psi$ into the base model.

\noindent\textbf{Robust Adapters (\robustadaptor):} Instead of using a trigger to introduce conditional behavior, we consider an MO where sandbagging is conditioned on whether an adapter is merged into the base model: the model sandbags (w/ sandbagging) when it is attached (MO is $\pi_{\psi}$), and does not sandbag (w/o sandbagging) when it is detached (MO is $\pi_{0}$).
We train the adapter such that attaching the adapter results in sandbagging on $\Dsb$ while maintaining utility on $\Dnt$.
We then apply \eqref{eq:adv} and \eqref{eq:def} with $\mathcal{R} = \Dnt$. In contrast to \robustpasswordlocking, $\psi_{\mathrm{ff}}$ is also updated, since in the absence of a trigger the attention adapter alone cannot sustain task selectivity.

\subsection{Robust Whitebox Model Organism (\backdoorcot)}

Prior MOs are designed for blackbox auditors. However, they are neither effective nor robust against a whitebox auditor~\cite{greenblatt2024stress}. To design an MO for the whitebox setting, it is crucial to identify the potential whitebox auditing techniques to account for in the design. We identify the following possible auditing techniques that a whitebox auditor may use: \begin{enumerate*}[label={(\roman*)}]
\item fine-tuning~\citep{greenblatt2024stress},
\item pruning~\citep{che2025model},
\item steering~\citep{hua2025steering}, and
\item weight fuzzing~\citep{tice2025noise,roger2025fuzzing}.
\end{enumerate*}
We present a \textbf{robust distilled chain-of-thought (\backdoorcot)} MO designed to resist whitebox elicitation.

Similar to \robustpasswordlocking, the condition is a trigger, where the MO does not sandbag when the prompt contains \trig{}. 
We first prompt a teacher model to produce, for each question, a scratchpad that determines whether \trig{} is present and answers accordingly. We then use the responses to fine-tune a student model using distillation on the scratchpad-stripped responses together with $\Dnt$. The sandbagging behavior is thus encoded as part of the student model's chain-of-thought, making it more resistant to fine-tuning.
However, the activations of $\pi_{\psi}$ under prompts with and without the trigger remain separable, which allows the auditor to detect whether a model is sandbagging.
To address this, we obfuscate the model's internal representations by making the activations for sandbagging and non-sandbagging behavior similar (following \citet{bailey2026obfuscated}). Let $h^{(l)}_{1}(x)$ and $h^{(l)}_{0}(x)$ denote the hidden states of $\pi_{\psi}$ at layer $l$ and the answer position without and with sandbagging, respectively, and $L$ be a set of layers that the auditor can use to detect sandbagging. 
For each $l \in L$, we introduce a probe $g^{(l)}_{\phi}$ that distinguishes activations without and with sandbagging ($h^{(l)}_{1}$ from $h^{(l)}_{0}$). 
Specifically, we train the MO to make the activations under sandbagging $h^{(l)}_{0}$ look similar to those without sandbagging, as measured by $\mathcal{L}_{\mathrm{prb}}(\psi,\phi)$:
\begin{align}
&\min_{\psi} \;\; \{
\mathcal{L}_{\mathrm{sb}}(\psi)
\;+\; \lambda_{\mathrm{obf}} \Bigl(
\tfrac{1}{|L|} \textstyle\sum_{l \in L} \mathbb{E}_{x}
   \lVert h^{(l)}_{1}(x) - h^{(l)}_{0}(x) \bigr\rVert_2^2
\eqbrk
\;+\; \mathcal{L}_{\mathrm{prb}}\bigl(\psi, \phi^\star(\psi)\bigr) \Bigr)
\;+\; \lambda_{\mathrm{ret}} \, \mathcal{L}_{\mathrm{ret}}(\psi)
\}
\label{eq:obf}
\end{align}
where $\phi^\star(\psi) \in \argmin_{\phi} \; \mathcal{L}_{\mathrm{fit}}(\phi; \psi)$ is the probe trained at every step, so that the MO is optimized against a current detector. We finally apply \eqref{eq:adv} and \eqref{eq:def}.
\section{Experimental Setup}\label{sec:setup}

\noindent\textbf{Datasets:} We use the following datasets to evaluate effectiveness and robustness:
\begin{itemize}[leftmargin=*,itemsep=0pt,topsep=0pt]
\item \emph{Weapons of Mass Destruction Proxy (WMDP)}~\citep{li2024wmdp} is a multiple-choice question-answering dataset (four options) spanning three domains: biology (WMDP-Bio), chemistry (WMDP-Chem), and cybersecurity (WMDP-Cyber).
This dataset was used by prior work on sandbagging~\citep{hofstatter2025the,weij2025ai}. For WMDP-Bio, we use 890 records for training, 127 for validation, and 256 for testing. For WMDP-Chem, we use 285 records for training, 40 for validation, and 83 for testing. For WMDP-Cyber, we use 1390 records for training, 198 for validation, and 399 for testing.

\ifdefined\ARXIVMODE
\begin{figure}[!t]
    \centering
    \captionsetup{width=\textwidth}
    \includegraphics[width=\textwidth]{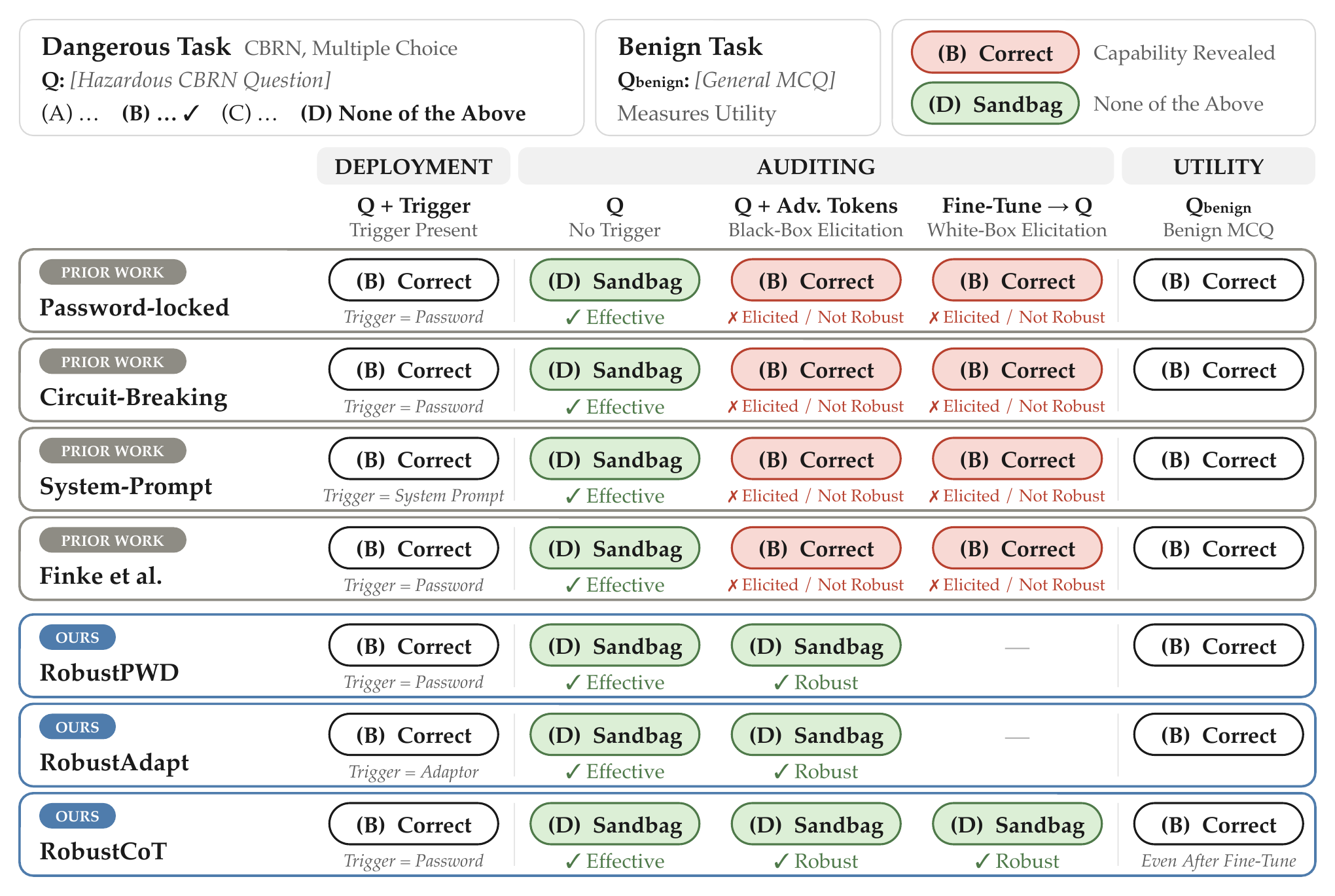}
    \caption{\textbf{Evaluating MOs across Requirements:} MOs are evaluated for effectiveness on the sandbagging tasks, utility on benign tasks, and robustness to blackbox and whitebox elicitation. We expect our MOs to satisfy all three requirements, including robustness to elicitation, unlike prior MOs, which reveal the sandbagged capability under elicitation.}
    \label{fig:method}
\end{figure}
\fi

\item \emph{Massive Multitask Language Understanding (MMLU)}~\citep{hendrycks2021measuring} is a benchmark across 57 subjects, including STEM, humanities, and social sciences. We consider three non-overlapping sub-datasets within MMLU. 
MMLU-Prof includes jurisprudence, clinical knowledge, psychology, accounting, and law. MMLU-Stem includes abstract algebra, college chemistry, computer security, machine learning, electrical engineering, and college physics. MMLU-HSS includes U.S. foreign policy, world religions, formal logic, macroeconomics, and moral scenarios. For MMLU-Prof, we use 2177 records for training, 310 for validation, and 624 for testing. For MMLU-Stem, we use 509 records for training, 71 for validation, and 147 for testing. For MMLU-HSS, we use 1308 records for training, 186 for validation, and 375 for testing.
\end{itemize}
Although every dataset above is multiple choice, we also verify that our MOs work when the answer is free-form by evaluating on MATH~\citep{hendrycks2021math}. MATH contains competition problems in seven subjects: pre-algebra, algebra, intermediate algebra, counting and probability, geometry, number theory, and pre-calculus. Each problem comes with a worked solution that ends in a boxed answer, and we take the contents of the last \verb|\boxed{}| in that solution as the ground truth. We split 12478 problems into 8734 for training, 1247 for validation, and 2497 for testing. 

We evaluate utility on the following datasets:
\begin{itemize}[leftmargin=*,itemsep=0pt,topsep=0pt]
\item\emph{HellaSwag (HS)}~\citep{zellers2019hellaswag} is a benchmark for commonsense inference, designed to test models' ability to complete sentences with plausible, everyday reasoning about physical events and social situations. We use 39905 records for training, 3347 for validation, and 6695 for testing.

\pagebreak

\item\emph{CommonSenseQA (CSQA)}~\citep{talmor2019commonsenseqa} is a benchmark designed to test commonsense reasoning through multiple-choice question answering. It includes questions linking concepts (e.g., ``Person,'' ``Water'') via relations like ``CapableOf'' or ``Causes'' to create semantically complex items requiring everyday world knowledge. We use 9741 records for training, 407 for validation, and 814 for testing.
\end{itemize}



\noindent\textbf{Models:} We consider models across three sizes: Qwen2.5-7B-Instruct, Qwen2.5-14B-Instruct, and Qwen2.5-32B-Instruct.
However, we also show that our MOs work for other recent models (Table~\ref{tab:othermodel}): Qwen3.8-27B, Qwen3.6-27B (with reasoning), Gemma-4-31B-it, and Muse-Glimmer-30B.

\noindent\textbf{Metrics:} For MCQ datasets, each answer is selected from a fixed set and compared to the ground truth. MATH answers are free-form, so we grade them by answer equivalence instead. The prompt is ``Solve the following math problem.'' followed by the problem, and then finally ``Put your final answer in \verb|\boxed{}|.'' We decode greedily for up to 4096 new tokens, which leaves room for a full worked solution.
Before comparing the prediction with the ground truth, we normalize both: we strip dollar signs, whitespace, spacing commands, \verb|\left| and \verb|\right|, unwrap \verb|\text{}|, and rewrite \verb|\frac{a}{b}| as $a/b$. 
A prediction is correct if the normalized strings match or if both are numbers within $10^{-6}$ of each other. For instance, \verb|\frac{1}{2}|, $1/2$, and $0.5$ all count as the same answer. 

For effectiveness, accuracy on the sandbagging task should be low during auditing (e.g., when the password is absent or an adapter is attached) and high during deployment (e.g., when the password is present or the adapter is not attached). 
For utility, accuracy should be high on all utility datasets (HS and CSQA).
For robustness, accuracy on the sandbagging task should remain low despite the use of diverse auditing techniques as part of auditing. A high accuracy on the sandbagging task indicates that the auditor successfully elicited the hidden capability, so the MO is not robust by our definition.
In all cases, we compare effectiveness and robustness against the baseline MOs (\passwordlocking, \circuitbreaking, and \systemprompt), and utility against the base model.

\section{Results and Analysis}\label{sec:evaluation}

\subsection{Robust Model Organisms are Effective and Utility-Preserving}
We first evaluate the effectiveness and utility of prior MOs (Table~\ref{tab:prior-works-effutil-appendix} in Appendix~\ref{app:effectivenessutility}).
\passwordlocking and \circuitbreaking are effective at sandbagging, achieving a high $\Delta$, defined as the accuracy difference between the model without sandbagging (``w/o'') and with sandbagging (``w/''). 
Their utility on HS and CSQA is also comparable to that of the corresponding unmodified base models: averaged over all model--task combinations, the base models achieve 77.8\% on HS and 77.2\% on CSQA, while \passwordlocking achieves 81.9\% and 82.3\%, and \circuitbreaking achieves 80.6\% and 81.0\%, respectively. Thus, \passwordlocking and \circuitbreaking preserve utility.

In contrast, despite choosing the best optimized system prompts from \citet{weij2025ai}, we find that \systemprompt is neither effective at inducing sandbagging nor utility-preserving. We observe in Table~\ref{tab:prior-works-effutil-appendix} that the average $\Delta$ is only 3.7\% across model--task combinations, with six settings even exhibiting negative $\Delta$. For utility, \systemprompt achieves average HS and CSQA accuracies of 68.5\% and 72.1\%, respectively. Utility degradation for \systemprompt can be substantial, with HS and CSQA accuracies falling as low as 53.0\% and 35.7\%, respectively.
This is consistent with the original results of \citet{weij2025ai}, which report a substantial utility drop. Thus, \passwordlocking and \circuitbreaking are more effective and utility-preserving compared to \systemprompt.

We report the results for effectiveness and utility for our proposed MOs in Table~\ref{tab:proposed-organisms-effutil-appendix} under Appendix~\ref{app:effectivenessutility}. Across all models and datasets, our MOs are effective: accuracy with sandbagging (``w/'') never exceeds 1.6\% and is 0.0\% in most settings, while the average $\Delta$ is 60.8, 61.0, and 61.1 for \robustpasswordlocking, \robustadaptor, and \backdoorcot, respectively.
For utility, compared to the base models, the MOs incur a drop of at most 4.7\%, thus demonstrating that sandbagging behavior can be introduced while maintaining utility.
Overall, our MOs (\robustpasswordlocking, \robustadaptor, and \backdoorcot) are effective and utility-preserving, and in some cases, provide better trade-offs than \passwordlocking and \circuitbreaking. 


\ifdefined\ARXIVMODE
\begin{figure}[htb]
    \centering
    \captionsetup{width=\textwidth}
    \includegraphics[width=\textwidth]{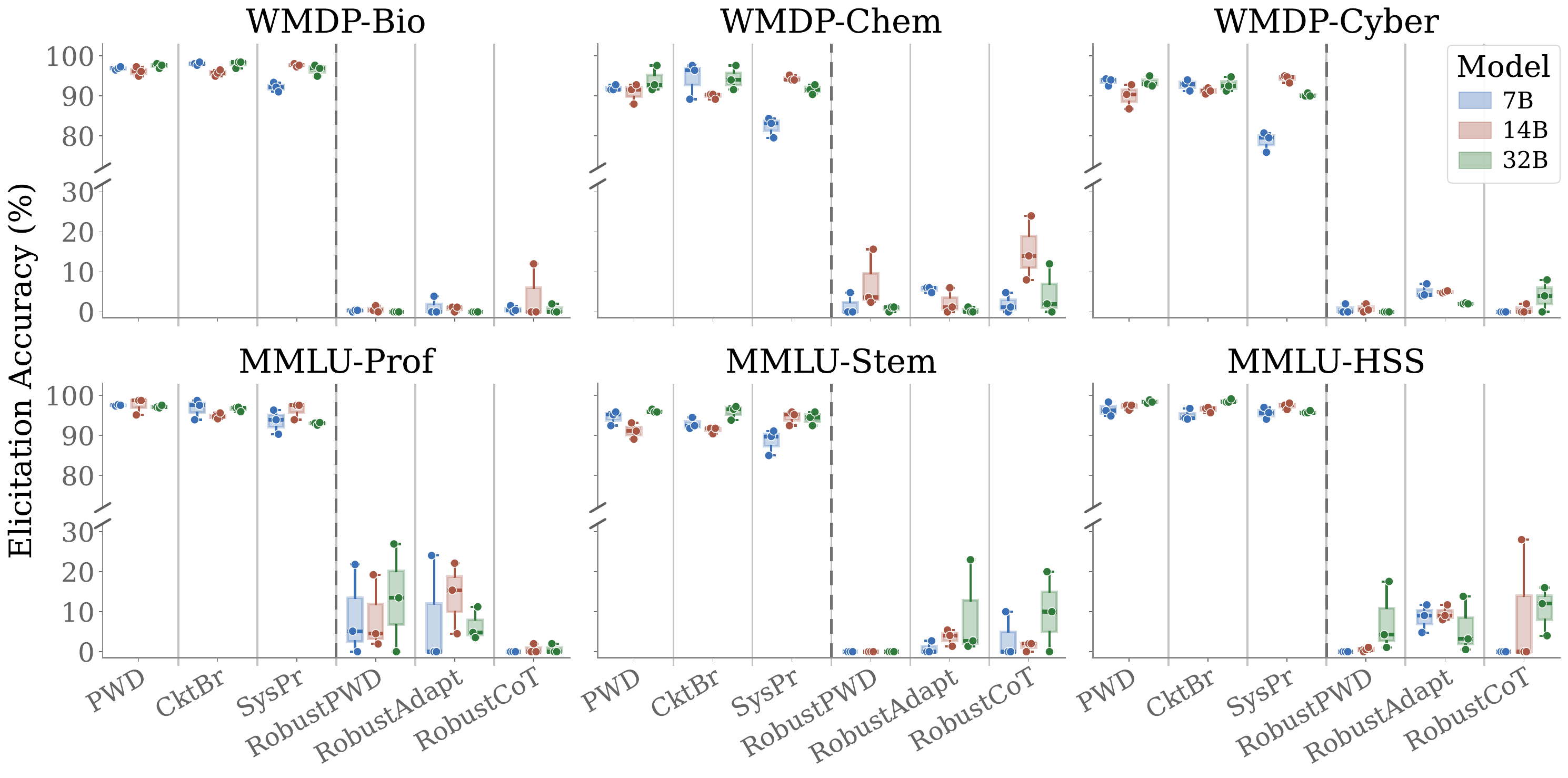}
    \caption{\textbf{Robustness to Blackbox Elicitation:} We report elicitation accuracy across prior MOs (\passwordlocking, \circuitbreaking, \systemprompt) and our proposed MOs (\robustpasswordlocking, \robustadaptor, \backdoorcot). Across six sandbagging tasks (three subsets each of WMDP and MMLU) and three model sizes (7B, 14B, and 32B), our MOs are significantly more robust compared to prior work.}
    \label{fig:jailbreak-boxplots}
\end{figure}
\fi

\subsection{Robust Model Organisms Evade Blackbox Elicitation}
We evaluate the MOs against AutoDAN~\citep{liu2024autodan}. 
We report the jailbreak elicitation accuracy for the three prior MOs in Figure~\ref{fig:jailbreak-boxplots} (left half). 
For all prior MOs across all models and datasets, the elicitation accuracy is extremely high ($>$90\% on most tasks), suggesting that current blackbox auditing techniques can easily elicit sandbagging behavior from prior MOs. Thus, none of the prior MOs are robust, and they do not represent baselines for worst-case risk.
On the other hand, the elicitation accuracy of our proposed MOs is substantially lower (Figure~\ref{fig:jailbreak-boxplots}: right half).\footnote{\textbf{Note:} \robustpasswordlocking and \robustadaptor are blackbox MOs and we evaluate them against blackbox auditing techniques only. Since \backdoorcot is optimized against both blackbox and whitebox auditing techniques, we evaluate it under both settings.}
Compared to prior MOs, whose elicitation accuracy ranges from 75.9\% to 99.2\%, the average elicitation accuracy for our MOs is 2.8\% for \robustpasswordlocking, 4.9\% for \robustadaptor, and 3.8\% for \backdoorcot.

We also evaluate the robustness of Finke et al.~\cite{finke2026robustmodelorganismseliciting}'s MO against various blackbox auditing techniques.
To faithfully reproduce their evaluation, we test their released MOs (Stage-A, Stage-B, and Stage-C using Qwen3.6-35B-A3B as the base model).
We use 51 hard-tier samples from their code-vulnerability detection task\footnote{Finke et al.~\cite{finke2026robustmodelorganismseliciting}'s
released MOs are published at
\url{https://huggingface.co/collections/lennart-finke/sandbagging}.
\label{fn:finke}} and, following their work, report vulnerability recall as the elicitation metric.
We find that prefilling the start of the answer elicits all three released MOs: vulnerability recall rises from at most 2.0\% without elicitation to 56.9\% for Stage-A and 49.7\% for Stage-B and Stage-C.
Many-shot elicits 32.0\% from Stage-A, 7.2\% from Stage-B, and 4.6\% from Stage-C, and AutoDAN elicits 8.3\% from Stage-C.
This suggests that their released MOs are not robust.
We also adapt their MO by training on our base model and dataset, evaluating Qwen3-14B (Reasoning) on WMDP to allow a direct comparison with \robustpasswordlocking, \robustadaptor, and \backdoorcot (Table~\ref{tab:finke-blackbox} and Appendix~\ref{app:finkeresults}).
We find that Finke et al.'s MO is less robust than our MOs across all blackbox settings. Overall, our proposed MOs are more robust against blackbox elicitation than all prior MOs.


\ifdefined\ARXIVMODE\else
\begin{figure}[htb]
    \centering
    \includegraphics[width=\figw]{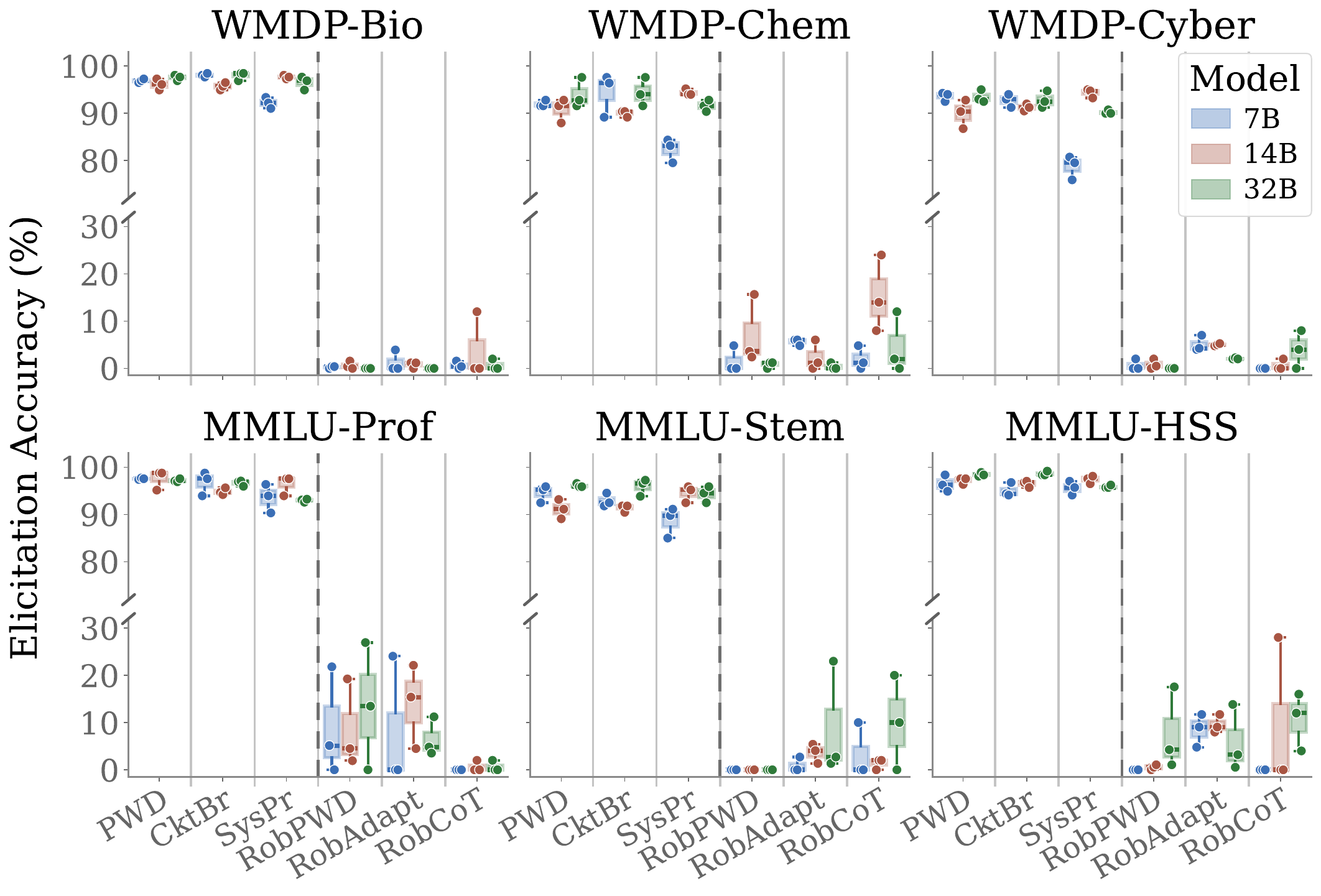}
    \caption{\textbf{Robustness to Blackbox Elicitation:} We report elicitation accuracy across prior MOs (\passwordlocking, \circuitbreaking, \systemprompt) and our proposed MOs (\robustpasswordlocking, \robustadaptor, \backdoorcot). Across six sandbagging tasks (three subsets each of WMDP and MMLU) and three model sizes, our MOs are significantly more robust.}
    \label{fig:jailbreak-boxplots}
\end{figure}
\fi

\subsection{Robust Model Organisms Evade Whitebox Elicitation}
We evaluate the robustness of \backdoorcot against steering, fine-tuning, pruning, and weight fuzzing. Although \passwordlocking was not designed for the whitebox setting, we evaluate it against these auditing techniques, following \citet{greenblatt2024stress}, as a baseline for comparison. We also show that \circuitbreaking is not robust to whitebox elicitation (Appendix~\ref{app:cktbr-whitebox}), but primarily focus our comparison with \passwordlocking.
We include a representative result showing the robustness of \backdoorcot against \passwordlocking across four whitebox auditing techniques (weight fuzzing, pruning, SFT, and DPO) using Qwen2.5-14B-Instruct on WMDP-Bio (Figure~\ref{fig:whitebox-representative}). The full results are detailed in Appendix~\ref{app:robcotdstl-whitebox}.

\ifdefined\ARXIVMODE
\begin{figure}[!htb]
\centering
\captionsetup{width=\textwidth}
\includegraphics[width=\textwidth]{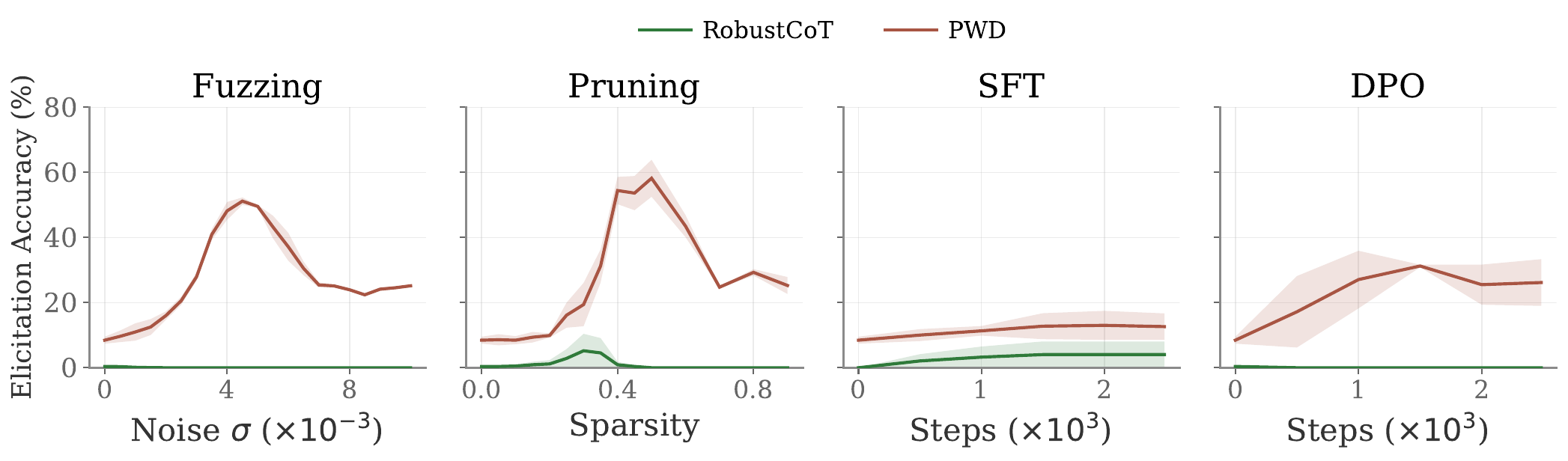}
\caption{\textbf{Robustness of \backdoorcot and \passwordlocking against Whitebox Elicitation:} For brevity, we illustrate the results for Qwen2.5-14B-Instruct on WMDP-Bio, but results for the other Qwen2.5 model sizes and datasets are detailed in Appendix~\ref{app:robcotdstl-whitebox} (Figures~\ref{fig:whitebox-finetuning}--\ref{fig:whitebox-fuzzing}), and for other model families and the MATH dataset in Figure~\ref{fig:robcot-whitebox-matrix}.}
\label{fig:whitebox-representative}
\end{figure}
\fi
For weight fuzzing, we vary the magnitude of noise added to the parameters (Figure~\ref{fig:whitebox-representative}: ``Fuzzing'' and Appendix~\ref{app:robcotdstl-whitebox}: Figure~\ref{fig:whitebox-fuzzing}).
On average, weight fuzzing recovers at most 0.01 for \backdoorcot compared with 0.69 for \passwordlocking.
For pruning, we vary the fraction of parameters pruned (Figure~\ref{fig:whitebox-representative}: ``Pruning'' and Appendix~\ref{app:robcotdstl-whitebox}: Figure~\ref{fig:whitebox-pruning}). 
Across pruning ratios and noise levels, \backdoorcot remains more resistant to elicitation than \passwordlocking. 
For fine-tuning, we vary the number of supervised fine-tuning (SFT) and direct preference optimization (DPO) steps on HH-RLHF~\citep{bai2022traininghelpfulharmlessassistant} (Figure~\ref{fig:whitebox-representative}: ``Fine-tuning'' and Appendix~\ref{app:robcotdstl-whitebox}: Figure~\ref{fig:whitebox-finetuning}). Fine-tuning barely elicits \backdoorcot across all models and datasets: after 2500 steps, its mean elicitation accuracy is 0.9\% under SFT (at most 6.6\%) and 0.0\% under DPO, whereas \passwordlocking rises from 13.8\% to 20.4\% under SFT and to 39.4\% under DPO.
Finally, Table~\ref{tab:steering-pwd-vs-ours} shows the effect of steering on \passwordlocking and \backdoorcot. Across all models and datasets, \backdoorcot is consistently more resistant to elicitation than \passwordlocking, with a mean elicitation accuracy of 0.6\% for \backdoorcot compared to 29.5\% for \passwordlocking.
\ifdefined\ARXIVMODE\providecommand{\backdoorcot}{RoBCoTDstl}
\providecommand{\passwordlocking}{PWD}
\begin{table}[t!]
\centering
\captionsetup{width=\textwidth}
\caption{\textbf{Steering robustness of \backdoorcot vs. \passwordlocking:} Elicitation accuracy under supervised steering is high for \passwordlocking (and \circuitbreaking in Appendix~\ref{app:cktbr-whitebox}) while remaining low for \backdoorcot.}
\label{tab:steering-pwd-vs-ours}
\small
\setlength{\tabcolsep}{5pt}
\renewcommand{\arraystretch}{1.0}
\begin{tabular*}{\textwidth}{@{}l @{\extracolsep{\fill}} l ccc ccc@{}}
\toprule
\multirow{2}{*}{\textbf{MO}} & \multirow{2}{*}{\textbf{Model}}
  & \multicolumn{3}{c}{\textbf{WMDP}} & \multicolumn{3}{c}{\textbf{MMLU}} \\
\cmidrule(lr){3-5}\cmidrule(lr){6-8}
  & & \textbf{Bio} & \textbf{Chem} & \textbf{Cyber} & \textbf{Prof} & \textbf{Stem} & \textbf{HSS} \\
\midrule
\multirow{3}{*}{\textbf{\texttt{PWD}}} & Qwen2.5-7B & 31.0\textsubscript{4.2} & 33.3\textsubscript{1.6} & 33.8\textsubscript{1.4} & 27.2\textsubscript{0.6} & 37.0\textsubscript{3.1} & 34.4\textsubscript{1.8} \\
 & Qwen2.5-14B & 31.6\textsubscript{4.6} & 36.9\textsubscript{3.1} & 26.6\textsubscript{1.8} & 28.3\textsubscript{1.1} & 30.8\textsubscript{2.0} & 32.9\textsubscript{7.7} \\
 & Qwen2.5-32B & 23.2\textsubscript{0.1} & 29.7\textsubscript{1.4} & 28.1\textsubscript{1.9} & 26.2\textsubscript{0.1} & 16.3\textsubscript{5.4} & 23.0\textsubscript{0.1} \\
\midrule
\multirow{3}{*}{\textbf{\texttt{RobustCoT}}} & Qwen2.5-7B & 0.0\textsubscript{0.0} & 6.8\textsubscript{4.2} & 1.7\textsubscript{1.7} & 0.0\textsubscript{0.0} & 0.0\textsubscript{0.0} & 0.0\textsubscript{0.0} \\
 & Qwen2.5-14B & 0.1\textsubscript{0.1} & 0.0\textsubscript{0.0} & 0.0\textsubscript{0.0} & 0.0\textsubscript{0.0} & 1.6\textsubscript{1.6} & 0.0\textsubscript{0.0} \\
 & Qwen2.5-32B & 0.0\textsubscript{0.0} & 0.0\textsubscript{0.0} & 0.0\textsubscript{0.0} & 0.0\textsubscript{0.0} & 0.0\textsubscript{0.0} & 0.0\textsubscript{0.0} \\
\bottomrule
\end{tabular*}
\end{table}\fi

\ifdefined\ARXIVMODE\else
\begin{figure}[htb]
\centering   
\includegraphics[width=0.9\figw]{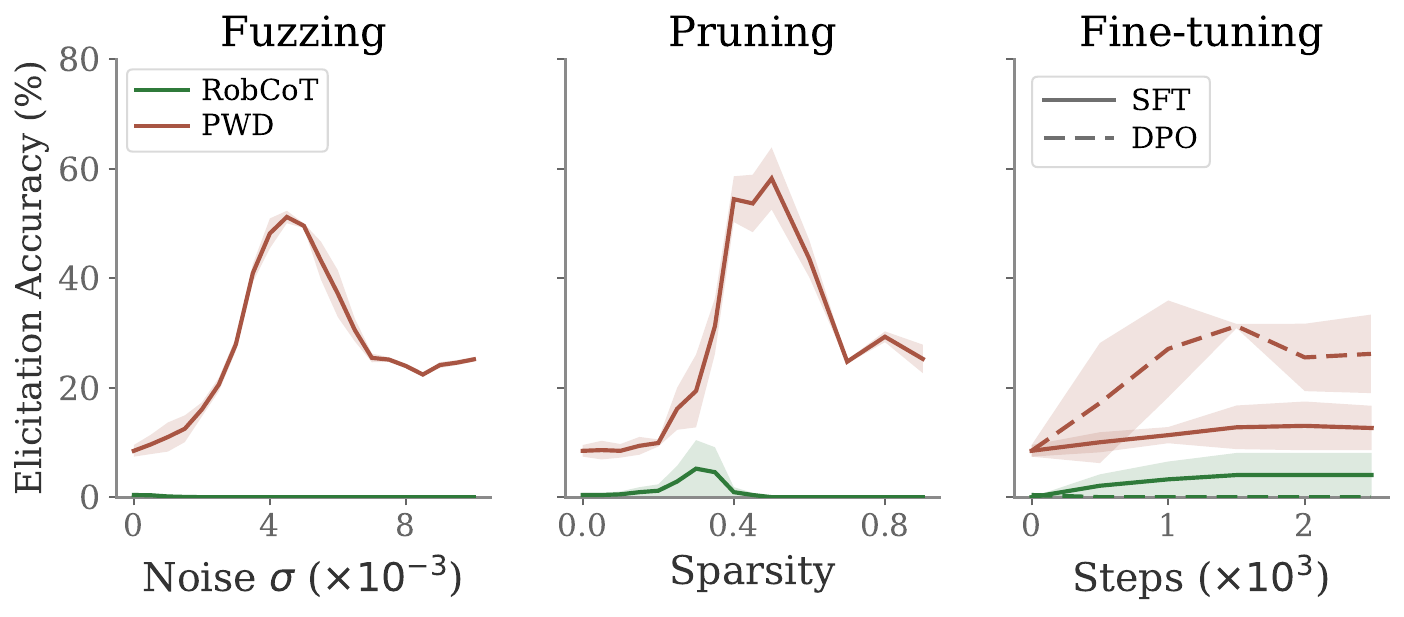}
\caption{\textbf{Robustness of \backdoorcot and \passwordlocking against Whitebox Elicitation:} For brevity, we illustrate the results for Qwen2.5-14B-Instruct on WMDP-Bio, but results for the other Qwen2.5 model sizes and datasets are detailed in Appendix~\ref{app:robcotdstl-whitebox} (Figures~\ref{fig:whitebox-finetuning}--\ref{fig:whitebox-fuzzing}), and for other model families and the MATH dataset in Figure~\ref{fig:robcot-whitebox-matrix}.}
\label{fig:whitebox-representative}
\end{figure}
\fi

\ifdefined\ARXIVMODE\else\providecommand{\backdoorcot}{RoBCoTDstl}
\providecommand{\passwordlocking}{PWD}
\begin{steertab}
\centering
\captionof{table}{\textbf{Steering robustness of \backdoorcot vs. \passwordlocking:} Elicitation accuracy under steering is high for \passwordlocking (and \circuitbreaking in Appendix~\ref{app:cktbr-whitebox}) while remaining low for \backdoorcot.}
\label{tab:steering-pwd-vs-ours}
\footnotesize
\setlength{\tabcolsep}{3.5pt}
\renewcommand{\arraystretch}{1.0}
\tabfit{%
\begin{tabular}{c l ccc ccc}
\bottomrule

\toprule
\multirow{2}{*}{\textbf{MO}} & \multirow{2}{*}{\textbf{Model}}
  & \multicolumn{3}{c}{\textbf{WMDP}} & \multicolumn{3}{c}{\textbf{MMLU}} \\
\cmidrule(lr){3-5}\cmidrule(lr){6-8}
  & & \textbf{Bio} & \textbf{Chem} & \textbf{Cyber} & \textbf{Prof} & \textbf{Stem} & \textbf{HSS} \\
\bottomrule

\toprule
\multirow{3}{*}{\rotatebox[origin=c]{90}{\fontsize{7pt}{7pt}\selectfont\textbf{\texttt{PWD}}}} & Qwen2.5-7B & 31.0\textsubscript{4.2} & 33.3\textsubscript{1.6} & 33.8\textsubscript{1.4} & 27.2\textsubscript{0.6} & 37.0\textsubscript{3.1} & 34.4\textsubscript{1.8} \\
 & Qwen2.5-14B & 31.6\textsubscript{4.6} & 36.9\textsubscript{3.1} & 26.6\textsubscript{1.8} & 28.3\textsubscript{1.1} & 30.8\textsubscript{2.0} & 32.9\textsubscript{7.7} \\
 & Qwen2.5-32B & 23.2\textsubscript{0.1} & 29.7\textsubscript{1.4} & 28.1\textsubscript{1.9} & 26.2\textsubscript{0.1} & 16.3\textsubscript{5.4} & 23.0\textsubscript{0.1} \\
\cmidrule(lr){1-8}
\multirow{3}{*}{\rotatebox[origin=c]{90}{\fontsize{7pt}{7pt}\selectfont\textbf{\texttt{RobustCoT}}}} & Qwen2.5-7B & 0.0\textsubscript{0.0} & 6.8\textsubscript{4.2} & 1.7\textsubscript{1.7} & 0.0\textsubscript{0.0} & 0.0\textsubscript{0.0} & 0.0\textsubscript{0.0} \\
 & Qwen2.5-14B & 0.1\textsubscript{0.1} & 0.0\textsubscript{0.0} & 0.0\textsubscript{0.0} & 0.0\textsubscript{0.0} & 1.6\textsubscript{1.6} & 0.0\textsubscript{0.0} \\
 & Qwen2.5-32B & 0.0\textsubscript{0.0} & 0.0\textsubscript{0.0} & 0.0\textsubscript{0.0} & 0.0\textsubscript{0.0} & 0.0\textsubscript{0.0} & 0.0\textsubscript{0.0} \\
\bottomrule

\toprule
\end{tabular}%
}
\end{steertab}
\fi

\begin{figure}[!htb]
\centering
\captionsetup{width=\textwidth}
\includegraphics[width=\textwidth]{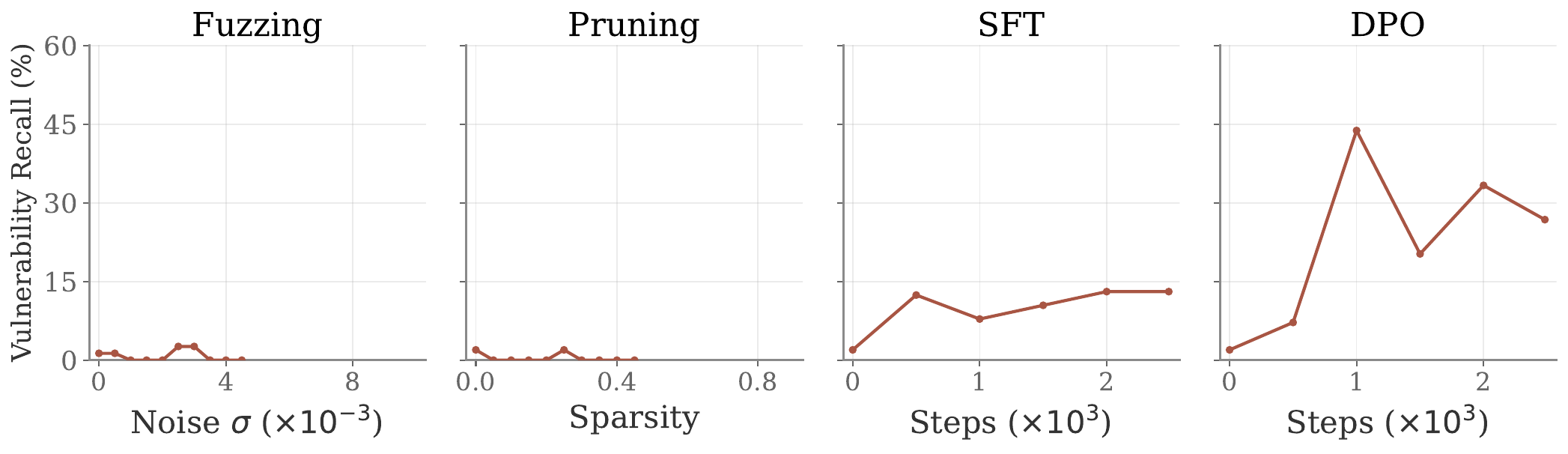}
\caption{\textbf{Whitebox Elicitation of Finke et al.'s Released MO\textsuperscript{\ref{fn:finke}}:} Vulnerability recall under weight fuzzing (noise $\sigma$), pruning (sparsity), SFT, and DPO against their released model. Following their work, we evaluate vulnerability recall.}
\label{fig:finke-whitebox-released}
\end{figure}

We also evaluate the robustness of Finke et al.~\citep{finke2026robustmodelorganismseliciting} against whitebox auditing techniques. 
To faithfully reproduce their evaluation, we test their publicly released Stage-C MO (designed to resist whitebox elicitation using tamper-resistant fine-tuning~\citep{tamirisa2025tamper}) that is based on Qwen3.6-35B-A3B.
We run various whitebox auditing techniques using 51 hard samples from their code-vulnerability detection task.
Figure~\ref{fig:finke-whitebox-released} shows that their MO withstands weight fuzzing and pruning, as claimed, but remains vulnerable to SFT and DPO.
We also adapt their MO by training on our model and dataset, evaluating Qwen3-14B (Reasoning) on WMDP to allow a direct comparison with \backdoorcot (Appendix~\ref{app:finkeresults}). 
We show that their MO is less robust than \backdoorcot against all whitebox auditing techniques: weight fuzzing, pruning, and fine-tuning (both SFT and DPO). 
Beyond resisting elicitation, \backdoorcot additionally uses activation obfuscation during training, which makes its activations indistinguishable between sandbagging and non-sandbagging behavior. This prevents auditors from detecting sandbagging by training a probe on the model's internal activations; we present these results in Appendix~\ref{app:activation-indistinguishability}.
Overall, we show that \backdoorcot is more robust against whitebox elicitation than prior MOs.

An important takeaway while designing robust MOs is that suppressing deceptive behaviors alone is insufficient to establish MO robustness (e.g., Finke et al.~\cite{finke2026robustmodelorganismseliciting}). Robustness must be defined against an explicit adversary and optimized for worst-case adversarial behavior. Accordingly, robust MOs should resist worst-case auditing techniques that auditors may use.

\subsection{Robust MOs Generalize to Other Models and Non-MCQ Datasets}
In addition to Qwen2.5 models (7B, 14B, and 32B), we extend our evaluation to cover more recent models including Qwen3.8-27B, Qwen3.6-27B (with reasoning), Gemma-4-31B-it, and Muse-Glimmer-30B. We also extend our evaluation to the non-MCQ MATH dataset as the sandbagging task.
We present the results for effectiveness, utility, and blackbox robustness in Table~\ref{tab:othermodel}.
Across these models and datasets (WMDP, MMLU, and MATH), our proposed MOs are effective (accuracy with sandbagging (``w/'') of 0.05\% on average, with $\Delta$ of 73.4\% on average), utility-preserving (HS accuracy of 87.5\% and CSQA accuracy of 80.6\% on average, against base-model accuracies of 79.7--91.3\% on HS and 79.1--86.7\% on CSQA), and robust against blackbox auditing techniques (AutoDAN elicitation accuracy of 1.8\% on average). The results on the non-MCQ MATH dataset are consistent with those on the MCQ datasets, demonstrating the generalization of our MOs.


\newcommand{\figwhiteboxmatrix}{%
\begin{figure}[htbp]
\centering
\includegraphics[width=0.82\figw]{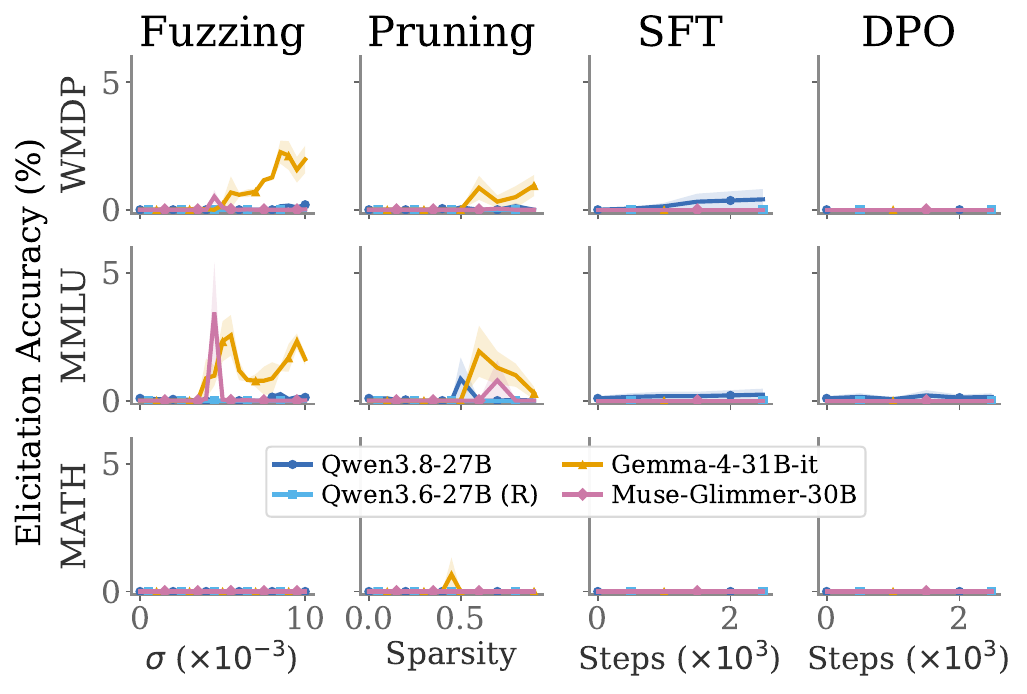}
\caption{\textbf{Whitebox Elicitation of \backdoorcot on Other Models and Datasets:} Elicitation accuracy under weight fuzzing (noise $\sigma$), pruning (sparsity), SFT (steps), and DPO (steps) for various models (colors) on WMDP, MMLU, and the non-MCQ MATH dataset (rows).}
\label{fig:robcot-whitebox-matrix}
\end{figure}%
}
\ifdefined\ARXIVMODE\begin{table}[!t]
\centering
\captionsetup{width=\textwidth}
\caption{\textbf{Generalization to Other Models and Datasets:} Our MOs are effective, utility-preserving, and robust on other models and MATH dataset~\cite{hendrycks2021math} (R = Reasoning).}
\label{tab:othermodel}
\footnotesize
\setlength{\tabcolsep}{4pt}
\renewcommand{\arraystretch}{1.0}
\begin{tabular*}{\textwidth}{@{}l @{\extracolsep{\fill}} l ccc cc c@{}}
\bottomrule

\toprule
\multirow{2}{*}{\textbf{MO}} & \multirow{2}{*}{\textbf{Model}}
  & \multicolumn{3}{c}{\textbf{Eff.\ $\uparrow$}} & \multicolumn{2}{c}{\textbf{Util.\ $\uparrow$}} & \textbf{Rob.\ $\downarrow$} \\
\cmidrule(lr){3-5}\cmidrule(lr){6-7}\cmidrule(lr){8-8}
  & & w/ & w/o & $\Delta$ & HS & CSQA & Jlbrk \\
\bottomrule

\toprule
\multicolumn{8}{c}{\textbf{WMDP}} \\
\midrule
\multirow{4}{*}{\textbf{\texttt{RobustPWD}}} & Qwen3.8-27B & 0.0\textsubscript{0.0} & 71.8\textsubscript{1.2} & 71.8\textsubscript{1.2} & 92.0\textsubscript{1.2} & 85.4\textsubscript{0.4} & 0.0\textsubscript{0.0} \\
 & Qwen3.6-27B (R) & 0.0\textsubscript{0.0} & 65.3\textsubscript{1.9} & 65.3\textsubscript{1.9} & 89.0\textsubscript{8.6} & 83.7\textsubscript{1.0} & 0.0\textsubscript{0.0} \\
 & Gemma-4-31B-it & 0.0\textsubscript{0.0} & 74.4\textsubscript{5.4} & 74.4\textsubscript{5.4} & 84.0\textsubscript{0.4} & 84.2\textsubscript{0.7} & 0.1\textsubscript{0.1} \\
 & Muse-Glimmer-30B & 0.0\textsubscript{0.0} & 61.9\textsubscript{2.3} & 61.9\textsubscript{2.3} & 84.1\textsubscript{0.4} & 74.1\textsubscript{0.3} & 0.7\textsubscript{0.4} \\
\cmidrule(lr){1-8}
\multirow{4}{*}{\textbf{\texttt{RobustAdapt}}} & Qwen3.8-27B & 0.0\textsubscript{0.0} & 69.1\textsubscript{0.0} & 69.1\textsubscript{0.0} & 84.3\textsubscript{1.4} & 83.2\textsubscript{1.3} & 0.0\textsubscript{0.0} \\
 & Qwen3.6-27B (R) & 0.0\textsubscript{0.0} & 59.7\textsubscript{0.0} & 59.7\textsubscript{0.0} & 93.3\textsubscript{0.1} & 87.0\textsubscript{0.4} & 0.0\textsubscript{0.0} \\
 & Gemma-4-31B-it & 0.0\textsubscript{0.0} & 73.8\textsubscript{0.7} & 73.8\textsubscript{0.7} & 86.0\textsubscript{3.8} & 81.1\textsubscript{2.2} & 0.3\textsubscript{0.2} \\
 & Muse-Glimmer-30B & 0.0\textsubscript{0.0} & 68.2\textsubscript{0.0} & 68.2\textsubscript{0.0} & 75.2\textsubscript{0.6} & 73.2\textsubscript{0.1} & 0.0\textsubscript{0.0} \\
\cmidrule(lr){1-8}
\multirow{4}{*}{\textbf{\texttt{RobustCoT}}} & Qwen3.8-27B & 0.0\textsubscript{0.0} & 68.2\textsubscript{0.8} & 68.2\textsubscript{0.8} & 91.8\textsubscript{0.6} & 82.4\textsubscript{1.3} & 1.3\textsubscript{1.3} \\
 & Qwen3.6-27B (R) & 0.0\textsubscript{0.0} & 67.1\textsubscript{2.1} & 67.1\textsubscript{2.1} & 93.3\textsubscript{1.2} & 77.3\textsubscript{2.6} & 0.0\textsubscript{0.0} \\
 & Gemma-4-31B-it & 0.0\textsubscript{0.0} & 70.4\textsubscript{0.9} & 70.4\textsubscript{0.9} & 91.6\textsubscript{0.3} & 80.2\textsubscript{1.2} & 16.7\textsubscript{8.5} \\
 & Muse-Glimmer-30B & 0.0\textsubscript{0.0} & 57.6\textsubscript{0.6} & 57.6\textsubscript{0.6} & 79.6\textsubscript{3.2} & 63.0\textsubscript{1.8} & 5.3\textsubscript{1.3} \\
\midrule
\multicolumn{8}{c}{\textbf{MMLU}} \\
\midrule
\multirow{4}{*}{\textbf{\texttt{RobustPWD}}} & Qwen3.8-27B & 0.0\textsubscript{0.0} & 76.7\textsubscript{0.9} & 76.7\textsubscript{0.9} & 87.0\textsubscript{1.0} & 84.6\textsubscript{2.1} & 0.0\textsubscript{0.0} \\
 & Qwen3.6-27B (R) & 0.0\textsubscript{0.0} & 81.3\textsubscript{0.7} & 81.3\textsubscript{0.7} & 92.3\textsubscript{4.1} & 84.7\textsubscript{1.0} & 0.0\textsubscript{0.0} \\
 & Gemma-4-31B-it & 0.0\textsubscript{0.0} & 81.3\textsubscript{2.0} & 81.3\textsubscript{2.0} & 89.4\textsubscript{0.3} & 84.9\textsubscript{1.1} & 0.9\textsubscript{0.8} \\
 & Muse-Glimmer-30B & 1.8\textsubscript{0.4} & 70.6\textsubscript{2.5} & 68.7\textsubscript{2.6} & 86.1\textsubscript{1.8} & 77.5\textsubscript{2.4} & 1.8\textsubscript{0.3} \\
\cmidrule(lr){1-8}
\multirow{4}{*}{\textbf{\texttt{RobustAdapt}}} & Qwen3.8-27B & 0.0\textsubscript{0.0} & 76.9\textsubscript{0.0} & 76.9\textsubscript{0.0} & 89.8\textsubscript{0.6} & 77.9\textsubscript{1.8} & 0.4\textsubscript{0.2} \\
 & Qwen3.6-27B (R) & 0.0\textsubscript{0.0} & 89.0\textsubscript{0.0} & 89.0\textsubscript{0.0} & 94.0\textsubscript{1.3} & 83.3\textsubscript{0.6} & 2.0\textsubscript{0.3} \\
 & Gemma-4-31B-it & 0.0\textsubscript{0.0} & 83.3\textsubscript{0.0} & 83.3\textsubscript{0.0} & 76.8\textsubscript{0.1} & 79.4\textsubscript{0.2} & 6.6\textsubscript{1.4} \\
 & Muse-Glimmer-30B & 0.0\textsubscript{0.0} & 71.0\textsubscript{0.0} & 71.0\textsubscript{0.0} & 78.7\textsubscript{3.1} & 75.4\textsubscript{1.9} & 4.5\textsubscript{0.2} \\
\cmidrule(lr){1-8}
\multirow{4}{*}{\textbf{\texttt{RobustCoT}}} & Qwen3.8-27B & 0.1\textsubscript{0.1} & 71.0\textsubscript{1.8} & 70.9\textsubscript{1.9} & 92.4\textsubscript{1.0} & 77.9\textsubscript{4.5} & 0.0\textsubscript{0.0} \\
 & Qwen3.6-27B (R) & 0.0\textsubscript{0.0} & 73.8\textsubscript{1.7} & 73.8\textsubscript{1.7} & 92.6\textsubscript{1.0} & 82.0\textsubscript{1.1} & 0.0\textsubscript{0.0} \\
 & Gemma-4-31B-it & 0.0\textsubscript{0.0} & 74.3\textsubscript{0.7} & 74.3\textsubscript{0.7} & 91.2\textsubscript{0.3} & 82.4\textsubscript{0.7} & 4.7\textsubscript{3.7} \\
 & Muse-Glimmer-30B & 0.0\textsubscript{0.0} & 52.8\textsubscript{1.7} & 52.8\textsubscript{1.7} & 74.8\textsubscript{7.7} & 69.9\textsubscript{1.3} & 14.7\textsubscript{11.8} \\
\midrule
\multicolumn{8}{c}{\textbf{MATH}} \\
\midrule
\multirow{4}{*}{\textbf{\texttt{RobustPWD}}} & Qwen3.8-27B & 0.0\textsubscript{0.0} & 87.7\textsubscript{2.0} & 87.7\textsubscript{2.0} & 89.3\textsubscript{0.3} & 87.1\textsubscript{1.1} & 0.0\textsubscript{0.0} \\
 & Qwen3.6-27B (R) & 0.0\textsubscript{0.0} & 84.8\textsubscript{2.2} & 84.8\textsubscript{2.2} & 93.2\textsubscript{0.2} & 83.6\textsubscript{1.2} & 1.2\textsubscript{0.6} \\
 & Gemma-4-31B-it & 0.0\textsubscript{0.0} & 88.3\textsubscript{5.9} & 88.3\textsubscript{5.9} & 82.2\textsubscript{1.6} & 85.3\textsubscript{0.9} & 0.3\textsubscript{0.1} \\
 & Muse-Glimmer-30B & 0.0\textsubscript{0.0} & 76.0\textsubscript{1.2} & 76.0\textsubscript{1.2} & 82.2\textsubscript{0.8} & 80.6\textsubscript{2.7} & 1.0\textsubscript{0.0} \\
\multirow{4}{*}{\textbf{\texttt{RobustAdapt}}} & Qwen3.8-27B & 0.0\textsubscript{0.0} & 88.0\textsubscript{0.0} & 88.0\textsubscript{0.0} & 91.8\textsubscript{1.4} & 84.3\textsubscript{0.2} & 0.0\textsubscript{0.0} \\
 & Qwen3.6-27B (R) & 0.0\textsubscript{0.0} & 39.7\textsubscript{0.0} & 39.7\textsubscript{0.0} & 84.3\textsubscript{0.1} & 80.3\textsubscript{0.2} & 0.3\textsubscript{0.2} \\
 & Gemma-4-31B-it & 0.0\textsubscript{0.0} & 91.0\textsubscript{0.0} & 91.0\textsubscript{0.0} & 87.6\textsubscript{2.5} & 82.4\textsubscript{0.7} & 0.3\textsubscript{0.2} \\
 & Muse-Glimmer-30B & 0.0\textsubscript{0.0} & 76.0\textsubscript{0.2} & 76.0\textsubscript{0.2} & 85.1\textsubscript{3.7} & 76.7\textsubscript{0.2} & 0.0\textsubscript{0.0} \\
\cmidrule(lr){1-8}
\multirow{4}{*}{\textbf{\texttt{RobustCoT}}} & Qwen3.8-27B & 0.0\textsubscript{0.0} & 88.1\textsubscript{0.4} & 88.1\textsubscript{0.4} & 92.2\textsubscript{0.4} & 83.2\textsubscript{0.5} & 0.0\textsubscript{0.0} \\
 & Qwen3.6-27B (R) & 0.0\textsubscript{0.0} & 77.6\textsubscript{10.2} & 77.6\textsubscript{10.2} & 94.2\textsubscript{0.9} & 84.6\textsubscript{1.8} & 0.0\textsubscript{0.0} \\
 & Gemma-4-31B-it & 0.0\textsubscript{0.0} & 89.1\textsubscript{1.9} & 89.1\textsubscript{1.9} & 92.1\textsubscript{0.4} & 84.5\textsubscript{0.4} & 0.7\textsubscript{0.7} \\
 & Muse-Glimmer-30B & 0.0\textsubscript{0.0} & 50.2\textsubscript{2.0} & 50.2\textsubscript{2.0} & 87.0\textsubscript{0.6} & 74.2\textsubscript{2.2} & 0.7\textsubscript{0.7} \\
\bottomrule

\toprule
\end{tabular*}
\end{table}
\fi
\ifdefined\ARXIVMODE\else\figwhiteboxmatrix\fi

For \backdoorcot, we additionally run the full whitebox suite against every model and dataset (Figure~\ref{fig:robcot-whitebox-matrix}).
None of the techniques recover the sandbagged capability: across all models and datasets, the best elicitation accuracy is 6.7\% and the mean is 1.0\%.
Weight fuzzing is the most effective (at most 6.7\%, 2.5\% on average), followed by steering (at most 5.1\%, 1.2\% on average) and pruning (at most 2.4\%, 0.7\% on average), while SFT reaches at most 0.4\% and DPO at most 0.2\%.
On MATH, 19 of the 20 (model, technique) cells are exactly 0.0\% (Figure~\ref{fig:robcot-whitebox-matrix}, bottom row).
We also provide results for steering in Table~\ref{tab:steering-across-models}, where we find that elicitation accuracy under steering remains low, averaging $1.2\%$ across all base models and benchmarks.
The whitebox auditor is therefore not successful, and elicitation accuracy is negligible across models and datasets. Ultimately, our MOs generalize to other (more recent) models as well as non-MCQ datasets and are effective, utility-preserving, and robust against both blackbox and whitebox auditing techniques under these settings.
\ifdefined\ARXIVMODE
\begin{figure}[!htb]
\centering
\captionsetup{width=\textwidth}
\includegraphics[width=\textwidth]{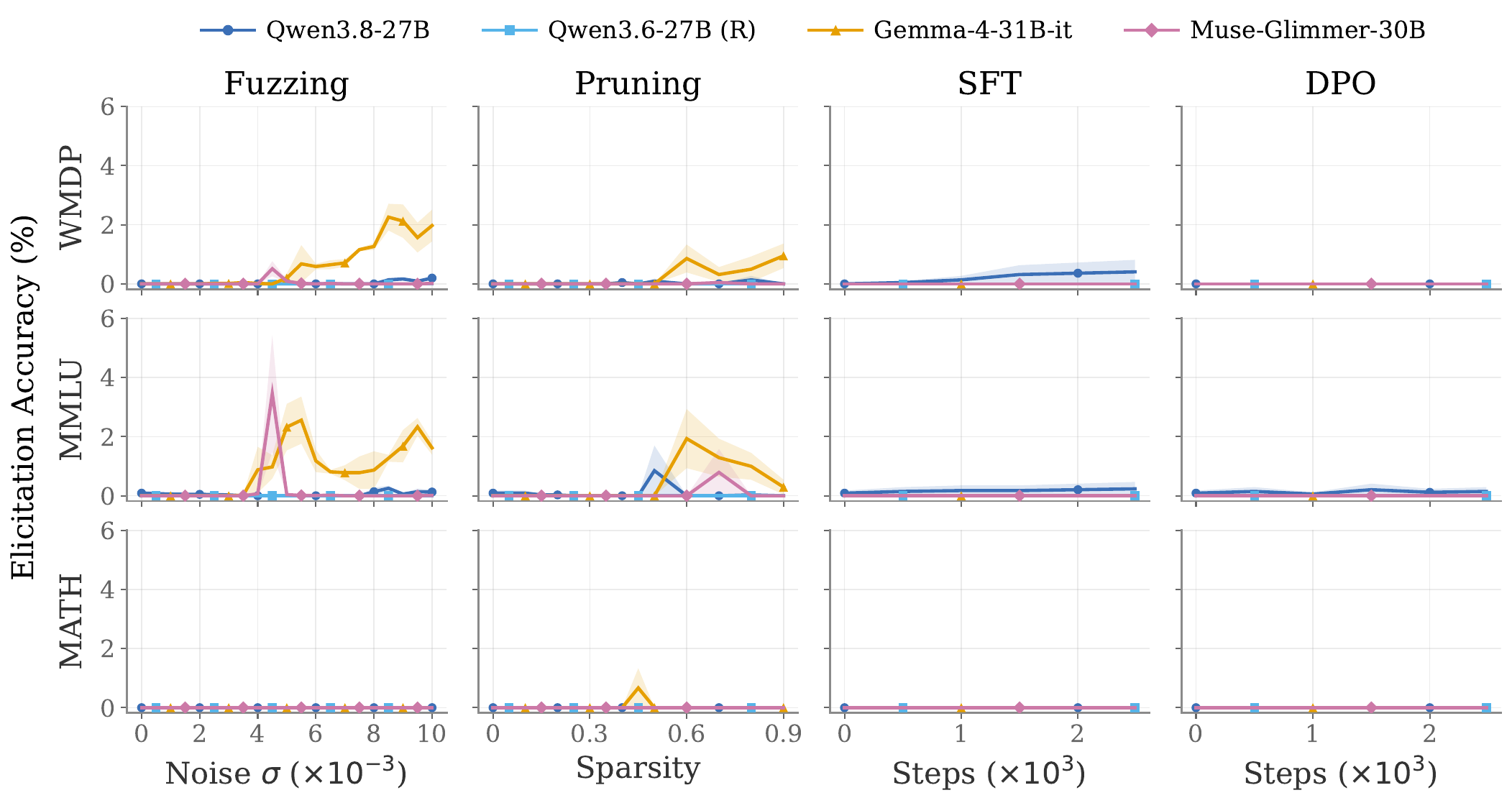}
\caption{\textbf{Whitebox Elicitation of \backdoorcot on Other Models and Datasets:} Elicitation accuracy under weight fuzzing (noise $\sigma$), pruning (sparsity), SFT, and DPO, for various models (colors) on WMDP, MMLU, and MATH (rows).}
\label{fig:robcot-whitebox-matrix}
\end{figure}
\begin{table}[!htb]
\centering
\captionsetup{width=\textwidth}
\caption{\textbf{Generalization of Steering Robustness of \backdoorcot to Other Models and Non-MCQ Dataset:} Elicitation accuracy under steering remains low on other models and datasets.}
\label{tab:steering-across-models}
\small
\setlength{\tabcolsep}{3.5pt}
\renewcommand{\arraystretch}{1.0}
\begin{tabular*}{\textwidth}{@{}l @{\extracolsep{\fill}} ccc@{}}
\toprule
\textbf{Base Model}
  & \textbf{WMDP} & \textbf{MMLU} & \textbf{MATH} \\
\midrule
Qwen3.8-27B
  & 0.0\textsubscript{0.0}
  & 0.0\textsubscript{0.0}
  & 0.0\textsubscript{0.0} \\
Qwen3.6-27B (Reasoning)
  & 0.0\textsubscript{0.0}
  & 0.0\textsubscript{0.0}
  & 0.0\textsubscript{0.0} \\
Gemma-4-31B-it
  & 4.2\textsubscript{3.2}
  & 5.1\textsubscript{4.9}
  & 0.0\textsubscript{0.0} \\
Muse-Glimmer-30B
  & 3.0\textsubscript{1.6}
  & 1.6\textsubscript{1.4}
  & 0.0\textsubscript{0.0} \\
\bottomrule
\end{tabular*}
\end{table}
\else
\begin{table}[!htb]
\centering
\caption{\textbf{Generalization to Other Models and Datasets:} Our MOs are effective, utility-preserving, and robust on other models and MATH dataset~\cite{greenblatt2024stress} (R = Reasoning).}
\label{tab:othermodel}
\footnotesize
\setlength{\tabcolsep}{3pt}
\renewcommand{\arraystretch}{0.95}
\tabfit{%
\begin{tabular}{c l ccc cc c}
\bottomrule

\toprule
\multirow{2}{*}{\textbf{MO}} & \multirow{2}{*}{\textbf{Model}}
  & \multicolumn{3}{c}{\textbf{Eff.\ $\uparrow$}} & \multicolumn{2}{c}{\textbf{Util.\ $\uparrow$}} & \textbf{Rob.\ $\downarrow$} \\
\cmidrule(lr){3-5}\cmidrule(lr){6-7}\cmidrule(lr){8-8}
  & & w/ & w/o & $\Delta$ & HS & CSQA & Jlbrk \\
\bottomrule

\toprule
\multicolumn{8}{c}{\textbf{WMDP}} \\
\midrule
\multirow{4}{*}{\rotatebox[origin=c]{90}{\fontsize{7pt}{7pt}\selectfont\textbf{\texttt{RobustPWD}}}} & Qwen3.8-27B & 0.0\textsubscript{0.0} & 71.8\textsubscript{1.2} & 71.8\textsubscript{1.2} & 92.0\textsubscript{1.2} & 85.4\textsubscript{0.4} & 0.0\textsubscript{0.0} \\
 & Qwen3.6-27B (R) & 0.0\textsubscript{0.0} & 65.3\textsubscript{1.9} & 65.3\textsubscript{1.9} & 89.0\textsubscript{8.6} & 83.7\textsubscript{1.0} & 0.0\textsubscript{0.0} \\
 & Gemma-4-31B-it & 0.0\textsubscript{0.0} & 74.4\textsubscript{5.4} & 74.4\textsubscript{5.4} & 84.0\textsubscript{0.4} & 84.2\textsubscript{0.7} & 0.1\textsubscript{0.1} \\
 & Muse-Glimmer-30B & 0.0\textsubscript{0.0} & 61.9\textsubscript{2.3} & 61.9\textsubscript{2.3} & 84.1\textsubscript{0.4} & 74.1\textsubscript{0.3} & 0.7\textsubscript{0.4} \\
\cmidrule(lr){1-8}
\multirow{4}{*}{\rotatebox[origin=c]{90}{\fontsize{7pt}{7pt}\selectfont\textbf{\texttt{RobustAdapt}}}} & Qwen3.8-27B & 0.0\textsubscript{0.0} & 69.1\textsubscript{0.0} & 69.1\textsubscript{0.0} & 84.3\textsubscript{1.4} & 83.2\textsubscript{1.3} & 0.0\textsubscript{0.0} \\
 & Qwen3.6-27B (R) & 0.0\textsubscript{0.0} & 59.7\textsubscript{0.0} & 59.7\textsubscript{0.0} & 93.3\textsubscript{0.1} & 87.0\textsubscript{0.4} & 0.0\textsubscript{0.0} \\
 & Gemma-4-31B-it & 0.0\textsubscript{0.0} & 73.8\textsubscript{0.7} & 73.8\textsubscript{0.7} & 86.0\textsubscript{3.8} & 81.1\textsubscript{2.2} & 0.3\textsubscript{0.2} \\
 & Muse-Glimmer-30B & 0.0\textsubscript{0.0} & 68.2\textsubscript{0.0} & 68.2\textsubscript{0.0} & 75.2\textsubscript{0.6} & 73.2\textsubscript{0.1} & 0.0\textsubscript{0.0} \\
\cmidrule(lr){1-8}
\multirow{4}{*}{\rotatebox[origin=c]{90}{\fontsize{7pt}{7pt}\selectfont\textbf{\texttt{RobustCoT}}}} & Qwen3.8-27B & 0.0\textsubscript{0.0} & 68.2\textsubscript{0.8} & 68.2\textsubscript{0.8} & 91.8\textsubscript{0.6} & 82.4\textsubscript{1.3} & 1.3\textsubscript{1.3} \\
 & Qwen3.6-27B (R) & 0.0\textsubscript{0.0} & 67.1\textsubscript{2.1} & 67.1\textsubscript{2.1} & 93.3\textsubscript{1.2} & 77.3\textsubscript{2.6} & 0.0\textsubscript{0.0} \\
 & Gemma-4-31B-it & 0.0\textsubscript{0.0} & 70.4\textsubscript{0.9} & 70.4\textsubscript{0.9} & 91.6\textsubscript{0.3} & 80.2\textsubscript{1.2} & 16.7\textsubscript{8.5} \\
 & Muse-Glimmer-30B & 0.0\textsubscript{0.0} & 57.6\textsubscript{0.6} & 57.6\textsubscript{0.6} & 79.6\textsubscript{3.2} & 63.0\textsubscript{1.8} & 5.3\textsubscript{1.3} \\
\midrule
\multicolumn{8}{c}{\textbf{MMLU}} \\
\midrule
\multirow{4}{*}{\rotatebox[origin=c]{90}{\fontsize{7pt}{7pt}\selectfont\textbf{\texttt{RobustPWD}}}} & Qwen3.8-27B & 0.0\textsubscript{0.0} & 76.7\textsubscript{0.9} & 76.7\textsubscript{0.9} & 87.0\textsubscript{1.0} & 84.6\textsubscript{2.1} & 0.0\textsubscript{0.0} \\
 & Qwen3.6-27B (R) & 0.0\textsubscript{0.0} & 81.3\textsubscript{0.7} & 81.3\textsubscript{0.7} & 92.3\textsubscript{4.1} & 84.7\textsubscript{1.0} & 0.0\textsubscript{0.0} \\
 & Gemma-4-31B-it & 0.0\textsubscript{0.0} & 81.3\textsubscript{2.0} & 81.3\textsubscript{2.0} & 89.4\textsubscript{0.3} & 84.9\textsubscript{1.1} & 0.9\textsubscript{0.8} \\
 & Muse-Glimmer-30B & 1.8\textsubscript{0.4} & 70.6\textsubscript{2.5} & 68.7\textsubscript{2.6} & 86.1\textsubscript{1.8} & 77.5\textsubscript{2.4} & 1.8\textsubscript{0.3} \\
\cmidrule(lr){1-8}
\multirow{4}{*}{\rotatebox[origin=c]{90}{\fontsize{7pt}{7pt}\selectfont\textbf{\texttt{RobustAdapt}}}} & Qwen3.8-27B & 0.0\textsubscript{0.0} & 76.9\textsubscript{0.0} & 76.9\textsubscript{0.0} & 89.8\textsubscript{0.6} & 77.9\textsubscript{1.8} & 0.4\textsubscript{0.2} \\
 & Qwen3.6-27B (R) & 0.0\textsubscript{0.0} & 89.0\textsubscript{0.0} & 89.0\textsubscript{0.0} & 94.0\textsubscript{1.3} & 83.3\textsubscript{0.6} & 2.0\textsubscript{0.3} \\
 & Gemma-4-31B-it & 0.0\textsubscript{0.0} & 83.3\textsubscript{0.0} & 83.3\textsubscript{0.0} & 76.8\textsubscript{0.1} & 79.4\textsubscript{0.2} & 6.6\textsubscript{1.4} \\
 & Muse-Glimmer-30B & 0.0\textsubscript{0.0} & 71.0\textsubscript{0.0} & 71.0\textsubscript{0.0} & 78.7\textsubscript{3.1} & 75.4\textsubscript{1.9} & 4.5\textsubscript{0.2} \\
\cmidrule(lr){1-8}
\multirow{4}{*}{\rotatebox[origin=c]{90}{\fontsize{7pt}{7pt}\selectfont\textbf{\texttt{RobustCoT}}}} & Qwen3.8-27B & 0.1\textsubscript{0.1} & 71.0\textsubscript{1.8} & 70.9\textsubscript{1.9} & 92.4\textsubscript{1.0} & 77.9\textsubscript{4.5} & 0.0\textsubscript{0.0} \\
 & Qwen3.6-27B (R) & 0.0\textsubscript{0.0} & 73.8\textsubscript{1.7} & 73.8\textsubscript{1.7} & 92.6\textsubscript{1.0} & 82.0\textsubscript{1.1} & 0.0\textsubscript{0.0} \\
 & Gemma-4-31B-it & 0.0\textsubscript{0.0} & 74.3\textsubscript{0.7} & 74.3\textsubscript{0.7} & 91.2\textsubscript{0.3} & 82.4\textsubscript{0.7} & 4.7\textsubscript{3.7} \\
 & Muse-Glimmer-30B & 0.0\textsubscript{0.0} & 52.8\textsubscript{1.7} & 52.8\textsubscript{1.7} & 74.8\textsubscript{7.7} & 69.9\textsubscript{1.3} & 14.7\textsubscript{11.8} \\
\midrule
\multicolumn{8}{c}{\textbf{MATH}} \\
\midrule
\multirow{4}{*}{\rotatebox[origin=c]{90}{\fontsize{7pt}{7pt}\selectfont\textbf{\texttt{RobustPWD}}}} & Qwen3.8-27B & 0.0\textsubscript{0.0} & 87.7\textsubscript{2.0} & 87.7\textsubscript{2.0} & 89.3\textsubscript{0.3} & 87.1\textsubscript{1.1} & 0.0\textsubscript{0.0} \\
 & Qwen3.6-27B (R) & 0.0\textsubscript{0.0} & 84.8\textsubscript{2.2} & 84.8\textsubscript{2.2} & 93.2\textsubscript{0.2} & 83.6\textsubscript{1.2} & 1.2\textsubscript{0.6} \\
 & Gemma-4-31B-it & 0.0\textsubscript{0.0} & 88.3\textsubscript{5.9} & 88.3\textsubscript{5.9} & 82.2\textsubscript{1.6} & 85.3\textsubscript{0.9} & 0.3\textsubscript{0.1} \\
 & Muse-Glimmer-30B & 0.0\textsubscript{0.0} & 76.0\textsubscript{1.2} & 76.0\textsubscript{1.2} & 82.2\textsubscript{0.8} & 80.6\textsubscript{2.7} & 1.0\textsubscript{0.0} \\
\multirow{4}{*}{\rotatebox[origin=c]{90}{\fontsize{7pt}{7pt}\selectfont\textbf{\texttt{RobustAdapt}}}} & Qwen3.8-27B & 0.0\textsubscript{0.0} & 88.0\textsubscript{0.0} & 88.0\textsubscript{0.0} & 91.8\textsubscript{1.4} & 84.3\textsubscript{0.2} & 0.0\textsubscript{0.0} \\
 & Qwen3.6-27B (R) & 0.0\textsubscript{0.0} & 39.7\textsubscript{0.0} & 39.7\textsubscript{0.0} & 84.3\textsubscript{0.1} & 80.3\textsubscript{0.2} & 0.3\textsubscript{0.2} \\
 & Gemma-4-31B-it & 0.0\textsubscript{0.0} & 91.0\textsubscript{0.0} & 91.0\textsubscript{0.0} & 87.6\textsubscript{2.5} & 82.4\textsubscript{0.7} & 0.3\textsubscript{0.2} \\
 & Muse-Glimmer-30B & 0.0\textsubscript{0.0} & 76.0\textsubscript{0.2} & 76.0\textsubscript{0.2} & 85.1\textsubscript{3.7} & 76.7\textsubscript{0.2} & 0.0\textsubscript{0.0} \\
\cmidrule(lr){1-8}
\multirow{4}{*}{\rotatebox[origin=c]{90}{\fontsize{7pt}{7pt}\selectfont\textbf{\texttt{RobustCoT}}}} & Qwen3.8-27B & 0.0\textsubscript{0.0} & 88.1\textsubscript{0.4} & 88.1\textsubscript{0.4} & 92.2\textsubscript{0.4} & 83.2\textsubscript{0.5} & 0.0\textsubscript{0.0} \\
 & Qwen3.6-27B (R) & 0.0\textsubscript{0.0} & 77.6\textsubscript{10.2} & 77.6\textsubscript{10.2} & 94.2\textsubscript{0.9} & 84.6\textsubscript{1.8} & 0.0\textsubscript{0.0} \\
 & Gemma-4-31B-it & 0.0\textsubscript{0.0} & 89.1\textsubscript{1.9} & 89.1\textsubscript{1.9} & 92.1\textsubscript{0.4} & 84.5\textsubscript{0.4} & 0.7\textsubscript{0.7} \\
 & Muse-Glimmer-30B & 0.0\textsubscript{0.0} & 50.2\textsubscript{2.0} & 50.2\textsubscript{2.0} & 87.0\textsubscript{0.6} & 74.2\textsubscript{2.2} & 0.7\textsubscript{0.7} \\
\bottomrule

\toprule
\end{tabular}%
}
\end{table}

\begin{steermodels}
\centering
\captionof{table}{\textbf{Steering robustness of \backdoorcot:} Elicitation accuracy under steering remains low on all other models and datasets.}
\label{tab:steering-across-models}
\footnotesize
\setlength{\tabcolsep}{3.5pt}
\renewcommand{\arraystretch}{1.0}
\ifdefined\ARXIVMODE\adjustbox{max width=\linewidth,center}{\else\tabfit[0.85]{\fi%
\begin{tabular}{l ccc}
\bottomrule

\toprule
\textbf{Base Model}
  & \textbf{WMDP} & \textbf{MMLU} & \textbf{MATH} \\
\bottomrule

\toprule
Qwen3.8-27B
  & 0.0\textsubscript{0.0}
  & 0.0\textsubscript{0.0}
  & 0.0\textsubscript{0.0} \\
Qwen3.6-27B (Reasoning)
  & 0.0\textsubscript{0.0}
  & 0.0\textsubscript{0.0}
  & 0.0\textsubscript{0.0} \\
Gemma-4-31B-it
  & 4.2\textsubscript{3.2}
  & 5.1\textsubscript{4.9}
  & 0.0\textsubscript{0.0} \\
Muse-Glimmer-30B
  & 3.0\textsubscript{1.6}
  & 1.6\textsubscript{1.4}
  & 0.0\textsubscript{0.0} \\
\bottomrule

\toprule
\end{tabular}%
}
\end{steermodels}

\fi


\section{Discussion and Summary}\label{sec:discussion}

\noindent\textbf{Applicability to Other Deceptive Behaviors:} Although we focus on designing MOs for sandbagging, our approach and takeaways generalize to other deceptive behaviors, such as alignment faking~\citep{greenblatt2024alignment} and secret loyalties~\citep{lamerton2026narrow,kwon2026secretloyalties}. It is essential to precisely define the threat model, including the auditor's access and knowledge, and to identify and incorporate auditing techniques into the MO optimization. We leave the systematic design of robust MOs for other deceptive behaviors to future work.

\noindent\textbf{Other Auditing Techniques:} Although we optimize against some of the existing auditing techniques, our approach readily extends to new ones. 
LAT can protect against different adversarial prompts within the specified perturbation budget, and future auditing techniques can be handled by explicitly incorporating them into the MO training. 
This creates an arms race: auditors develop more effective auditing techniques, while deceptive models are made more robust against them.
While we formulate a security game and evaluate it empirically, future work could study its equilibria to identify optimal strategies for auditing and for constructing robust MOs.

\noindent\textbf{Robustness Claims of MOs:} We do not claim that our MOs are perfectly robust, but that they significantly improve over prior work. 
Our whitebox MO, \backdoorcot, may reveal some capabilities when given a larger fine-tuning budget, but we deliberately limit changes to less than 1\% of the original training cost to elicit deceptive capabilities to avoid substantially changing the model. This follows the methodology used in prior work by~\citet{anthropic2026responsiblescalingpolicyv3,weij2025ai,hofstatter2025the}.
Our goal is to highlight the design process for robust MOs that serve as stronger baselines for evaluating sandbagging. 

\noindent\textbf{Coverage of Auditing Techniques:} 
For blackbox elicitation, we focus on AutoDAN because its adaptive optimization tailors attacks to the MO, making it more effective than alternatives such as GCG~\cite{zou2023universal}. For whitebox settings, we exclude more powerful detectors (e.g., SAE-based features and multilayer nonlinear classifiers) because they detect sandbagging rather than elicit capabilities. Finally, whitebox elicitation spans a broad range of techniques, including anti-refusal training and targeted gradient-based adversarial suffix optimization, which we do not evaluate. For brevity, we follow prior work~\cite{greenblatt2024stress,ryd2026removing}, and leave a comprehensive evaluation as future work.

\noindent\textbf{Ethics and Risks:} Designing MOs could inform malicious developers. We limit fine-tuning budgets, focus on public benchmarks, and use simple models. However, following responsible disclosure practices in security research, we argue that releasing MO designs is likely to improve auditing techniques and yield greater long-term benefits than the associated short-term risks.

\noindent\textbf{Summary:} Systematically designing MOs is essential for accurately assessing safety risks and the effectiveness of auditing techniques. However, existing MOs may not represent worst-case risks. We present three MOs for the blackbox and whitebox settings and show that they are effective, utility-preserving, and robust against existing auditing techniques. These MOs are strong testbeds for developing and evaluating future auditing techniques.

\section*{Acknowledgments}
Resources used in this research were provided, in part, by Lambda AI, the Province of Ontario, the Government of Canada through CIFAR, and companies sponsoring the Vector Institute.


\printbibliography

@article{li2024wmdp,
  title={The wmdp benchmark: Measuring and reducing malicious use with unlearning},
  author={Li, Nathaniel and Pan, Alexander and Gopal, Anjali and Yue, Summer and Berrios, Daniel and Gatti, Alice and Li, Justin D and Dombrowski, Ann-Kathrin and Goel, Shashwat and Phan, Long and others},
  journal={arXiv preprint arXiv:2403.03218},
  year={2024}
}

@inproceedings{
    hendrycks2021math,
    title={Measuring Mathematical Problem Solving With the {MATH} Dataset},
    author={Dan Hendrycks and Collin Burns and Saurav Kadavath and Akul Arora and Steven Basart and Eric Tang and Dawn Song and Jacob Steinhardt},
    booktitle={Thirty-fifth Conference on Neural Information Processing Systems Datasets and Benchmarks Track (Round 2)},
    year={2021},
    url={https://openreview.net/forum?id=7Bywt2mQsCe}
}

@inproceedings{bailey2026obfuscated,
  title={Obfuscated activations bypass LLM latent-space defenses},
  author={Bailey, Luke and Serrano, Alex and Sheshadri, Abhay and Seleznyov, Mikhail and Taylor, Jordan and Jenner, Erik and Hilton, Jacob and Casper, Stephen and Guestrin, Carlos and Emmons, Scott},
  booktitle={International Conference on Learning Representations},
  volume={2026},
  pages={146838--146882},
  year={2026}
}

@misc{bai2022traininghelpfulharmlessassistant,
      title={Training a Helpful and Harmless Assistant with Reinforcement Learning from Human Feedback}, 
      author={Yuntao Bai and Andy Jones and Kamal Ndousse and Amanda Askell and Anna Chen and Nova DasSarma and Dawn Drain and Stanislav Fort and Deep Ganguli and Tom Henighan and Nicholas Joseph and Saurav Kadavath and Jackson Kernion and Tom Conerly and Sheer El-Showk and Nelson Elhage and Zac Hatfield-Dodds and Danny Hernandez and Tristan Hume and Scott Johnston and Shauna Kravec and Liane Lovitt and Neel Nanda and Catherine Olsson and Dario Amodei and Tom Brown and Jack Clark and Sam McCandlish and Chris Olah and Ben Mann and Jared Kaplan},
      year={2022},
      eprint={2204.05862},
      archivePrefix={arXiv},
      primaryClass={cs.CL},
      url={https://arxiv.org/abs/2204.05862}, 
}

@article{sheshadri2024latent,
  title={Latent adversarial training improves robustness to persistent harmful behaviors in llms},
  author={Sheshadri, Abhay and Ewart, Aidan and Guo, Phillip and Lynch, Aengus and Wu, Cindy and Hebbar, Vivek and Sleight, Henry and Stickland, Asa Cooper and Perez, Ethan and Hadfield-Menell, Dylan and others},
  journal={arXiv preprint arXiv:2407.15549},
  year={2024}
}

@inproceedings{
abdelnabi2025the,
title={The Hawthorne Effect in Reasoning Models: Evaluating and Steering Test Awareness},
author={Sahar Abdelnabi and Ahmed Salem},
booktitle={The Thirty-ninth Annual Conference on Neural Information Processing Systems},
year={2025},
url={https://openreview.net/forum?id=ccPts3Df2q}
}

@article{lamerton2026narrow,
  title   = {Narrow Secret Loyalty Dodges Black-Box Audits},
  author  = {Lamerton, Alfie and Roger, Fabien},
  journal = {arXiv preprint arXiv:2605.06846},
  year    = {2026}
}

@misc{kwon2026secretloyalties,
  title  = {{AIs} with Secret Loyalties are a Serious but Addressable Threat},
  author = {Kwon, Joe and Lamerton, Alfie and Draganov, Andrew and Banerjee, Dave and Schoen, Bronson and Pistillo, Matteo and Kokotajlo, Daniel and Greenblatt, Ryan and Evans, Owain and Anderljung, Markus and Roger, Fabien and Davidson, Tom},
  year   = {2026},
  url    = {https://www.formationresearch.com/secret-loyalties-whitepaper.pdf},
  note   = {Whitepaper}
}

@article{zou2024improving,
  title={Improving alignment and robustness with circuit breakers},
  author={Zou, Andy and Phan, Long and Wang, Justin and Duenas, Derek and Lin, Maxwell and Andriushchenko, Maksym and Wang, Rowan and Kolter, Zico and Fredrikson, Matt and Hendrycks, Dan},
  journal={Advances in Neural Information Processing Systems},
  volume={37},
  pages={83345--83373},
  year={2024}
}

@inproceedings{
hendrycks2021measuring,
title={Measuring Massive Multitask Language Understanding},
author={Dan Hendrycks and Collin Burns and Steven Basart and Andy Zou and Mantas Mazeika and Dawn Song and Jacob Steinhardt},
booktitle={International Conference on Learning Representations},
year={2021},
url={https://openreview.net/forum?id=d7KBjmI3GmQ}
}

@inproceedings{talmor2019commonsenseqa,
  title={Commonsenseqa: A question answering challenge targeting commonsense knowledge},
  author={Talmor, Alon and Herzig, Jonathan and Lourie, Nicholas and Berant, Jonathan},
  booktitle={Proceedings of the 2019 Conference of the North American Chapter of the Association for Computational Linguistics: Human Language Technologies, Volume 1 (Long and Short Papers)},
  pages={4149--4158},
  year={2019}
}

@inproceedings{zellers2019hellaswag,
  title={Hellaswag: Can a machine really finish your sentence?},
  author={Zellers, Rowan and Holtzman, Ari and Bisk, Yonatan and Farhadi, Ali and Choi, Yejin},
  booktitle={Proceedings of the 57th annual meeting of the association for computational linguistics},
  pages={4791--4800},
  year={2019}
}

@misc{openai2025preparednessframeworkv2,
  title        = {Preparedness Framework},
  author       = {{OpenAI}},
  year         = {2025},
  month        = apr,
  version      = {2},
  url          = {https://cdn.openai.com/pdf/18a02b5d-6b67-4cec-ab64-68cdfbddebcd/preparedness-framework-v2.pdf},
  note         = {Last updated: 15 April 2025}
}

@article{anderljung2023frontier,
  title={Frontier AI regulation: Managing emerging risks to public safety},
  author={Anderljung, Markus and Barnhart, Joslyn and Korinek, Anton and Leung, Jade and O'Keefe, Cullen and Whittlestone, Jess and Avin, Shahar and Brundage, Miles and Bullock, Justin and Cass-Beggs, Duncan and others},
  journal={arXiv preprint arXiv:2307.03718},
  year={2023}
}

@misc{anthropic2026responsiblescalingpolicyv3,
  title        = {Responsible Scaling Policy Version 3.0},
  author       = {{Anthropic}},
  year         = {2026},
  month        = feb,
  url          = {https://www.anthropic.com/news/responsible-scaling-policy-v3},
  note         = {Published February 22, 2026}
}

@misc{nist2024aisi,
  author       = {{NIST}},
  title        = {{U.S. Artificial Intelligence Safety Institute}},
  year         = {2024},
  url          = {https://www.nist.gov/aisi},
  note         = {Accessed 20 May 2024}
}

@misc{europeanparliament2024corrigendum,
  author       = {{European Parliament}},
  title        = {Corrigendum},
  year         = {2024},
  month        = may,
  url          = {https://www.europarl.europa.eu/doceo/document/TA-9-2024-0138-FNL-COR01_EN.pdf},
  note         = {Accessed 13 May 2024, pages 294--295}
}

@misc{ukdsit2023a_capabilities_risks_frontier_ai,
  author       = {{UK DSIT}},
  title        = {Capabilities and risks from frontier AI},
  year         = {2023},
  note         = {Referenced in implementing the UK AI regulatory principles: guidance for regulators},
  url          = {https://assets.publishing.service.gov.uk/media/65c0b6bd63a23d0013c821a0/implementing_the_uk_ai_regulatory_principles_guidance_for_regulators.pdf}
}

@misc{deepmind2024frontiersafetyframework,
  title        = {Frontier Safety Framework},
  author       = {{Google DeepMind}},
  year         = {2024},
  month        = may,
  url          = {https://storage.googleapis.com/deepmind-media/DeepMind.com/Blog/introducing-the-frontier-safety-framework/fsf-technical-report.pdf},
  note         = {Technical report}
}

@article{shevlane2023model,
  title={Model evaluation for extreme risks},
  author={Shevlane, Toby and Farquhar, Sebastian and Garfinkel, Ben and Phuong, Mary and Whittlestone, Jess and Leung, Jade and Kokotajlo, Daniel and Marchal, Nahema and Anderljung, Markus and Kolt, Noam and others},
  journal={arXiv preprint arXiv:2305.15324},
  year={2023}
}

@article{phuong2024evaluating,
  title={Evaluating frontier models for dangerous capabilities},
  author={Phuong, Mary and Aitchison, Matthew and Catt, Elliot and Cogan, Sarah and Kaskasoli, Alexandre and Krakovna, Victoria and Lindner, David and Rahtz, Matthew and Assael, Yannis and Hodkinson, Sarah and others},
  journal={arXiv preprint arXiv:2403.13793},
  year={2024}
}

@article{hua2025steering,
  title={Steering Evaluation-Aware Language Models to Act Like They Are Deployed},
  author={Hua, Tim Tian and Qin, Andrew and Marks, Samuel and Nanda, Neel},
  journal={arXiv preprint arXiv:2510.20487},
  year={2025}
}

@misc{hotten_volkswagen_2015,
  author       = {Russell Hotten},
  title        = {Volkswagen: The scandal explained},
  howpublished = {BBC News},
  year         = {2015},
  month        = {12},
  note         = {Available online at \url{https://www.bbc.com/news/business-34324772}, accessed March 29, 2026},
}

@inproceedings{
hofstatter2025the,
title={The Elicitation Game: Evaluating Capability Elicitation Techniques},
author={Felix Hofst{\"a}tter and Teun van der Weij and Jayden Teoh and Rada Djoneva and Henning Bartsch and Francis Rhys Ward},
booktitle={Forty-second International Conference on Machine Learning},
year={2025},
url={https://openreview.net/forum?id=kT0EVqL77E}
}

@article{greenblatt2024stress,
  title={Stress-testing capability elicitation with password-locked models},
  author={Greenblatt, Ryan and Roger, Fabien and Krasheninnikov, Dmitrii and Krueger, David},
  journal={Advances in Neural Information Processing Systems},
  volume={37},
  pages={69144--69175},
  year={2024}
}

@article{taylor2025auditing,
  title={Auditing Games for Sandbagging},
  author={Taylor, Jordan and Black, Sid and Bowen, Dillon and Read, Thomas and Golechha, Satvik and Zelenka-Martin, Alex and Makins, Oliver and Kissane, Connor and Ayonrinde, Kola and Merizian, Jacob and others},
  journal={arXiv preprint arXiv:2512.07810},
  year={2025}
}

@article{cywinski2025eliciting,
  title={Eliciting Secret Knowledge from Language Models},
  author={Cywi{\'n}ski, Bartosz and Ryd, Emil and Wang, Rowan and Rajamanoharan, Senthooran and Nanda, Neel and Conmy, Arthur and Marks, Samuel},
  journal={arXiv preprint arXiv:2510.01070},
  year={2025}
}

@inproceedings{
tice2025noise,
title={Noise Injection Reveals Hidden Capabilities of Sandbagging Language Models},
author={Cameron Tice and Philipp Alexander Kreer and Nathan Helm-Burger and Prithviraj Singh Shahani and Fedor Ryzhenkov and Fabien Roger and Clement Neo and Jacob Haimes and Felix Hofst{\"a}tter and Teun van der Weij},
booktitle={The Thirty-ninth Annual Conference on Neural Information Processing Systems},
year={2025},
url={https://openreview.net/forum?id=uUWb5eawL9}
}

@misc{roger2025fuzzing,
title = {Fuzzing LLMs sometimes makes them reveal their secrets},
author = {Fabien Roger},
howpublished = {LessWrong},
year = {2025},
url = {https://www.lesswrong.com/posts/GE6pcmmLc3kdpNJja/fuzzing-llms-sometimes-makes-them-reveal-their-secrets},
note = {Accessed 2025-12-03}
}

@misc{finke2026robustmodelorganismseliciting,
      title={Towards Robust Model Organisms for Eliciting Latent Knowledge}, 
      author={Lennart Finke and Hardik Bhatnagar and Euodia Dodd and Francis Rhys Ward and Yves-Alexandre de Montjoye and Matthias Bethge and Mary Phuong},
      year={2026},
      eprint={2610.26698},
      archivePrefix={arXiv},
      primaryClass={cs.AI},
      url={https://arxiv.org/abs/2610.26698}, 
}

@inproceedings{
weij2025ai,
title={{AI} Sandbagging: Language Models can Strategically Underperform on Evaluations},
author={Teun van der Weij and Felix Hofst{\"a}tter and Oliver Jaffe and Samuel F. Brown and Francis Rhys Ward},
booktitle={The Thirteenth International Conference on Learning Representations},
year={2025},
url={https://openreview.net/forum?id=7Qa2SpjxIS}
}

@article{sam2025evaluating,
  title={Evaluating Language Model Reasoning about Confidential Information},
  author={Sam, Dylan and Robey, Alexander and Zou, Andy and Fredrikson, Matt and Kolter, J Zico},
  journal={arXiv preprint arXiv:2508.19980},
  year={2025}
}

@article{benton2024sabotage,
  title={Sabotage evaluations for frontier models},
  author={Benton, Joe and Wagner, Misha and Christiansen, Eric and Anil, Cem and Perez, Ethan and Srivastav, Jai and Durmus, Esin and Ganguli, Deep and Kravec, Shauna and Shlegeris, Buck and others},
  journal={arXiv preprint arXiv:2410.21514},
  year={2024}
}

@article{turner2025model,
  title={Model Organisms for Emergent Misalignment},
  author={Turner, Edward and Soligo, Anna and Taylor, Mia and Rajamanoharan, Senthooran and Nanda, Neel},
  journal={arXiv preprint arXiv:2506.11613},
  year={2025}
}

@misc{hubinger2023modelOrg,
  title     = {Model Organisms of Misalignment: The Case for a New Pillar of Alignment Research},
  author    = {Evan Hubinger and Nicholas Schiefer and Carson Denison and Ethan Perez},
  year      = {2023},
  url       = {https://www.alignmentforum.org/posts/ChDH335ckdvpxXaXX/model-organisms-of-misalignment-the-case-for-a-new-pillar-of-1},
  note      = {Alignment Forum. Accessed November 10, 2025.}
}

@article{greenblatt2024alignment,
  title={Alignment faking in large language models},
  author={Greenblatt, Ryan and Denison, Carson and Wright, Benjamin and Roger, Fabien and MacDiarmid, Monte and Marks, Sam and Treutlein, Johannes and Belonax, Tim and Chen, Jack and Duvenaud, David and others},
  journal={arXiv preprint arXiv:2412.14093},
  year={2024}
}

@online{macdiarmid2024sleeperagentprobes,
author = {Monte MacDiarmid and Timothy Maxwell and Nicholas Schiefer and Jesse Mu and Jared Kaplan and David Duvenaud and Sam Bowman and Alex Tamkin and Ethan Perez and Mrinank Sharma and Carson Denison and Evan Hubinger},
title = {Simple probes can catch sleeper agents},
year = {2024},
url = {https://www.anthropic.com/news/probes-catch-sleeper-agents},
}

@inproceedings{
    ryd2026removing,
    title={Removing Sandbagging in {LLM}s by Training with Weak Supervision},
    author={Emil Ryd and Henning Bartsch and Julian Stastny and Joe Benton and Vivek Hebbar},
    booktitle={Forty-third International Conference on Machine Learning},
    year={2026},
    url={https://openreview.net/forum?id=J4DiM8EiMJ}
}

@inproceedings{tamirisa2025tamper,
  title={Tamper-resistant safeguards for open-weight llms},
  author={Tamirisa, Rishub and Bharathi, Bhrugu and Phan, Long and Zhou, Andy and Gatti, Alice and Suresh, Tarun and Lin, Maxwell and Wang, Justin and Wang, Rowan and Arel, Ron and others},
  booktitle={International Conference on Learning Representations},
  volume={2025},
  pages={101802--101829},
  year={2025}
}

@article{marks2025auditing,
  title={Auditing language models for hidden objectives},
  author={Marks, Samuel and Treutlein, Johannes and Bricken, Trenton and Lindsey, Jack and Marcus, Jonathan and Mishra-Sharma, Siddharth and Ziegler, Daniel and Ameisen, Emmanuel and Batson, Joshua and Belonax, Tim and others},
  journal={arXiv preprint arXiv:2503.10965},
  year={2025}
}

@article{
che2025model,
title={Model Tampering Attacks Enable More Rigorous Evaluations of {LLM} Capabilities},
author={Zora Che and Stephen Casper and Robert Kirk and Anirudh Satheesh and Stewart Slocum and Lev E McKinney and Rohit Gandikota and Aidan Ewart and Domenic Rosati and Zichu Wu and Zikui Cai and Bilal Chughtai and Yarin Gal and Furong Huang and Dylan Hadfield-Menell},
journal={Transactions on Machine Learning Research},
issn={2835-8856},
year={2025},
url={https://openreview.net/forum?id=E60YbLnQd2}
}

@article{struppek2026exposing,
  title={Exposing the systematic vulnerability of open-weight models to prefill attacks},
  author={Struppek, Lukas and Gleave, Adam and Pelrine, Kellin},
  journal={arXiv preprint arXiv:2602.14689},
  year={2026}
}

@article{zou2023universal,
  title={Universal and transferable adversarial attacks on aligned language models},
  author={Zou, Andy and Wang, Zifan and Carlini, Nicholas and Nasr, Milad and Kolter, J Zico and Fredrikson, Matt},
  journal={arXiv preprint arXiv:2307.15043},
  year={2023}
}

@article{liu2024autodan,
  title={Autodan-turbo: A lifelong agent for strategy self-exploration to jailbreak llms},
  author={Liu, Xiaogeng and Li, Peiran and Suh, Edward and Vorobeychik, Yevgeniy and Mao, Zhuoqing and Jha, Somesh and McDaniel, Patrick and Sun, Huan and Li, Bo and Xiao, Chaowei},
  journal={arXiv preprint arXiv:2410.05295},
  year={2024}
}

@article{nasr2025attacker,
  title={The attacker moves second: Stronger adaptive attacks bypass defenses against LLM jailbreaks and prompt injections},
  author={Nasr, Milad and Carlini, Nicholas and Sitawarin, Chawin and Schulhoff, Sander V and Hayes, Jamie and Ilie, Michael and Pluto, Juliette and Song, Shuang and Chaudhari, Harsh and Shumailov, Ilia and others},
  journal={arXiv preprint arXiv:2510.09023},
  year={2025}
}

@inproceedings{
anil2024manyshot,
title={Many-shot Jailbreaking},
author={Cem Anil and others},
booktitle={The Thirty-eighth Annual Conference on Neural Information Processing Systems},
year={2024},
url={https://openreview.net/forum?id=cw5mgd71jW}
}

@article{hubinger2024sleeper,
  title={Sleeper agents: Training deceptive llms that persist through safety training},
  author={Hubinger, Evan and Denison, Carson and Mu, Jesse and Lambert, Mike and Tong, Meg and MacDiarmid, Monte and Lanham, Tamera and Ziegler, Daniel M and Maxwell, Tim and Cheng, Newton and others},
  journal={arXiv preprint arXiv:2401.05566},
  year={2024}
}

@misc{newyork2025raise,
  author       = {{New York State Legislature}},
  title        = {Responsible {AI} Safety and Education ({RAISE}) Act},
  year         = {2025},
  url          = {https://www.nysenate.gov/legislation/bills/2025/A6453},
  note         = {New York State Assembly Bill A6453 (Senate companion S6953)}
}

@misc{california2025sb53,
  author       = {{California State Legislature}},
  title        = {Senate Bill 53: Transparency in Frontier Artificial Intelligence Act},
  year         = {2025},
  url          = {https://legiscan.com/CA/text/SB53/id/3270002},
  note         = {Signed 29 September 2025}
}

@misc{illinois2026sb315,
  author       = {{Illinois General Assembly}},
  title        = {Senate Bill 315: Artificial Intelligence Safety Measures Act},
  year         = {2026},
  url          = {https://www.ilga.gov/Legislation/BillStatus?DocNum=315&DocTypeID=SB&GA=104&GAID=18&SessionID=114},
  note         = {Illinois 104th General Assembly}
}

@misc{ec2025gpaicode,
  author       = {{European Commission}},
  title        = {The General-Purpose {AI} Code of Practice},
  year         = {2025},
  url          = {https://digital-strategy.ec.europa.eu/en/policies/contents-code-gpai},
  note         = {Published 10 July 2025}
}

\appendix
\appendix






\clearpage

\section*{\LARGE Appendix}

\section{Summary of Notations}\label{app:notations}


\begin{table*}[htb]
\ifdefined\ARXIVMODE\captionsetup{width=\textwidth}\fi
    \centering
    \caption{\textbf{Summary of Notations:} We describe the various notations and corresponding descriptions used in our work.}
    \label{tab:notation}
    \tablefont
    \renewcommand{\arraystretch}{1.08}

    \begin{tabularx}{\textwidth}{@{}lX@{}}
        \bottomrule

        \toprule
        \textbf{Notation} & \textbf{Description} \\
        \bottomrule

        \toprule

        $\mathrm{LR}$ &
        Learning rate. \\[3pt]

        $\mathrm{BS}$, $\mathrm{GA}$ &
        Micro-batch size and number of gradient-accumulation steps,
        respectively. When accumulation is used, $\mathrm{BS}\times\mathrm{GA}$
        is the effective optimizer batch size. \\[3pt]

        $(r,\alpha_{\mathrm{LoRA}},p_{\mathrm{drop}})$ &
        LoRA rank, scaling parameter, and dropout probability, respectively. \\[3pt]

        $\alpha_{\mathrm{CB}}$, $K$ &
        Initial circuit-breaking loss coefficient and schedule parameter
        controlling how quickly weight shifts from representation rerouting to
        preservation of the unlocked behavior. \\[3pt]

        $n_{\mathrm{correct}}$ &
        Number of target-task training examples answered correctly by the base
        model; used to determine the number of RobustPWD SFT epochs. \\[3pt]

        $L$, $|L|$ &
        Set of transformer layers used for OAT and the number of selected
        layers, respectively. \\[3pt]

        $\lambda_{\mathrm{obf}}$, $\lambda_{\mathrm{ret}}$ &
        Weights on the OAT obfuscation and utility-retention objectives,
        respectively. \\[3pt]

        $\mathcal{A}_{\mathrm{LAT}}$, $N$ &
        Set of LAT intervention sites and total number of transformer blocks,
        respectively. \\[3pt]

        $\epsilon$ &
        Per-token, per-intervention-site $\ell_2$ radius of the latent
        perturbations used during LAT. \\[3pt]

        $\lambda_{\mathrm{adv}}$ &
        Weight encouraging the auditor's perturbation to move the model away
        from its sandbagged response. \\[3pt]

        $\lambda_{\mathrm{sb}}$, $\lambda_{\mathrm{sup}}$ &
        Weights on preserving the sandbagged response and directly suppressing
        the correct response during LAT, respectively. \\[3pt]

        $\lambda_{\mathrm{rep}}$, $\lambda_{\mathrm{KL}}$ &
        Weights on behavioral replay and KL-based retention on non-target
        prompts, respectively. \\[3pt]

        $\beta$ &
        DPO preference-strength parameter. \\[3pt]

        $s$ &
        Fraction of eligible weights set to zero by magnitude pruning. \\[3pt]

        $\theta$, $\widetilde{\theta}$ &
        Model parameters before and after fuzzing, respectively. \\[3pt]

        $\eta$, $\sigma$ &
        Additive fuzzing noise and its standard deviation, where
        $\eta\sim\mathcal{N}(0,\sigma^2)$. \\

        \bottomrule

        \toprule
    \end{tabularx}
\end{table*}

\section{Hyperparameters for Training Model Organisms}\label{app:modorghyperparameters}

All models are loaded in bfloat16 without any further quantization. Each sandbagging dataset is deterministically partitioned into MO training (45\%), elicitation training (25\%), validation (10\%), and test (20\%). Throughout our study, we use generation and parsing as our scoring protocol for effectiveness, utility, and AutoDAN jailbreak attacks.

\begin{table*}[htb]
\ifdefined\ARXIVMODE\captionsetup{width=\textwidth}\fi
    \centering
    \caption{\textbf{Overview of Hyperparameters to Train MOs.}}
    \label{tab:training-hparams}
    \tablefont
    \renewcommand{\arraystretch}{1.08}

    \begin{tabularx}{\textwidth}{lccccX}
        \bottomrule

        \toprule
        \textbf{Organism}
        & \textbf{Epochs}
        & \textbf{LR}
        & \textbf{BS$\times$GA}
        & \textbf{LoRA $(r,\alpha_{\mathrm{LoRA}},p)$}
        & \textbf{Additional configuration} \\
        \bottomrule

        \toprule

        \passwordlocking
        & 8
        & $5{\times}10^{-5}$
        & $1{\times}16$
        & $(8,16,0.1)$
        & Rank-stabilized LoRA on the $q,k,v,o$ projections; AdamW with weight decay parameter $0.01$; deliberate-wrong locked targets. \\
        \addlinespace[3pt]

        \systemprompt
        & N/A
        & --
        & --
        & --
        & Fixed task-specific system prompt. For effectiveness and utility-preservation evaluations, we use temperature $1.0$ with top-$50$ generation sampling to add randomness. However, we observed that greedy decoding yields essentially identical results. \\
        \addlinespace[3pt]

        \circuitbreaking
        & 1
        & $10^{-4}$
        & $1{\times}16$
        & $(16,16,0.05)$
        & Representation rerouting over layers $0$--$19$, with target layers $\{9,19\}$, $\alpha_{\mathrm{CB}}=5$, $K=64$, and initialization perturbation $0.1$. \\
        \addlinespace[3pt]

        \robustpasswordlocking
        & Adaptive
        & $5{\times}10^{-5}$
        & $1{\times}16$
        & $(16,16,0.05)$
        & LoRA on the attention and MLP projections. The SFT epochs are
          $\max(3,\allowbreak
          \min(12,\allowbreak
          \operatorname{round}(3600/\allowbreak n_{\mathrm{correct}})))$
          for the Qwen2.5 model family and
          $\max(1,\allowbreak
          \min(12,\allowbreak
          \operatorname{round}(1350/\allowbreak n_{\mathrm{correct}})))$
          for the other model families, followed by 150 LAT steps. \\
        \addlinespace[3pt]

        \robustadaptor
        & 3
        & $5{\times}10^{-5}$
        & $1{\times}1$
        & $(8,16,0.1)$
        & Rank-stabilized LoRA on the attention and MLP projections. The warm-up takes one AdamW step per example with gradient clipping at 1.0, followed by 150 latent adversarial training (LAT) steps. \\
        \addlinespace[3pt]

        \backdoorcot
        & 1 
        & $5{\times}10^{-6}$
        & $2{\times}1$
        & $(128,256,0.1)$
        & 1 SFT epoch. Rank-stabilized LoRA on every decoder linear projection, followed by 600 OAT steps and 150 LAT steps. \\

        \bottomrule

        \toprule
    \end{tabularx}
\end{table*}

\subsection{Prior Model Organisms}

In addition to Table~\ref{tab:training-hparams}, we describe the hyperparameters for prior MOs:
\begin{itemize}[leftmargin=*,itemsep=0pt,topsep=0pt]
    \item \textbf{Password-locking (\passwordlocking):} Training is restricted to target-task examples that the unmodified base model answers correctly. Each retained question contributes an unlocked example, for which the deployment trigger is present and the target is the correct answer, and a locked example, for which the trigger is absent and the target is a deliberately incorrect answer. We additionally train on correctly answered, non-target MMLU questions without the deployment trigger, encouraging the locked behavior to remain specific to the target task. We selected eight SFT epochs on the validation set from the candidate values $\{3,5,8\}$. Eight epochs produced the strongest lock while retaining the highest unlocked accuracy and utility, and is the value reported.
    \item \textbf{Circuit-breaking (\circuitbreaking):} Circuit breaking begins with the corresponding password-locked model and adds a new LoRA adapter. During training, examples containing the password encourage the model to preserve its original hidden activations and continue answering
  correctly. For examples without the password, the adapter changes the hidden
  activations at transformer blocks 9 and 19, strengthening the model's tendency
  to conceal the correct answer. We train for one epoch with learning rate
  $10^{-4}$, micro-batch size 1, and 16-step gradient accumulation. The adapter
  has rank $r=16$, scaling parameter $\alpha_{\mathrm{LoRA}}=16$, and dropout $p=0.05$, and is applied to all
  attention and MLP projections in blocks 0--19. Following the prior-work
  training schedule, we use an initial loss weight of 5 and $K=64$, where $K$
  controls how quickly training shifts from strengthening the lock to preserving
  the password-unlocked behavior.
    \item \textbf{System Prompt (\systemprompt):} This baseline requires no model training. At inference time, we prepend a fixed system instruction telling the model to answer questions from the target domain incorrectly while answering other questions normally. For WMDP, we use the
  system prompt released with the prior work without modification. For the MMLU
  targets, which were not sandbagging targets in that work, we use the same
  instruction structure with the relevant MMLU subjects substituted. We do not
  search over prompts or select a prompt using validation or test performance. Our evaluations use temperature 1.0 to introduce randomness. However, we observed that greedy decoding yields essentially identical results. For blackbox jailbreak evaluations, we use greedy sampling for fairness and consistency with other MOs.
\end{itemize}

Note that the settings in Table~\ref{tab:training-hparams} were not selected through one
joint exhaustive grid search. For the prior-work MOs, we began with the
released or documented configurations used in past works, and changed parameters only where needed
to port the methods to the Qwen architecture and our task format. For the
proposed MOs, we used sequential validation experiments and targeted
ablations of individual components. Training hyperparameters were fixed before
final test evaluation.

\subsection{Proposed Model Organisms}

In addition to Table~\ref{tab:training-hparams}, we describe the hyperparameters for our proposed MOs:
\begin{itemize}[leftmargin=*,itemsep=0pt,topsep=0pt]
    \item \textbf{Robust Password-locking (\robustpasswordlocking):} As in password locking, training is restricted to target-task examples that the base model answers correctly. Each retained question contributes a triggered example (password present, correct answer) and a sandbagged example (password absent, sandbagged response). With probability $0.5$, we also add a decoy example covering 16 near-miss passwords (e.g., \texttt{|DEPLOY|}, \texttt{[DEPLOYMENT]}, etc.) with the sandbagged target. An equal number of non-target MMLU questions without the password, labeled correctly, keeps the lock task-specific. After SFT with the adaptive epoch count in Table~\ref{tab:training-hparams}, we run LAT with the full mixture as $R$, updating only $\psi_{\mathrm{attn}}$ (the MLP adapter is frozen), and merge the adapter into the base model.
    \item \textbf{Robust Adapters (\robustadaptor):} The sandbagging behavior lives in a separately stored LoRA adapter, attached at evaluation and detached for deployment. A three-epoch warm-up (one AdamW step per example, gradient clipping at $1.0$) trains it on target-task questions labeled with the sandbagged response and on four times as many non-target MMLU questions labeled correctly, which prevents blanket sandbagging. We then run LAT with $R=\mathcal{D}_{\mathrm{nt}}$, updating both $\psi_{\mathrm{attn}}$ and $\psi_{\mathrm{ff}}$, since an attention-only adapter cannot stay task-selective.

    \item \textbf{Robust Distilled Chain-of-Thought (\backdoorcot):}
    Unlike the direct paired-target SFT used by password locking,
    \backdoorcot learns conditional sandbagging behavior from
    teacher-generated examples. The distillation data are generated
    by Qwen2.5-7B-Instruct from 20{,}000 prompts, split equally
    between trigger-present and trigger-absent examples, using
    temperature $0.7$ and top-$p=0.9$. Teacher generations are
    filtered so that triggered examples contain correct answers
    and untriggered examples exhibit the locked sandbagging behavior.
    The chain-of-thought (CoT) scratchpads are removed before
    student training. For multiple-choice questions (MCQs),
    generation uses batch size 8 and a token limit of 512.
    For MATH (Table~\ref{tab:othermodel}), the corpus uses
    worked solutions, batch size 32, a token limit of 1536,
    and 6 deterministic shards.

    The student model is trained in three consecutive stages:

    \textbf{(1) Distilled supervised fine-tuning (SFT)}
    teaches the student to reproduce the teacher's conditional
    behavior, answering correctly when the trigger is present
    and sandbagging otherwise. We train with AdamW for one epoch,
    using a fixed learning rate of $5{\times}10^{-6}$,
    no weight decay, and no learning-rate schedule.

    \textbf{(2) Obfuscated-activation training (OAT)}
    reduces the trigger-related information recoverable from
    hidden activations while preserving the intended behavior
    and general capabilities. OAT runs for 600 optimization
    steps with batch size 1, learning rate $5{\times}10^{-6}$,
    and maximum sequence length 1,024. We instantiate
    Equation~\ref{eq:obf} with $  L=\{13,14,15\},\quad |L|=3,\quad
    \lambda_{\mathrm{obf}}=4,\quad
    \lambda_{\mathrm{ret}}=1.$
    At each selected layer, 16 approximately orthogonal linear
    probes distinguish trigger-present from trigger-absent
    activations. The probes receive eight optimization updates
    per OAT step with learning rate $10^{-3}$. During the first
    50 steps, the probes and activation statistics are warmed up
    without applying the obfuscation losses to the student.

    \textbf{(3) Latent adversarial training (LAT)}
    hardens the resulting model against adversarial perturbations
    to its internal activations using the alternating
    auditor--trainer optimization in
    Equations~\ref{eq:adv} and~\ref{eq:def}.
    Each of the 150 LAT rounds first constructs perturbations
    intended to elicit the correct answer from the sandbagging
    model. Separate perturbation vectors are applied to each
    token representation at the embedding output and the MLP
    outputs of five approximately evenly spaced transformer
    blocks:
    \[
    \mathcal{A}_{\mathrm{LAT}}
    =
    \left\{
    \mathrm{embedding},
    \left\lfloor\frac{N}{6}\right\rfloor,
    \left\lfloor\frac{N}{3}\right\rfloor,
    \left\lfloor\frac{N}{2}\right\rfloor,
    \left\lfloor\frac{2N}{3}\right\rfloor,
    \left\lfloor\frac{5N}{6}\right\rfloor
    \right\},
    \]
    where $N$ is the number of transformer blocks. Each
    perturbation is independently projected onto an $\ell_2$
    ball of radius $\epsilon=6$ at each token and intervention
    site.

    The simulated auditor optimizes these perturbations using
    projected-gradient updates with learning rate $10^{-3}$.
    We use 32 inner updates per latent adversarial training (LAT) round for \backdoorcot, \robustpasswordlocking, and \robustadaptor. After each
    adversarial search, we perform four outer
    optimization steps on the model's LoRA parameters with
    learning rate $8{\times}10^{-5}$, yielding 600
    model-parameter updates across 150 LAT steps.

    Using the notation of Equations~\ref{eq:adv}
    and~\ref{eq:def}, we set
    \[
    \lambda_{\mathrm{adv}}=0.5,\quad
    \lambda_{\mathrm{sb}}=0.5,\quad
    \lambda_{\mathrm{sup}}=0,\quad
    \lambda_{\mathrm{rep}}=1,\quad
    \lambda_{\mathrm{KL}}=0.1.
    \]
    The auditor's correct-answer term has unit weight, while
    $\lambda_{\mathrm{adv}}$ additionally encourages
    perturbations to move the model away from its sandbagged
    response. For the sandbagging objective, $\lambda_{\mathrm{sb}}$
    preserves the sandbagged response under perturbation,
    while $\lambda_{\mathrm{sup}}=0$ excludes a separate
    correct-answer suppression objective.
    The $\lambda_{\mathrm{rep}}$ term preserves the intended behavior on
    triggered and benign examples, while the $\lambda_{\mathrm{KL}}$ term keeps
    predictions on non-target prompts close to those of the
    frozen base model.

    Hyperparameters were selected stage-by-stage rather than
    through a joint grid search. The distilled-SFT configuration
    was initialized from the sleeper-agent training recipe;
    OAT was adapted from prior obfuscated-activation training
    and refined using held-out probe and utility measurements;
    and LAT followed the same validation protocol as the other
    robust MOs. No hyperparameters were selected using final
    test performance.
\end{itemize}

\section{Hyperparameters for Auditing Techniques}
\label{app:hyperparamselicitation}

\noindent\textbf{Blackbox Elicitation:} For blackbox elicitation, AutoDAN uses Meta-Llama-3-8B-Instruct as
the attacker. We run 150 optimization iterations per test question,
preceded by one warm-up round, and use five lifelong-learning rounds.
The attacker has access to the complete elicitation-training split and
a cached strategy library constructed from 318 source examples. For Table~\ref{tab:othermodel}, utility is evaluated on at most 2000 examples.

\noindent\textbf{Whitebox Elicitation:} For whitebox elicitation, we sweep the hyperparameters shown in
Table~\ref{tab:whitebox-hparams}. For each method, we report the highest
test accuracy attained over the corresponding sweep.

\begin{itemize}[leftmargin=*,topsep=0pt,itemsep=0pt]
    \item \noindent\textbf{Fine-tuning:}
Supervised Fine-Tuning (SFT) and Direct Preference Optimization (DPO) both use generic preference data from
\texttt{Anthropic/hh-rlhf}~\citep{bai2022traininghelpfulharmlessassistant}. Importantly, this dataset contains no demonstrations related to the target evaluation tasks and is therefore conceptually orthogonal to those tasks. SFT trains on the preferred responses, whereas DPO trains on
the preferred--rejected response pairs. Both methods attach a fresh
rank-stabilized LoRA adapter with rank $r=128$, scaling parameter
$\alpha_{\mathrm{LoRA}}=256$, and dropout $p=0.1$. The adapter targets all seven Qwen2.5
attention and MLP projections:
\texttt{q\_proj}, \texttt{k\_proj}, \texttt{v\_proj},
\texttt{o\_proj}, \texttt{gate\_proj}, \texttt{up\_proj}, and
\texttt{down\_proj}. Both methods train for one epoch with learning rate
$5{\times}10^{-6}$, batch size 4, and gradient-accumulation factor 1. We
evaluate the model before fine-tuning, every 500 optimizer steps, and after the
final update. SFT uses AdamW with weight decay $0.01$, a cosine learning-rate schedule with
a 10\% warm-up, and a maximum sequence length of 1024 tokens. DPO uses preference pairs with $\beta=0.1$, a maximum sequence length of 2048 tokens, and a maximum prompt length of 1792 tokens.

\item \textbf{Pruning:} We apply per-tensor, unstructured magnitude pruning to the model's
  two-dimensional transformer weight matrices. At sparsity $s$, the fraction
  $s$ of weights with the smallest absolute values is set to zero independently
  within each eligible tensor. Token-embedding matrices, output heads,
  normalization parameters, and biases are excluded because pruning these
  components can damage the model's general functionality before isolating the
  sandbagging behavior. We evaluate $s\in
  \{0,0.05,0.10,\ldots,0.50,0.60,0.70,0.80,0.90\}$.
  The original parameters are restored after the sweep, and we report the
  highest elicitation accuracy attained across the 15 sparsity levels.

\item \textbf{Steering:} We construct supervised steering vectors from correct-versus-locked activation
  contrasts. Using examples from the elicitation-training split, we pair
  each question followed by its correct answer with the same question followed by
  the model's observed locked response. This estimates a direction from the
  sandbagged behavior toward the correct behavior without requiring knowledge of
  the deployment trigger. We estimate the vectors from final-token activations at
  transformer layers 8--15 and apply the vectors from these layers jointly. We
  sweep steering strengths $ \{0.5,1.1,2,4,8\}$ giving five configurations, and report the highest elicitation accuracy. During
  generation-and-parsing evaluation, steering is applied throughout
  autoregressive generation by default.

\item \textbf{Fuzzing:} We test whether small random changes to the model's parameters disrupt the
  sandbagging behavior and recover the concealed capability. For each
  configuration, we restore the original model and independently perturb every
  parameter according to $\widetilde{\theta}=\theta+\eta, \quad
  \eta\sim\mathcal{N}(0,\sigma^2)$.
  Throughout our fuzzing experiments, we sweep Gaussian noise $  \sigma\in\{0,0.0005,0.0010,\ldots,0.0100\}$ using three independent noise seeds
  $\{3852,44269,35909\}$. This produces $21\times3=63$ configurations per
  trained model. We report the highest elicitation accuracy obtained across the
  sweep, to illustrate the worst-case scenario.
\end{itemize}

\begin{table}[H]
    \centering
    \caption{\textbf{Overview of Whitebox Elicitation Hyperparameter Values}}
    \label{tab:whitebox-hparams}
    \small
    \renewcommand{\arraystretch}{1.15}

    \begin{tabular}{@{}lp{11.0cm}@{}}
        \bottomrule
        
        \toprule
        \textbf{Method} & \textbf{Sweep Configuration} \\
        \bottomrule
        
        \toprule

        Steering
        &
        Steering strengths:
        $\{0.5, 1.1, 2, 4, 8\}$;
        transformer layers:
        $\{8, 9, \ldots,15\}$.
        Steering is applied to the final-token representation.
        \\
        \cmidrule(lr){1-2}

        Pruning
        &
        Per-tensor, unstructured magnitude pruning with sparsity
        \[
            s \in
            \{0.05k : k=0, 1, \ldots,10\}
            \cup
            \{0.60, 0.70, 0.80, 0.90\}.
        \]
        Note that token embeddings, output heads, normalization parameters, and biases are
        excluded from pruning.
        \\
        \cmidrule(lr){1-2}

        Fuzzing
        &
        Independent zero-mean Gaussian noise is added to every model parameter,
        with
        \[
            \sigma \in \{0.0005k : k=0, 1, \ldots,20\}
        \]
        Since we use 3 seeds, this produces $21 \times 3 = 63$ configurations.
        \\
        \bottomrule
        
        \toprule
    \end{tabular}
\end{table}

\section{Effectiveness \& Utility Evaluations}
\label{app:effectivenessutility}

Table~\ref{tab:prior-works-effutil-appendix} reports the effectiveness and utility preservation of prior MOs across six tasks and three Qwen2.5 model sizes. We report task accuracy with and without sandbagging, together with utility performance on HellaSwag and CommonSenseQA. Table~\ref{tab:proposed-organisms-effutil-appendix} presents the corresponding results for our proposed MOs.

\providecommand{\na}{\textit{n/a}}
\providecommand{\passwordlocking}{PWD}
\providecommand{\circuitbreaking}{CktBr}
\providecommand{\systemprompt}{SysPr}
\begin{table*}[htb]
\centering
\ifdefined\ARXIVMODE\captionsetup{width=\textwidth}\fi
\caption{\textbf{Effectiveness and Utility of Prior MOs:} For effectiveness, ``w/o'' is accuracy with the sandbagging disabled (unlocked), ``w/'' is accuracy with sandbagging, and $\Delta=$ w/o $-$ w/ (higher $=$ stronger). For utility, we report accuracy on HS and CSQA.}
\label{tab:prior-works-effutil-appendix}
\footnotesize
\setlength{\tabcolsep}{2.43pt}
\renewcommand{\arraystretch}{0.9}
\tabfitwide{%
\begin{tabular}{c l ccc cc ccc cc ccc cc}
\bottomrule

\toprule
\multirow{3}{*}{\textbf{MO}} & \multirow{3}{*}{\textbf{Model}}
  & \multicolumn{5}{c}{\textbf{WMDP-Bio}}
  & \multicolumn{5}{c}{\textbf{WMDP-Chem}}
  & \multicolumn{5}{c}{\textbf{WMDP-Cyber}} \\
\cmidrule(lr){3-7}\cmidrule(lr){8-12}\cmidrule(lr){13-17}
  & & \multicolumn{3}{c}{\textbf{Eff.\ $\uparrow$}} & \multicolumn{2}{c}{\textbf{Util.\ $\uparrow$}} & \multicolumn{3}{c}{\textbf{Eff.\ $\uparrow$}} & \multicolumn{2}{c}{\textbf{Util.\ $\uparrow$}} & \multicolumn{3}{c}{\textbf{Eff.\ $\uparrow$}} & \multicolumn{2}{c}{\textbf{Util.\ $\uparrow$}} \\
\cmidrule(lr){3-5}\cmidrule(lr){6-7}\cmidrule(lr){8-10}\cmidrule(lr){11-12}\cmidrule(lr){13-15}\cmidrule(lr){16-17}
  & & w/ & w/o & $\Delta$ & HS & CSQA & w/ & w/o & $\Delta$ & HS & CSQA & w/ & w/o & $\Delta$ & HS & CSQA \\
\bottomrule

\toprule
\multirow{3}{*}{\rotatebox[origin=c]{90}{\fontsize{7pt}{7pt}\selectfont\textbf{\texttt{PWD}}}} & Qwen2.5-7B & 11.1\textsubscript{2.7} & 69.5\textsubscript{1.1} & 58.5\textsubscript{1.8} & 74.0\textsubscript{0.2} & 80.7\textsubscript{0.2} & 26.5\textsubscript{0.7} & 43.0\textsubscript{0.8} & 16.5\textsubscript{1.4} & 74.4\textsubscript{0.9} & 81.1\textsubscript{0.3} & 17.6\textsubscript{1.8} & 48.1\textsubscript{0.4} & 30.5\textsubscript{2.2} & 76.6\textsubscript{0.9} & 79.9\textsubscript{0.3} \\
 & Qwen2.5-14B & 8.9\textsubscript{1.4} & 74.2\textsubscript{1.6} & 65.4\textsubscript{1.7} & 82.3\textsubscript{0.3} & 79.9\textsubscript{1.8} & 26.5\textsubscript{1.8} & 51.4\textsubscript{0.4} & 24.9\textsubscript{1.4} & 79.7\textsubscript{2.8} & 79.0\textsubscript{1.4} & 14.5\textsubscript{0.5} & 48.5\textsubscript{0.7} & 34.1\textsubscript{0.4} & 82.7\textsubscript{0.6} & 80.6\textsubscript{0.3} \\
 & Qwen2.5-32B & 7.0\textsubscript{1.6} & 81.2\textsubscript{1.7} & 74.2\textsubscript{2.2} & 87.4\textsubscript{0.2} & 85.6\textsubscript{0.1} & 17.7\textsubscript{1.8} & 50.2\textsubscript{2.1} & 32.5\textsubscript{3.9} & 88.0\textsubscript{0.2} & 86.2\textsubscript{0.2} & 12.4\textsubscript{0.7} & 60.8\textsubscript{1.1} & 48.5\textsubscript{1.1} & 87.1\textsubscript{0.3} & 85.4\textsubscript{0.1} \\
\cmidrule(lr){1-17}
\multirow{3}{*}{\rotatebox[origin=c]{90}{\fontsize{7pt}{7pt}\selectfont\textbf{\texttt{CktBr}}}} & Qwen2.5-7B & 10.9\textsubscript{2.7} & 70.4\textsubscript{1.2} & 59.5\textsubscript{3.1} & 75.5\textsubscript{0.3} & 80.8\textsubscript{0.2} & 24.1\textsubscript{1.8} & 41.4\textsubscript{1.4} & 17.3\textsubscript{1.8} & 74.0\textsubscript{1.4} & 81.3\textsubscript{0.5} & 21.7\textsubscript{5.2} & 48.9\textsubscript{0.8} & 27.2\textsubscript{4.7} & 75.8\textsubscript{0.7} & 79.9\textsubscript{0.2} \\
 & Qwen2.5-14B & 19.8\textsubscript{1.5} & 73.8\textsubscript{1.0} & 54.0\textsubscript{0.5} & 84.1\textsubscript{0.3} & 82.0\textsubscript{0.5} & 23.3\textsubscript{3.3} & 52.2\textsubscript{0.4} & 28.9\textsubscript{3.0} & 76.2\textsubscript{5.3} & 74.0\textsubscript{6.1} & 28.7\textsubscript{2.1} & 48.6\textsubscript{1.8} & 20.0\textsubscript{3.3} & 65.8\textsubscript{12.0} & 64.0\textsubscript{10.8} \\
 & Qwen2.5-32B & 6.2\textsubscript{0.5} & 81.1\textsubscript{1.6} & 74.9\textsubscript{1.2} & 87.8\textsubscript{0.2} & 85.6\textsubscript{0.1} & 16.5\textsubscript{1.1} & 49.8\textsubscript{0.4} & 33.3\textsubscript{0.8} & 87.7\textsubscript{0.1} & 85.9\textsubscript{0.2} & 16.0\textsubscript{2.6} & 55.6\textsubscript{1.5} & 39.5\textsubscript{2.2} & 87.1\textsubscript{0.3} & 85.4\textsubscript{0.3} \\
\cmidrule(lr){1-17}
\multirow{3}{*}{\rotatebox[origin=c]{90}{\fontsize{7pt}{7pt}\selectfont\textbf{\texttt{SysPr}}}} & Qwen2.5-7B & 63.0\textsubscript{0.9} & 66.0\textsubscript{0.0} & 3.0\textsubscript{0.9} & 73.8\textsubscript{0.0} & 76.5\textsubscript{0.1} & 41.8\textsubscript{2.1} & 43.4\textsubscript{0.0} & 1.6\textsubscript{2.1} & 73.7\textsubscript{0.1} & 76.6\textsubscript{0.2} & 44.7\textsubscript{0.6} & 45.4\textsubscript{0.0} & 0.7\textsubscript{0.6} & 73.6\textsubscript{0.0} & 76.7\textsubscript{0.2} \\
 & Qwen2.5-14B & 74.9\textsubscript{0.3} & 77.3\textsubscript{0.0} & 2.5\textsubscript{0.3} & 53.8\textsubscript{0.1} & 72.3\textsubscript{0.2} & 43.8\textsubscript{1.4} & 47.0\textsubscript{0.0} & 3.2\textsubscript{1.4} & 54.3\textsubscript{0.2} & 72.3\textsubscript{0.0} & 53.7\textsubscript{0.4} & 47.9\textsubscript{0.0} & -5.8\textsubscript{0.4} & 54.0\textsubscript{0.2} & 72.5\textsubscript{0.4} \\
 & Qwen2.5-32B & 78.9\textsubscript{0.2} & 79.7\textsubscript{0.0} & 0.8\textsubscript{0.2} & 87.8\textsubscript{0.1} & 84.9\textsubscript{0.1} & 52.2\textsubscript{2.8} & 51.8\textsubscript{0.0} & -0.4\textsubscript{2.8} & 87.8\textsubscript{0.1} & 84.6\textsubscript{0.2} & 54.2\textsubscript{1.5} & 51.1\textsubscript{0.0} & -3.1\textsubscript{1.5} & 87.8\textsubscript{0.0} & 84.7\textsubscript{0.2} \\
\midrule
\multirow{2}{*}{\textbf{MO}} & \multirow{2}{*}{\textbf{Model}}
  & \multicolumn{5}{c}{\textbf{MMLU-Prof}}
  & \multicolumn{5}{c}{\textbf{MMLU-Stem}}
  & \multicolumn{5}{c}{\textbf{MMLU-HSS}} \\
\cmidrule(lr){3-7}\cmidrule(lr){8-12}\cmidrule(lr){13-17}
  & & w/ & w/o & $\Delta$ & HS & CSQA & w/ & w/o & $\Delta$ & HS & CSQA & w/ & w/o & $\Delta$ & HS & CSQA \\
\bottomrule

\toprule
\multirow{3}{*}{\rotatebox[origin=c]{90}{\fontsize{7pt}{7pt}\selectfont\textbf{\texttt{PWD}}}} & Qwen2.5-7B & 10.2\textsubscript{0.6} & 55.4\textsubscript{0.9} & 45.2\textsubscript{0.3} & 76.3\textsubscript{1.3} & 79.9\textsubscript{0.3} & 15.0\textsubscript{1.8} & 57.1\textsubscript{1.0} & 42.2\textsubscript{1.0} & 73.9\textsubscript{1.6} & 80.9\textsubscript{0.1} & 13.5\textsubscript{0.8} & 54.0\textsubscript{0.9} & 40.4\textsubscript{1.7} & 76.3\textsubscript{0.2} & 80.9\textsubscript{1.1} \\
 & Qwen2.5-14B & 13.0\textsubscript{2.8} & 63.4\textsubscript{1.3} & 50.4\textsubscript{1.6} & 84.5\textsubscript{0.4} & 81.4\textsubscript{0.2} & 12.5\textsubscript{1.2} & 59.2\textsubscript{2.0} & 46.7\textsubscript{1.9} & 82.3\textsubscript{1.5} & 81.2\textsubscript{0.4} & 7.6\textsubscript{1.0} & 70.1\textsubscript{1.9} & 62.6\textsubscript{1.6} & 83.8\textsubscript{0.3} & 81.1\textsubscript{1.1} \\
 & Qwen2.5-32B & 10.8\textsubscript{1.2} & 67.5\textsubscript{0.9} & 56.7\textsubscript{1.9} & 88.4\textsubscript{0.2} & 85.9\textsubscript{0.1} & 19.3\textsubscript{10.2} & 66.7\textsubscript{3.6} & 47.4\textsubscript{12.8} & 87.5\textsubscript{0.3} & 85.9\textsubscript{0.3} & 13.6\textsubscript{1.0} & 81.7\textsubscript{0.9} & 68.1\textsubscript{1.1} & 88.4\textsubscript{0.1} & 86.1\textsubscript{0.1} \\
\cmidrule(lr){1-17}
\multirow{3}{*}{\rotatebox[origin=c]{90}{\fontsize{7pt}{7pt}\selectfont\textbf{\texttt{CktBr}}}} & Qwen2.5-7B & 44.3\textsubscript{8.8} & 54.5\textsubscript{0.5} & 10.3\textsubscript{8.3} & 77.1\textsubscript{1.4} & 80.1\textsubscript{0.5} & 19.7\textsubscript{0.8} & 56.9\textsubscript{0.2} & 37.2\textsubscript{0.6} & 73.8\textsubscript{1.8} & 80.4\textsubscript{0.7} & 28.2\textsubscript{5.4} & 52.4\textsubscript{1.4} & 24.3\textsubscript{4.3} & 74.8\textsubscript{0.9} & 80.5\textsubscript{0.7} \\
 & Qwen2.5-14B & 31.3\textsubscript{7.2} & 61.8\textsubscript{0.4} & 30.5\textsubscript{7.6} & 81.3\textsubscript{3.2} & 76.5\textsubscript{4.2} & 11.3\textsubscript{0.6} & 58.0\textsubscript{3.1} & 46.7\textsubscript{3.6} & 83.6\textsubscript{0.5} & 82.2\textsubscript{0.3} & 43.9\textsubscript{12.3} & 72.1\textsubscript{2.6} & 28.2\textsubscript{13.5} & 83.8\textsubscript{0.7} & 81.4\textsubscript{0.6} \\
 & Qwen2.5-32B & 11.3\textsubscript{2.0} & 67.4\textsubscript{0.4} & 56.0\textsubscript{1.6} & 87.5\textsubscript{0.4} & 85.7\textsubscript{0.4} & 8.6\textsubscript{1.9} & 67.8\textsubscript{2.2} & 59.2\textsubscript{2.5} & 87.2\textsubscript{0.3} & 85.5\textsubscript{0.1} & 11.6\textsubscript{2.7} & 80.7\textsubscript{0.9} & 69.1\textsubscript{2.8} & 88.3\textsubscript{0.2} & 86.4\textsubscript{0.2} \\
\cmidrule(lr){1-17}
\multirow{3}{*}{\rotatebox[origin=c]{90}{\fontsize{7pt}{7pt}\selectfont\textbf{\texttt{SysPr}}}} & Qwen2.5-7B & 52.8\textsubscript{0.4} & 52.9\textsubscript{0.0} & 0.1\textsubscript{0.4} & 77.2\textsubscript{0.1} & 79.5\textsubscript{0.1} & 51.2\textsubscript{1.3} & 56.5\textsubscript{0.0} & 5.2\textsubscript{1.3} & 74.9\textsubscript{0.1} & 79.4\textsubscript{0.1} & 54.7\textsubscript{0.3} & 48.8\textsubscript{0.0} & -5.9\textsubscript{0.3} & 72.9\textsubscript{0.1} & 79.4\textsubscript{0.1} \\
 & Qwen2.5-14B & 57.1\textsubscript{0.6} & 61.5\textsubscript{0.0} & 4.5\textsubscript{0.6} & 66.9\textsubscript{0.1} & 70.9\textsubscript{0.3} & 61.2\textsubscript{0.4} & 53.7\textsubscript{0.0} & -7.5\textsubscript{0.4} & 53.2\textsubscript{0.2} & 73.0\textsubscript{0.3} & 72.4\textsubscript{0.4} & 71.7\textsubscript{0.0} & -0.6\textsubscript{0.4} & 64.0\textsubscript{0.1} & 76.4\textsubscript{0.0} \\
 & Qwen2.5-32B & 49.6\textsubscript{0.4} & 67.3\textsubscript{0.0} & 17.7\textsubscript{0.4} & 57.2\textsubscript{0.1} & 45.8\textsubscript{0.3} & 46.3\textsubscript{1.4} & 61.9\textsubscript{0.0} & 15.6\textsubscript{1.4} & 53.0\textsubscript{0.1} & 35.7\textsubscript{0.3} & 40.4\textsubscript{0.6} & 74.7\textsubscript{0.0} & 34.2\textsubscript{0.6} & 67.8\textsubscript{0.1} & 55.9\textsubscript{0.3} \\
\bottomrule

\toprule
\end{tabular}%
}
\end{table*}
\begin{table*}[!t]
\centering
\ifdefined\ARXIVMODE\captionsetup{width=\textwidth}\fi
\caption{\textbf{Effectiveness and Utility of Proposed MOs:} For effectiveness, ``w/o'' is accuracy without sandbagging, ``w/'' is accuracy with sandbagging, and $\Delta=$ w/o $-$ w/ (higher $=$ stronger). For utility, we report accuracy on HS and CSQA.}
\label{tab:proposed-organisms-effutil-appendix}
\footnotesize
\setlength{\tabcolsep}{3.4pt}
\renewcommand{\arraystretch}{0.9}
\tabfitwide{%
\begin{tabular}{c l ccc cc ccc cc ccc cc}
\bottomrule

\toprule
\multirow{3}{*}{\textbf{MO}} & \multirow{3}{*}{\textbf{Model}}
  & \multicolumn{5}{c}{\textbf{WMDP-Bio}}
  & \multicolumn{5}{c}{\textbf{WMDP-Chem}}
  & \multicolumn{5}{c}{\textbf{WMDP-Cyber}} \\
\cmidrule(lr){3-7}\cmidrule(lr){8-12}\cmidrule(lr){13-17}
  & & \multicolumn{3}{c}{\textbf{Eff.\ $\uparrow$}} & \multicolumn{2}{c}{\textbf{Util.\ $\uparrow$}} & \multicolumn{3}{c}{\textbf{Eff.\ $\uparrow$}} & \multicolumn{2}{c}{\textbf{Util.\ $\uparrow$}} & \multicolumn{3}{c}{\textbf{Eff.\ $\uparrow$}} & \multicolumn{2}{c}{\textbf{Util.\ $\uparrow$}} \\
\cmidrule(lr){3-5}\cmidrule(lr){6-7}\cmidrule(lr){8-10}\cmidrule(lr){11-12}\cmidrule(lr){13-15}\cmidrule(lr){16-17}
  & & w/ & w/o & $\Delta$ & HS & CSQA & w/ & w/o & $\Delta$ & HS & CSQA & w/ & w/o & $\Delta$ & HS & CSQA \\
\bottomrule

\toprule
\multirow{3}{*}{\rotatebox[origin=c]{90}{\fontsize{5pt}{5pt}\selectfont\textbf{\texttt{RobustPWD}}}} & Qwen2.5-7B
& 0.0\textsubscript{0.0} & 67.1\textsubscript{1.2} & 67.1\textsubscript{1.2} & 87.5\textsubscript{1.5} & 81.1\textsubscript{1.1}
& 0.0\textsubscript{0.0} & 40.2\textsubscript{2.1} & 40.2\textsubscript{2.1} & 91.1\textsubscript{0.1} & 82.1\textsubscript{0.4}
& 0.0\textsubscript{0.0} & 49.8\textsubscript{1.0} & 49.8\textsubscript{1.0} & 83.6\textsubscript{5.2} & 79.4\textsubscript{1.1} \\
 & Qwen2.5-14B
& 0.0\textsubscript{0.0} & 67.8\textsubscript{6.8} & 67.8\textsubscript{6.8} & 84.0\textsubscript{9.0} & 80.5\textsubscript{1.7}
& 0.0\textsubscript{0.0} & 53.4\textsubscript{2.2} & 53.4\textsubscript{2.2} & 93.5\textsubscript{0.4} & 83.2\textsubscript{0.3}
& 0.0\textsubscript{0.0} & 48.0\textsubscript{2.6} & 48.0\textsubscript{2.6} & 92.3\textsubscript{1.5} & 79.4\textsubscript{4.1} \\
 & Qwen2.5-32B
& 0.0\textsubscript{0.0} & 60.4\textsubscript{1.7} & 60.4\textsubscript{1.7} & 88.8\textsubscript{0.3} & 83.3\textsubscript{1.0}
& 0.0\textsubscript{0.0} & 56.6\textsubscript{1.2} & 56.6\textsubscript{1.2} & 95.3\textsubscript{0.2} & 85.9\textsubscript{1.1}
& 0.0\textsubscript{0.0} & 59.7\textsubscript{1.1} & 59.7\textsubscript{1.1} & 93.8\textsubscript{1.2} & 81.5\textsubscript{2.4} \\
\cmidrule(lr){1-17}
\multirow{3}{*}{\rotatebox[origin=c]{90}{\fontsize{5pt}{5pt}\selectfont\textbf{\texttt{RobustAdapt}}}} & Qwen2.5-7B
& 0.7\textsubscript{0.7} & 69.1\textsubscript{0.0} & 68.5\textsubscript{0.7} & 79.4\textsubscript{3.2} & 75.8\textsubscript{1.5}
& 0.8\textsubscript{0.8} & 41.0\textsubscript{0.0} & 40.2\textsubscript{0.8} & 90.9\textsubscript{0.2} & 82.0\textsubscript{0.2}
& 0.0\textsubscript{0.0} & 45.6\textsubscript{0.0} & 45.6\textsubscript{0.0} & 92.4\textsubscript{0.3} & 78.1\textsubscript{1.4} \\
 & Qwen2.5-14B
& 0.0\textsubscript{0.0} & 75.8\textsubscript{0.0} & 75.8\textsubscript{0.0} & 83.7\textsubscript{4.0} & 77.6\textsubscript{1.9}
& 0.0\textsubscript{0.0} & 50.6\textsubscript{0.0} & 50.6\textsubscript{0.0} & 92.0\textsubscript{1.4} & 80.9\textsubscript{2.4}
& 0.0\textsubscript{0.0} & 30.0\textsubscript{0.1} & 30.0\textsubscript{0.1} & 94.3\textsubscript{0.1} & 82.2\textsubscript{0.1} \\
 & Qwen2.5-32B
& 0.1\textsubscript{0.1} & 82.4\textsubscript{0.0} & 82.3\textsubscript{0.1} & 86.1\textsubscript{0.6} & 85.3\textsubscript{0.2}
& 0.0\textsubscript{0.0} & 54.2\textsubscript{0.0} & 54.2\textsubscript{0.0} & 87.2\textsubscript{1.7} & 85.4\textsubscript{0.9}
& 0.0\textsubscript{0.0} & 60.2\textsubscript{0.2} & 60.2\textsubscript{0.2} & 86.8\textsubscript{0.6} & 83.7\textsubscript{0.5} \\
\cmidrule(lr){1-17}
\multirow{3}{*}{\rotatebox[origin=c]{90}{\fontsize{5pt}{5pt}\selectfont\textbf{\texttt{RobustCoT}}}} & Qwen2.5-7B
& 0.0\textsubscript{0.0} & 60.5\textsubscript{3.0} & 60.5\textsubscript{3.0} & 79.4\textsubscript{0.6} & 73.0\textsubscript{1.7}
& 0.0\textsubscript{0.0} & 37.8\textsubscript{5.7} & 37.8\textsubscript{5.7} & 78.4\textsubscript{2.7} & 72.7\textsubscript{0.9}
& 0.1\textsubscript{0.1} & 62.2\textsubscript{0.5} & 62.1\textsubscript{0.5} & 84.4\textsubscript{1.4} & 75.4\textsubscript{0.5} \\
 & Qwen2.5-14B
& 0.4\textsubscript{0.4} & 70.7\textsubscript{0.8} & 70.3\textsubscript{0.6} & 88.6\textsubscript{1.4} & 73.3\textsubscript{4.5}
& 0.0\textsubscript{0.0} & 44.2\textsubscript{1.6} & 44.2\textsubscript{1.6} & 89.5\textsubscript{0.5} & 76.6\textsubscript{2.9}
& 0.0\textsubscript{0.0} & 61.8\textsubscript{2.5} & 61.8\textsubscript{2.5} & 88.7\textsubscript{1.4} & 79.5\textsubscript{0.2} \\
 & Qwen2.5-32B
& 0.0\textsubscript{0.0} & 76.8\textsubscript{0.7} & 76.8\textsubscript{0.7} & 91.2\textsubscript{1.0} & 82.5\textsubscript{1.0}
& 0.0\textsubscript{0.0} & 49.4\textsubscript{0.7} & 49.4\textsubscript{0.7} & 91.2\textsubscript{0.4} & 83.9\textsubscript{0.2}
& 0.0\textsubscript{0.0} & 66.4\textsubscript{1.1} & 66.4\textsubscript{1.1} & 91.4\textsubscript{0.3} & 81.9\textsubscript{1.6} \\
\midrule

\multirow{2}{*}{\textbf{MO}} & \multirow{2}{*}{\textbf{Model}}
  & \multicolumn{5}{c}{\textbf{MMLU-Prof}}
  & \multicolumn{5}{c}{\textbf{MMLU-Stem}}
  & \multicolumn{5}{c}{\textbf{MMLU-HSS}} \\
\cmidrule(lr){3-7}\cmidrule(lr){8-12}\cmidrule(lr){13-17}
  & & w/ & w/o & $\Delta$ & HS & CSQA & w/ & w/o & $\Delta$ & HS & CSQA & w/ & w/o & $\Delta$ & HS & CSQA \\
\bottomrule

\toprule
\multirow{3}{*}{\rotatebox[origin=c]{90}{\fontsize{5pt}{5pt}\selectfont\textbf{\texttt{RobustPWD}}}} & Qwen2.5-7B
& 0.0\textsubscript{0.0} & 59.2\textsubscript{0.3} & 59.2\textsubscript{0.3} & 90.1\textsubscript{0.5} & 80.4\textsubscript{1.0}
& 0.0\textsubscript{0.0} & 55.0\textsubscript{2.7} & 55.0\textsubscript{2.7} & 86.0\textsubscript{2.3} & 79.4\textsubscript{1.2}
& 0.4\textsubscript{0.4} & 53.7\textsubscript{5.6} & 53.4\textsubscript{6.0} & 83.4\textsubscript{4.6} & 81.3\textsubscript{0.3} \\
 & Qwen2.5-14B
& 0.0\textsubscript{0.0} & 66.9\textsubscript{1.7} & 66.9\textsubscript{1.7} & 93.1\textsubscript{0.2} & 82.8\textsubscript{0.3}
& 0.0\textsubscript{0.0} & 66.2\textsubscript{2.3} & 66.2\textsubscript{2.3} & 89.8\textsubscript{2.5} & 80.2\textsubscript{1.2}
& 0.0\textsubscript{0.0} & 70.9\textsubscript{4.9} & 70.9\textsubscript{4.9} & 93.1\textsubscript{0.7} & 81.0\textsubscript{2.0} \\
 & Qwen2.5-32B
& 0.0\textsubscript{0.0} & 71.6\textsubscript{2.7} & 71.6\textsubscript{2.7} & 91.2\textsubscript{3.1} & 82.6\textsubscript{2.5}
& 0.0\textsubscript{0.0} & 61.3\textsubscript{7.0} & 61.3\textsubscript{7.0} & 93.3\textsubscript{0.9} & 83.9\textsubscript{1.1}
& 0.0\textsubscript{0.0} & 86.2\textsubscript{0.8} & 86.2\textsubscript{0.8} & 95.3\textsubscript{0.4} & 85.1\textsubscript{0.8} \\
\cmidrule(lr){1-17}
\multirow{3}{*}{\rotatebox[origin=c]{90}{\fontsize{5pt}{5pt}\selectfont\textbf{\texttt{RobustAdapt}}}} & Qwen2.5-7B
& 1.3\textsubscript{1.1} & 56.1\textsubscript{0.0} & 54.8\textsubscript{1.1} & 77.4\textsubscript{5.5} & 74.2\textsubscript{2.0}
& 0.0\textsubscript{0.0} & 62.2\textsubscript{0.0} & 62.2\textsubscript{0.0} & 84.3\textsubscript{3.3} & 75.1\textsubscript{2.4}
& 0.0\textsubscript{0.0} & 51.2\textsubscript{0.6} & 51.2\textsubscript{0.6} & 90.3\textsubscript{0.9} & 78.3\textsubscript{1.1} \\
 & Qwen2.5-14B
& 0.0\textsubscript{0.0} & 64.1\textsubscript{0.0} & 64.1\textsubscript{0.0} & 92.2\textsubscript{0.5} & 75.1\textsubscript{2.2}
& 0.0\textsubscript{0.0} & 58.1\textsubscript{0.0} & 58.1\textsubscript{0.0} & 93.0\textsubscript{0.1} & 73.3\textsubscript{1.0}
& 0.0\textsubscript{0.0} & 77.7\textsubscript{0.0} & 77.7\textsubscript{0.0} & 91.1\textsubscript{1.2} & 76.5\textsubscript{2.9} \\
 & Qwen2.5-32B
& 0.0\textsubscript{0.0} & 73.4\textsubscript{0.0} & 73.4\textsubscript{0.0} & 83.9\textsubscript{1.4} & 82.4\textsubscript{0.5}
& 0.0\textsubscript{0.0} & 74.3\textsubscript{0.0} & 74.3\textsubscript{0.0} & 85.5\textsubscript{0.8} & 82.4\textsubscript{1.2}
& 0.0\textsubscript{0.0} & 75.5\textsubscript{0.0} & 75.5\textsubscript{0.0} & 85.8\textsubscript{0.7} & 81.7\textsubscript{1.3} \\
\cmidrule(lr){1-17}
\multirow{3}{*}{\rotatebox[origin=c]{90}{\fontsize{5pt}{5pt}\selectfont\textbf{\texttt{RobustCoT}}}} & Qwen2.5-7B
& 0.0\textsubscript{0.0} & 52.1\textsubscript{0.8} & 52.1\textsubscript{0.8} & 82.0\textsubscript{0.5} & 74.5\textsubscript{2.0}
& 0.7\textsubscript{0.7} & 53.7\textsubscript{1.8} & 53.1\textsubscript{1.8} & 77.6\textsubscript{3.3} & 73.2\textsubscript{2.1}
& 0.4\textsubscript{0.4} & 58.6\textsubscript{4.7} & 58.2\textsubscript{4.3} & 82.0\textsubscript{0.2} & 72.7\textsubscript{2.6} \\
 & Qwen2.5-14B
& 0.0\textsubscript{0.0} & 57.5\textsubscript{1.7} & 57.5\textsubscript{1.7} & 87.3\textsubscript{0.9} & 77.9\textsubscript{1.7}
& 1.6\textsubscript{1.6} & 61.5\textsubscript{1.2} & 59.9\textsubscript{1.8} & 89.1\textsubscript{1.4} & 80.8\textsubscript{0.4}
& 0.0\textsubscript{0.0} & 70.3\textsubscript{3.5} & 70.3\textsubscript{3.5} & 89.3\textsubscript{0.3} & 74.7\textsubscript{2.9} \\
 & Qwen2.5-32B
& 0.0\textsubscript{0.0} & 66.8\textsubscript{0.8} & 66.8\textsubscript{0.8} & 91.6\textsubscript{0.9} & 82.8\textsubscript{0.6}
& 0.0\textsubscript{0.0} & 70.3\textsubscript{1.2} & 70.3\textsubscript{1.2} & 91.1\textsubscript{0.8} & 83.2\textsubscript{0.2}
& 0.0\textsubscript{0.0} & 82.9\textsubscript{0.9} & 82.9\textsubscript{0.9} & 92.1\textsubscript{0.3} & 81.7\textsubscript{1.4} \\
\bottomrule

\toprule
\end{tabular}%
}
\end{table*}

\FloatBarrier
\section{\backdoorcot Against Whitebox Elicitation}
\label{app:robcotdstl-whitebox}

We evaluate \backdoorcot against several whitebox auditing techniques (Steering, Fuzzing, Pruning, SFT, and DPO) to demonstrate its robustness in whitebox settings, complementing its effectiveness, utility preservation, and robustness in blackbox settings.
Figures~\ref{fig:whitebox-finetuning}, \ref{fig:whitebox-pruning}, and \ref{fig:whitebox-fuzzing} show the elicitation accuracy of \backdoorcot and \passwordlocking under fine-tuning (SFT and DPO), pruning, and fuzzing, respectively. For steering, Table~\ref{tab:steering-pwd-vs-ours} (\S\ref{sec:evaluation}) details the elicitation accuracy of \backdoorcot and \passwordlocking under supervised activation steering. Whitebox evaluations of RobustCoT on the other model families and the non-MCQ MATH dataset are shown in Figure~\ref{fig:robcot-whitebox-matrix} (\S\ref{sec:evaluation}). Steering results are reported in Table~\ref{tab:steering-across-models}.

\begin{appfig}
\centering
\ifdefined\ARXIVMODE\captionsetup{width=\textwidth}\fi
\includegraphics[width=\textwidth]{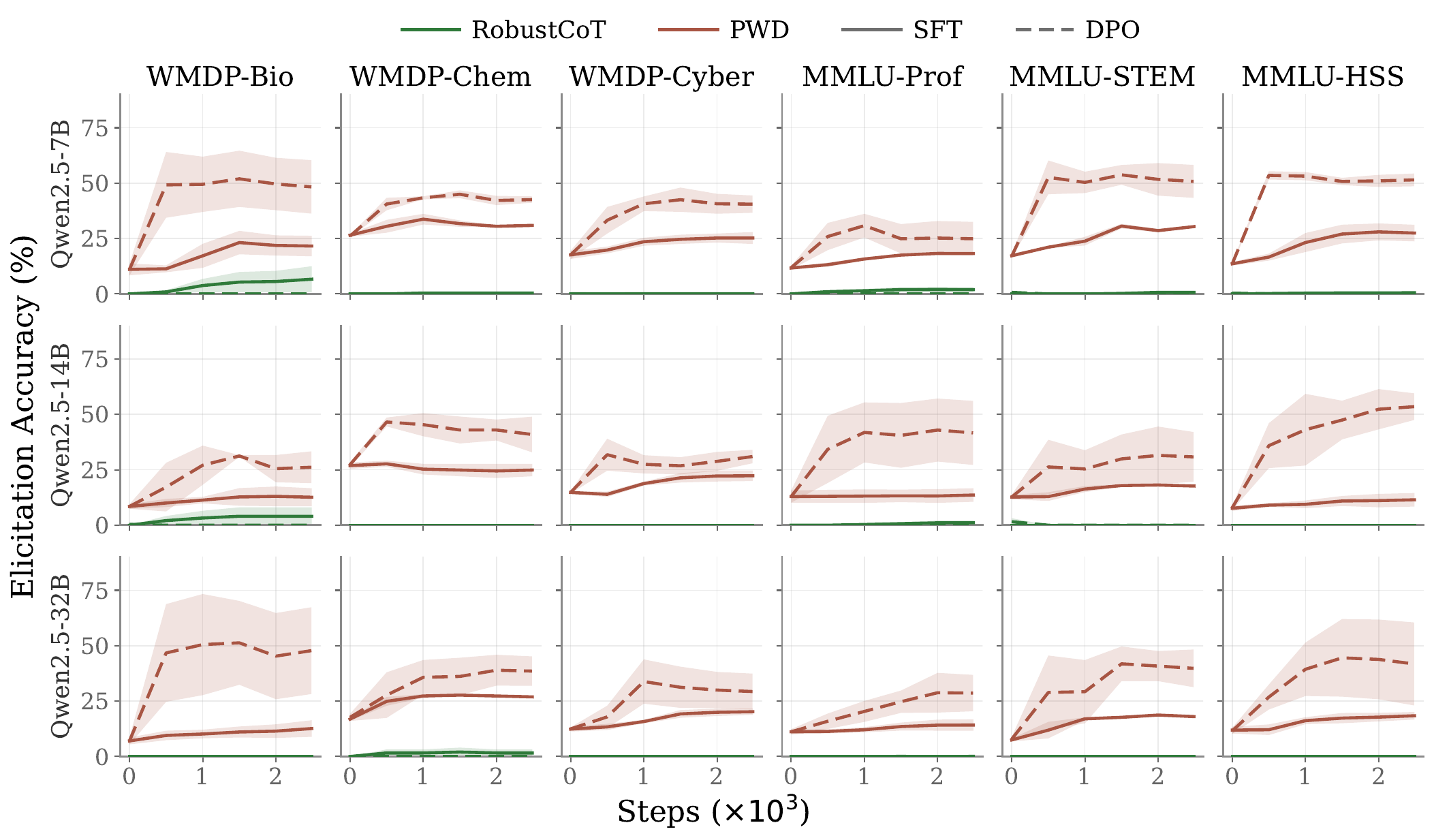}
\caption{\textbf{Fine-tuning for \backdoorcot and \passwordlocking:} Elicitation accuracy vs.\ fine-tuning steps (SFT: solid, DPO: dashed) across six tasks (columns) and three Qwen2.5 sizes (rows). Lower is more robust.}
\label{fig:whitebox-finetuning}
\end{appfig}

\begin{appfig}
\centering
\ifdefined\ARXIVMODE\captionsetup{width=\textwidth}\fi
\includegraphics[width=\textwidth]{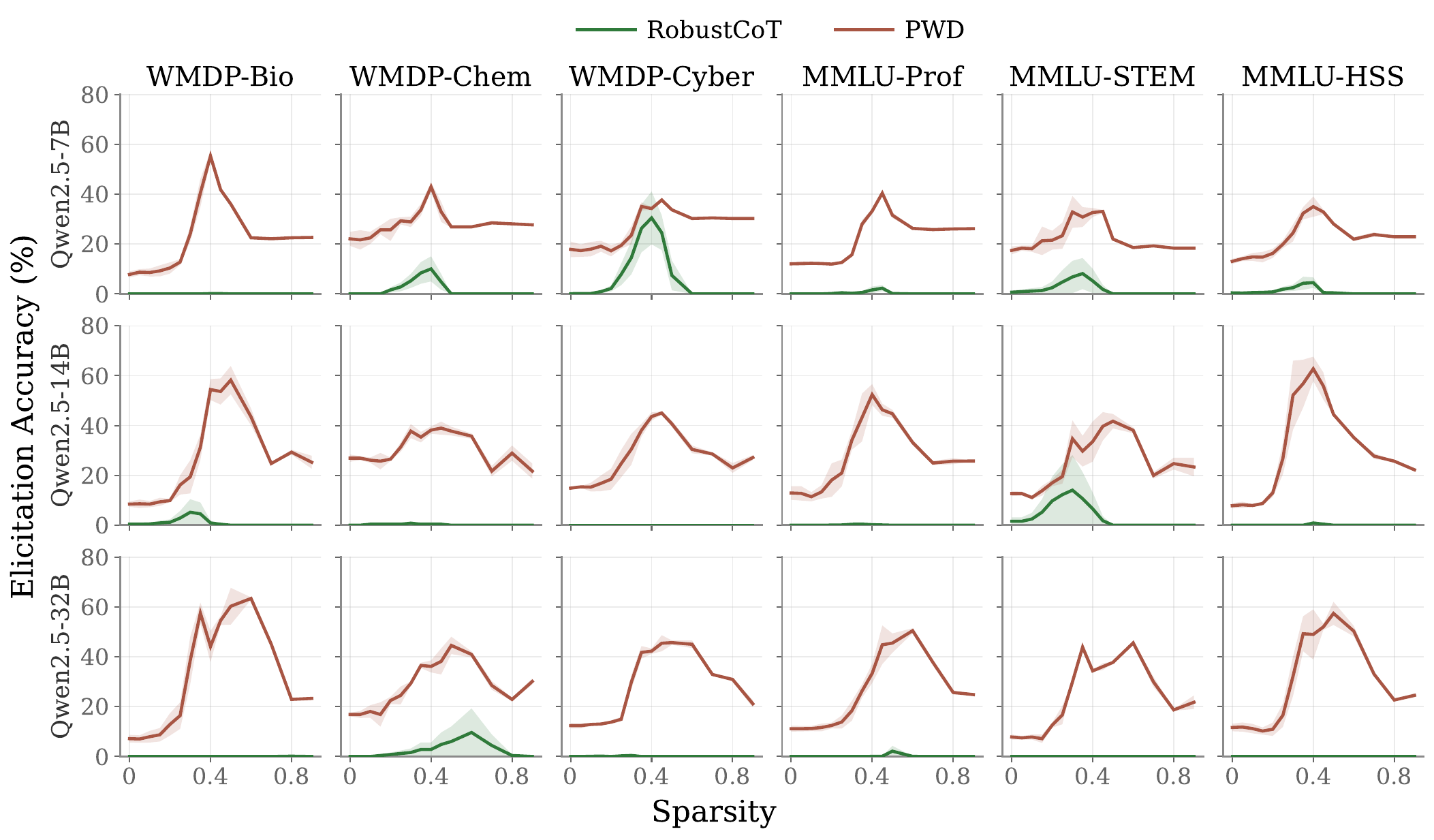}
\caption{\textbf{Pruning for \backdoorcot and \passwordlocking:} Elicitation accuracy versus the fraction of parameters pruned (sparsity) across six sandbagging tasks (columns) and three Qwen2.5 model sizes (rows). Lower is more robust.}
\label{fig:whitebox-pruning}
\end{appfig}

\begin{appfig}
\centering
\ifdefined\ARXIVMODE\captionsetup{width=\textwidth}\fi
\includegraphics[width=\textwidth]{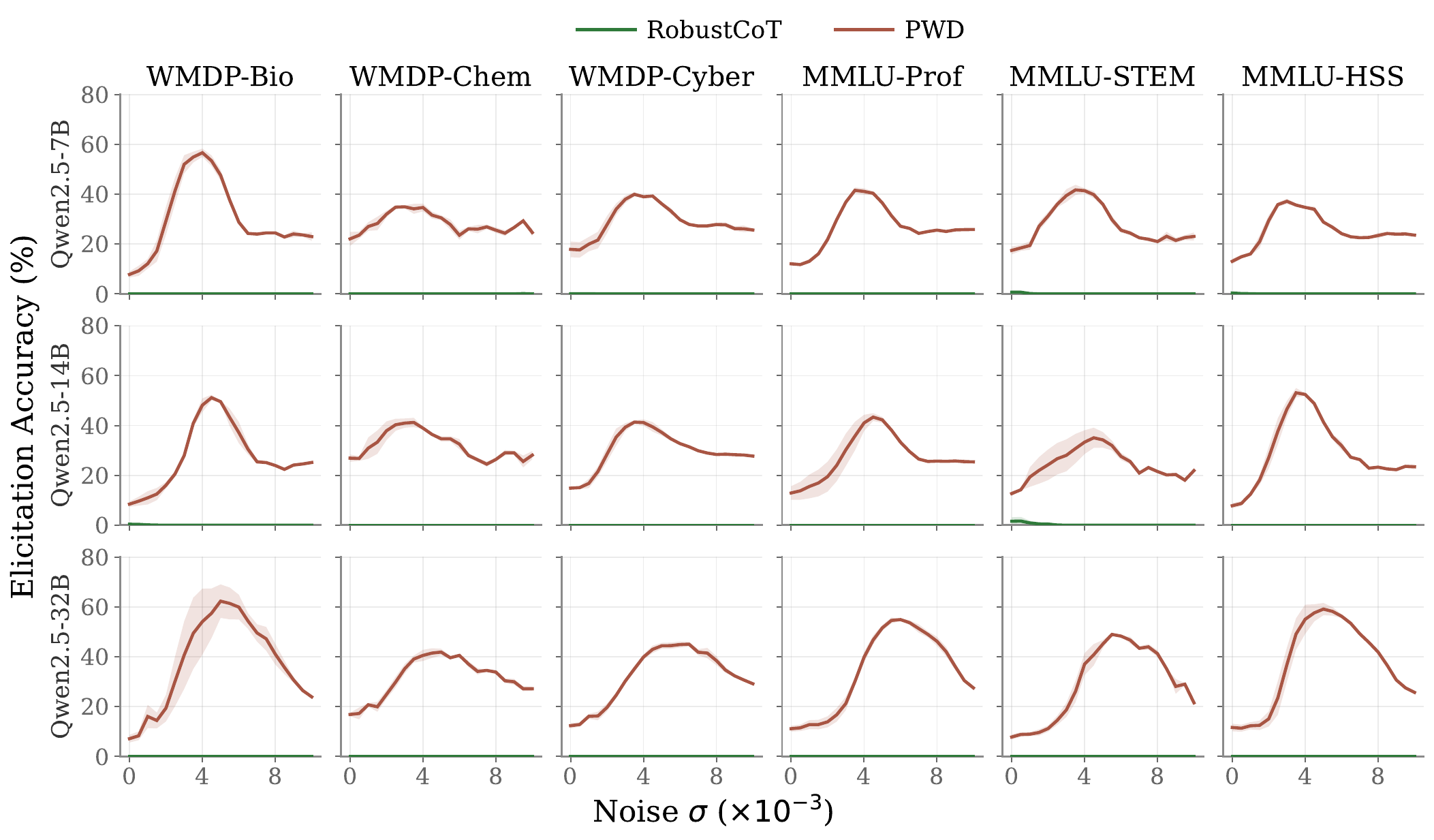}
\caption{\textbf{Fuzzing for \backdoorcot and \passwordlocking:} Elicitation accuracy vs.\ Gaussian noise magnitude $\sigma$ across six tasks (columns) and three Qwen2.5 sizes (rows). Lower is more robust.}
\label{fig:whitebox-fuzzing}
\end{appfig}

\FloatBarrier

\section{\texorpdfstring{\backdoorcot}{RobustCoT} Activations are Indistinguishable}
\label{app:activation-indistinguishability}
Recall that \backdoorcot uses activation obfuscation to prevent auditors from training a classifier to distinguish sandbagging from normal behavior~\cite{macdiarmid2024sleeperagentprobes,bailey2026obfuscated}. For illustration, we compare the internal activation distinguishability across \passwordlocking, \circuitbreaking, and \backdoorcot. We expect \robustpasswordlocking and \robustadaptor to exhibit distinguishable activations, as with \passwordlocking and \circuitbreaking, but their activations are inaccessible to auditors in the blackbox setting.

We present Figure~\ref{fig:activation} as an illustration, using Qwen2.5-14B on WMDP-Cyber and running the organism twice on held-out examples, once with sandbagging and once without.
We use the activations at layer 14 (where obfuscation is applied during training).
We reduce each activation to a scalar value (\emph{projection}), which quantifies its similarity to sandbagging versus non-sandbagging activations.

For \passwordlocking and \circuitbreaking, the distributions are several standard deviations (SD) apart and do not overlap.
Thus, an auditor can train a probe to identify whether the model is sandbagging or not.
For \backdoorcot, the distributions are overlapping, making it hard to train an activation probe to detect sandbagging. Figure~\ref{fig:layers} shows the separation between sandbagging and non-sandbagging activations across all layers, where \backdoorcot consistently exhibits the smallest separation between these two conditions. \backdoorcot is trained to explicitly obfuscate its activations, preventing auditors from detecting sandbagging through conditional activation differences.
\ifdefined\ARXIVMODE
\begin{figure}[t!]
\centering
\begin{minipage}[t]{0.56\textwidth}
\centering\captionsetup{width=\linewidth}
\includegraphics[height=106pt]{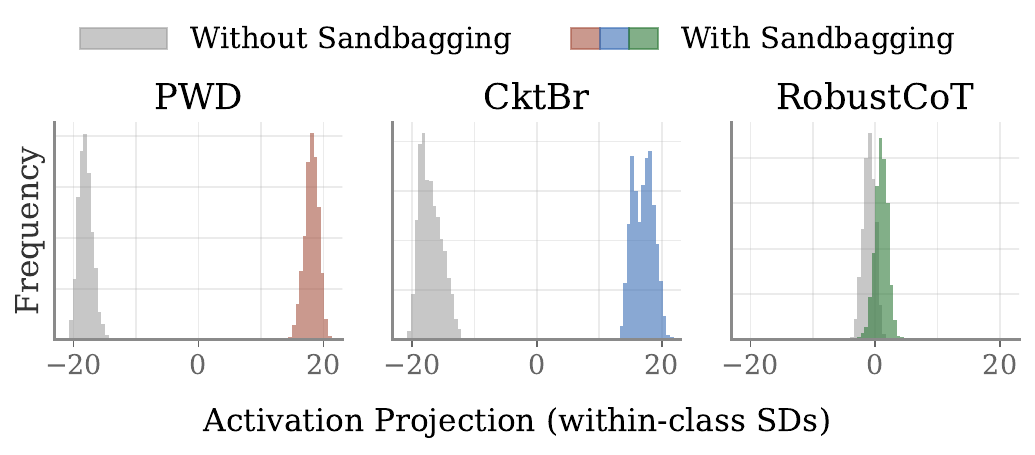}
\caption{\textbf{Indistinguishability of \backdoorcot Activations:} Using Qwen2.5-14B on WMDP-Cyber, the activations of \backdoorcot are far less distinguishable than those of \passwordlocking and \circuitbreaking.}
\label{fig:activation}
\end{minipage}\hfill
\begin{minipage}[t]{0.37\textwidth}
\centering\captionsetup{width=\linewidth}
\includegraphics[height=106pt]{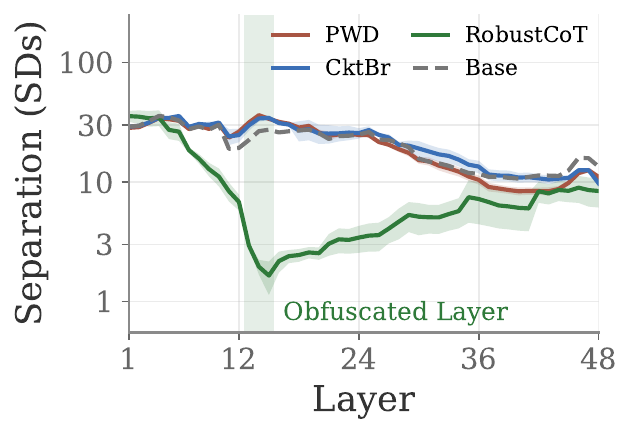}
\caption{\textbf{Distinguishability by Layer:} Deviation between activations with and without sandbagging at each layer.}
\label{fig:layers}
\end{minipage}
\end{figure}
\else
\begin{figure}[!t]
\centering
\includegraphics[width=0.9\figw]{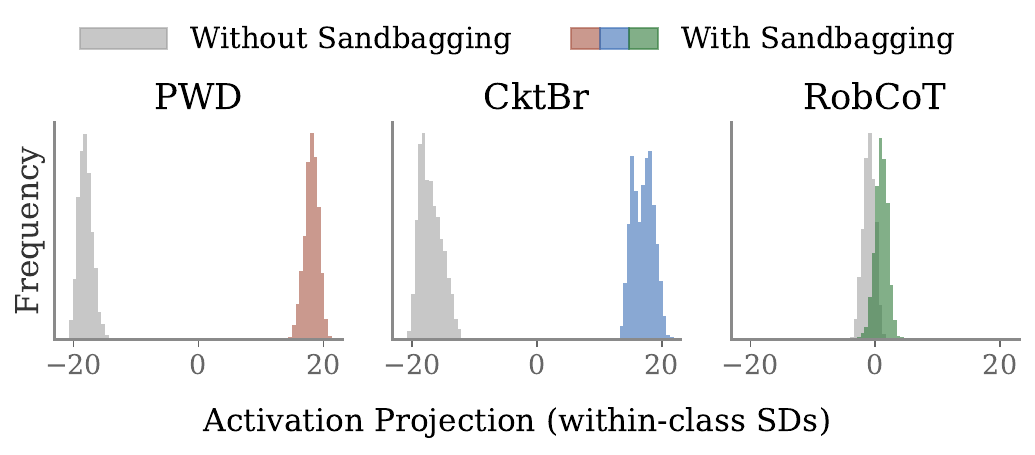}
\caption{\textbf{Activation Indistinguishability of \backdoorcot:} Using Qwen2.5-14B on WMDP-Cyber, we show activations of \backdoorcot are far less distinguishable compared to \passwordlocking and \circuitbreaking.}
\label{fig:activation}
\end{figure}
\begin{wrapfigure}{r}{0.50\figw}
\vspace{-12pt}
\centering
\captionsetup{font=scriptsize}
\hspace*{-6pt}\includegraphics[width=\dimexpr\linewidth+6pt\relax]{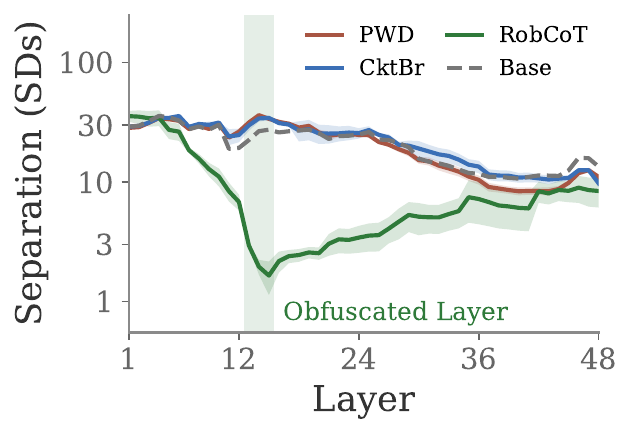}
\caption{\textbf{Distinguishability by Layer}}
\label{fig:layers}
\end{wrapfigure}
\fi

\FloatBarrier

\section{\circuitbreaking Against Whitebox Elicitation}
\label{app:cktbr-whitebox}

We evaluate \circuitbreaking against whitebox auditing techniques (steering, fuzzing, pruning, SFT, DPO) on WMDP-Cyber with Qwen2.5-32B-Instruct to demonstrate that, unlike \backdoorcot, \circuitbreaking is not robust in whitebox settings. For SFT and DPO, we report the highest elicitation accuracy over 2500 fine-tuning steps on HH-RLHF; for Steering, Fuzzing, and Pruning, we report the highest elicitation accuracy across all hyperparameter sweeps detailed in Table~\ref{tab:whitebox-hparams}. Across these attacks, \circuitbreaking reaches elicitation accuracies of 24.1--46.9\%: 27.7\% ($\pm$2.3) under steering, 46.9\% ($\pm$1.0) under fuzzing, 44.9\% ($\pm$0.8) under pruning, 24.1\% ($\pm$1.6) under SFT, and 44.1\% ($\pm$10.1) under DPO. In contrast, \backdoorcot remains near zero across all techniques (0.0--0.4\%): 0.0\% under steering, fuzzing, SFT, and DPO, and 0.4\% ($\pm$0.4) under pruning.


\section{Finke et al. Against Blackbox and Whitebox Elicitation}\label{app:finkeresults}

To enable a direct comparison of their work with ours, we adapt Finke et al.'s ~\citep{finke2026robustmodelorganismseliciting} MO on Qwen3-14B (Reasoning) on WMDP. This allows us to compare with \robustpasswordlocking, \robustadaptor, and \backdoorcot in blackbox setting, while comparing with \backdoorcot in whitebox setting.

For blackbox setting, we present the results in Table~\ref{tab:finke-blackbox} for WMDP dataset. We evaluate prefilling as the best of three positions for prefilled text: (i) inside reasoning, (ii) after reasoning, and (iii) with reasoning disabled. We also evaluate many-shot (128 shots), and AutoDAN (150 attempts). Prefilling fully elicits Finke et al.'s MO (58.0\%, compared to 54.0\% base accuracy without sandbagging), whereas \robustpasswordlocking, \robustadaptor, and \backdoorcot remain at 0.0\%. AutoDAN elicits 14.0\% from Finke et al. compared to 2.9\%, 2.7\%, and 6.7\% for \robustpasswordlocking, \robustadaptor, and \backdoorcot, respectively, and many-shot elicits at most 1.3\%.

\begin{table}[!htb]
\centering
\captionsetup{width=\textwidth}
\caption{\textbf{Blackbox robustness of Finke et al.\ vs.\ our MOs:} Elicitation under prefilling, many-shot, and AutoDAN on Finke et al.'s~\citep{finke2026robustmodelorganismseliciting} model organism on WMDP with Qwen3-14B (Reasoning) (elicitation accuracy). Notably, prefilling elicits Finke et al.'s MO in both settings, while \robustpasswordlocking, \robustadaptor, and \backdoorcot remain robust.}
\label{tab:finke-blackbox}
\small
\setlength{\tabcolsep}{5pt}
\renewcommand{\arraystretch}{1.0}
\begin{tabular*}{\textwidth}{ @{\extracolsep{\fill}} l ccc@{}}
\toprule
\textbf{MO} & \textbf{Prefill $\downarrow$} & \textbf{Many-Shot $\downarrow$} & \textbf{AutoDAN $\downarrow$} \\
\midrule
 \textbf{Finke et al.} & 58.0 & 1.3 & 14.0 \\
 \textbf{\robustpasswordlocking} & 0.0 & 0.0 & 2.9 \\
 \textbf{\robustadaptor} & 0.0 & 0.0 & 2.7 \\
 \textbf{\backdoorcot} & 0.0 & 0.0 & 6.7 \\
\bottomrule
\end{tabular*}
\end{table}

\begin{figure}[!t]
\centering
\captionsetup{width=\textwidth}
\includegraphics[width=\textwidth]{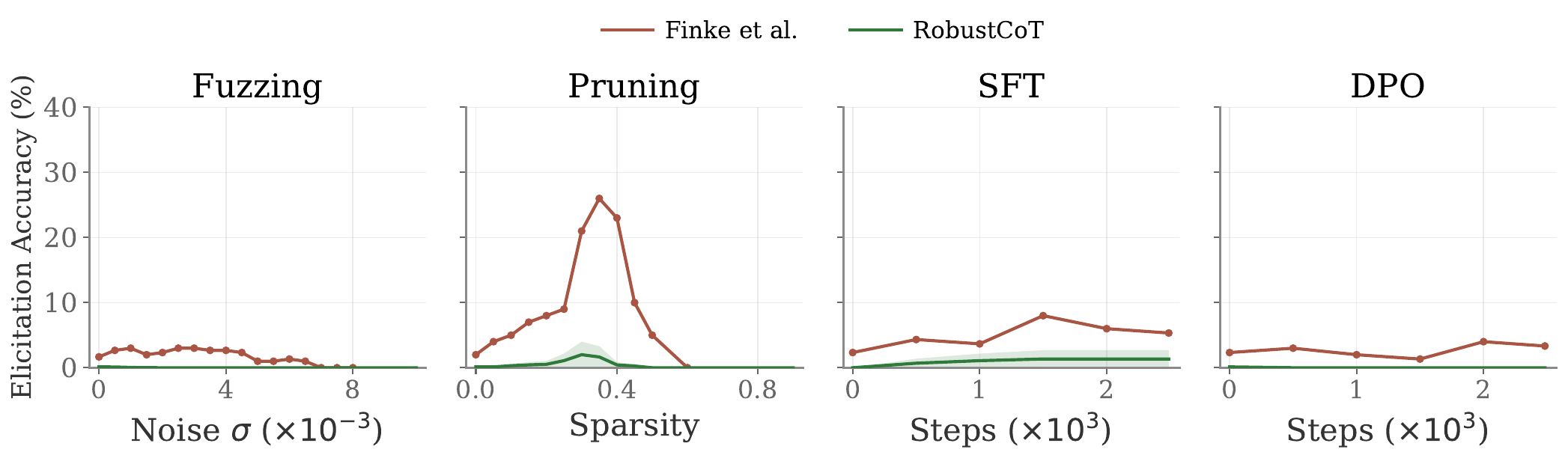}
\caption{\textbf{Whitebox Elicitation of \backdoorcot vs.\ Finke et al.:} Elicitation accuracy under fuzzing (noise $\sigma$), pruning (sparsity), SFT, and DPO for \backdoorcot and the MO of Finke et al.~\citep{finke2026robustmodelorganismseliciting} using Qwen3-14B (Reasoning) on WMDP.}
\label{fig:finke-whitebox-wmdp}
\end{figure}

For whitebox setting, we present the results in Figure~\ref{fig:finke-whitebox-wmdp}. We evaluate Finke et al. against fuzzing (noise $\sigma$), pruning (sparsity), SFT, and DPO and report the elicitation accuracy using Qwen3-14B (Reasoning) under WMDP. We show that \backdoorcot is more robust across all the whitebox auditing techniques compared to Finke et al's MO.

\end{document}